\documentclass{emulateapj}

\shorttitle{Infrared Observation toward SNR IC 443}
\shortauthors{Shinn et al.}

\begin{document}

\newcommand{\cu}{ph s$^{-1}$ cm$^{-2}$ sr$^{-1}$ \AA$^{-1}$}
\newcommand{\lu}{ph s$^{-1}$ cm$^{-2}$ sr$^{-1}$}
\newcommand{\luerg}{erg s$^{-1}$ cm$^{-2}$ sr$^{-1}$}
\newcommand{\ncm}{cm$^{-3}$}
\newcommand{\ncmK}{cm$^{-3}$ K}
\newcommand{\Ncm}{cm$^{-2}$}
\newcommand{\kms}{km s$^{-1}$}
\newcommand{\um}{$\mu$m}

\newcommand{\astH}[1]{H {\small #1}}
\newcommand{\Ha}{H${\alpha}$}
\newcommand{\Htwo}{H$_2$}
\newcommand{\HtwolineK}{H$_2$ $\upsilon=1\rightarrow0$ S(1)}
\newcommand{\HtwolineN}{H$_2$ $\upsilon=0\rightarrow0$ S(2)}
\newcommand{\astHe}[1]{He {\small #1}}
\newcommand{\astC}[1]{C {\small #1}}
\newcommand{\astN}[1]{N {\small #1}]}
\newcommand{\astO}[1]{O {\small #1}]}
\newcommand{\astAl}[1]{Al {\small #1}}
\newcommand{\astSi}[1]{Si {\small #1}}
\newcommand{\astSiext}[1]{Si {\small #1}$^*$}
\newcommand{\astFe}[1]{Fe {\small #1}}
\newcommand{\COline}{$^{12}$CO $J=2\rightarrow1$}

\newcommand{\ionH}[1]{H$^{#1}$}
\newcommand{\ionHthree}[1]{H$_3^{#1}$}
\newcommand{\ionC}[1]{C$^{#1}$}
\newcommand{\ionO}[1]{O$^{#1}$}
\newcommand{\ionAl}[1]{Al$^{#1}$}
\newcommand{\ionSi}[1]{Si$^{#1}$}

\newcommand{\dl}{$\lambda\lambda$}
\newcommand{\EBV}{$E(B-V)$}
\newcommand{\Rv}{$R_V$}
\newcommand{\Av}{$A_V$}
\newcommand{\Eion}{$E_{ion}$}
\newcommand{\Xco}{$X_{\textrm{\tiny{CO}}}$}
\newcommand{\Bzero}{$B_0$}
\newcommand{\Phis}{$\Phi_s$}
\newcommand{\vs}{$\upsilon_s$}

\newcommand{\nHone}{$n(\textrm{H I})$}
\newcommand{\nHtwo}{$n(\textrm{H}_2)$}
\newcommand{\lognHtwo}{log[$n(\textrm{H}_2)$]}
\newcommand{\nHtwoini}{$n_{\textrm{\tiny{H}}_2,\tiny{0}}$}
\newcommand{\nH}{$n_{\textrm{\tiny{H}}}$}
\newcommand{\defnH}{$n_{\textrm{\tiny{H}}}$=$n$(\astH{I})+2$n$(\Htwo)}
\newcommand{\nHe}{$n(\textrm{He})$}

\newcommand{\NHone}{$N(\textrm{H I})$}
\newcommand{\logNHone}{log[$N(\textrm{H I})$]}
\newcommand{\NHtwolte}{$N_{LTE}(\textrm{H}_2)$}
\newcommand{\NHtwo}{$N(\textrm{H}_2)$}
\newcommand{\logNHtwo}{log[$N(\textrm{H}_2)$]}
\newcommand{\NCO}{$N(\textrm{CO})$}
\newcommand{\NvJ}{$N(\upsilon,J)$}
\newcommand{\NH}{$N_{\textrm{\tiny{H}}}$}
\newcommand{\defNH}{$N_{\textrm{\tiny{H}}}$=$N$(\astH{I})+2$N$(\Htwo)}

\newcommand{\EvJ}{$E(\upsilon,J)$}
\newcommand{\IvJ}{$I(\upsilon,J \rightarrow \upsilon',J')$}
\newcommand{\AvJ}{$A(\upsilon,J \rightarrow \upsilon',J')$}
\newcommand{\vJ}{$\upsilon,J$}
\newcommand{\bb}{$b$}

\newcommand{\akari}{\textit{AKARI}}
\newcommand{\chandra}{\textit{Chandra}}
\newcommand{\iso}{\textit{ISO}}
\newcommand{\isofull}{\textit{Infrared Space Observatory}}
\newcommand{\spitzer}{\textit{Spitzer}}
\newcommand{\herschel}{\textit{Herschel}}

\title{\akari{} Near-infrared Spectral Observations of Shocked \Htwo{} Gas of the Supernova Remnant IC 443}

\author{Jong-Ho Shinn\altaffilmark{1}, Bon-Chul Koo\altaffilmark{2}, Kwang-Il Seon\altaffilmark{1}, Ho-Gyu Lee\altaffilmark{3}}

\email{jhshinn@kasi.re.kr}
\altaffiltext{1}{Korea Astronomy and Space Science Institute, 776 Daedeok-daero, Yuseong-gu, Daejeon, 305-348, Republic of Korea}
\altaffiltext{2}{Dept. of Physics and Astronomy, FPRD, Seoul National University, 599 Gwanangno, Gwanak-gu, Seoul, 151-747, Republic of Korea}
\altaffiltext{3}{Dept. of Astronomy and Astrophysics, University of Toronto, Toronto, ON M5S 3H4, Canada}

\begin{abstract}
We present near-infrared ($2.5-5.0$ \um) spectra of shocked \Htwo{} gas in the supernova remnant IC 443, obtained with the satellite \akari.
Three shocked clumps---known as B, C, and G---and one background region were observed, and only \Htwo{} emission lines were detected.
Except the clump B, the extinction-corrected level population shows the ortho-to-para ratio of $\sim3.0$.
From the level population of the clumps C and G---both \akari's \emph{only} and the one extended with previous mid-infrared observations---we found that the $\upsilon=0$ levels are more populated than the $\upsilon>0$ levels at a fixed level energy, which cannot be reproduced by any combination of \Htwo{} gas in Local Thermodynamic Equilibrium.
The populations are described by the two-density power-law thermal admixture model, revised to include the collisions with H atoms.
We attributed the lower (\nHtwo{}=$10^{2.8-3.8}$ \ncm) and higher (\nHtwo{}=$10^{5.4-5.8}$ \ncm) density gases to the shocked \Htwo{} gas behind C-type and J-type shocks, respectively, based on several arguments including the obtained high \astH{I} abundance \nHone/\nHtwo=0.01.
Under the hierarchical picture of molecular clouds, the C-type and J-type shocks likely propagate into ``clumps'' and ``clouds'' (interclump media), respectively.
The power-law index $b$ of 1.6 and 3.5, mainly determined by the lower density gas, is attributed to the shock-velocity diversity, which may be a natural result during shock-cloud interactions.
According to our results, \HtwolineK{} emission is mainly from J-shocks propagating into interclump media.
The \Htwo{} emission was also detected at the background region, and this diffuse \Htwo{} emission may originate from collisional process in addition to the ultraviolet photon pumping.
\end{abstract}


\section{Introduction} \label{intro}
IC 443 (G189.1+3.0) is an extensively studied supernova remnant (SNR) and famous for its interaction with nearby molecular clouds.
The interaction has been widely observed over diverse wavelengths: the $\gamma$-ray from hardronic collisions \citep{Esposito(1996)ApJ_461_820,Albert(2007)ApJ_664_L87,Acciari(2009)ApJ_698_L133,Abdo(2010)ApJ_712_459,Tavani(2010)ApJ_710_L151}, the X-ray absorption by foreground clouds \citep{Troja(2006)ApJ_649_258}, the infrared forbidden lines \citep{Burton(1988)MNRAS_231_617,Burton(1990)ApJ_355_197,Inoue(1993)PASJ_45_539,Richter(1995)ApJ_454_277,Cesarsky(1999)A&A_348_945,Rho(2001)ApJ_547_885,Rosado(2007)AJ_133_89}, the enhanced CO line ratio \citep{Seta(1998)ApJ_505_286,Xu(2011)ApJ_727_81}, the broad molecular emission lines \citep{White(1987)A&A_173_337,Wang(1992)ApJ_386_158,Dickman(1992)ApJ_400_203,vanDishoeck(1993)A&A_279_541,Snell(2005)ApJ_620_758,Zhang(2009)arxiv_11_4815}, the bow-like feature in position-velocity diagrams of molecular line \citep{Tauber(1994)ApJ_421_570}, and the OH maser line \citep{Claussen(1997)ApJ_489_143,Hoffman(2003)ApJ_583_272,Hewitt(2006)ApJ_652_1288}.
Therefore, IC 443 is usually observed when studying the shock-cloud interaction.

Its age ranges from $\sim3-4$ kyr \citep{Petre(1988)ApJ_335_215,Wang(1992)PASJ_44_303,Troja(2008)A&A_485_777} to $\sim20-30$ kyr \citep{Chevalier(1999)ApJ_511_798,Olbert(2001)ApJ_554_L205,Gaensler(2006)ApJ_648_1037,Bykov(2008)ApJ_676_1050,Lee(2008)AJ_135_796}, and its distance is thought to be about 1.5 kpc based on several arguments, such as the contact with Gem OB 1 association \citep{Poveda(1968)AJ_73_65}, the empirical $\Sigma-D$ relation \citep{Milne(1979)AuJPh_32_83,Caswell(1979)MNRAS_187_201}, the total remnant energy \citep{Fesen(1980)ApJ_242_1023,Fesen(1984)ApJ_281_658}, and the high-velocity absorption lines observed against background stars \citep{Welsh(2003)A&A_408_545}.
It extends $\sim45'$ (cf.~\citealt{Gaensler(2006)ApJ_648_1037}) and overlaps in the sky with another more extended SNR G189.6+3.3 \citep{Asaoka(1994)A&A_284_573}.
IC 443 consists of two half shells, and another large shell of the SNR G189.6+3.3 overlaps with the former two shells; these three shells were named as A, B, and C, respectively, by \cite{Braun(1986)A&A_164_193}.
The overall picture of the remnant region is well outlined in \cite{Troja(2006)ApJ_649_258} and \cite{Lee(2008)AJ_135_796}.
In the middle of the two shells A and B, there is a W-shaped ridge which shows strong \Htwo{} emission lines \citep[cf.~Fig.~\ref{fig-slit} and][]{Rho(2001)ApJ_547_885}; this ridge is thought to be a torus-type molecular clouds overrun by the SNR shock. 

Infrared \Htwo{} emission lines are useful to study shocked molecular clouds, since \Htwo{} is the most abundant molecule and its quantum levels cover a wide energy range enough to study shocked gas, whose temperature ranges from a few hundred to a few thousand kelvin.
Toward the W-shaped \Htwo{} ridge, infrared spectral observations have already been performed from ground \citep{Richter(1995)ApJ_454_277} and space by \isofull{} \citep[ISO,][]{Cesarsky(1999)A&A_348_945} and \spitzer{} \citep{Neufeld(2007)ApJ_664_890}.
However, the \Htwo{} emission lines which falls within $2.5-5.0$ \um{} have not been observed completely; the ground observations missed several lines because of the atmospheric absorption, and this wavelength range is not covered by \spitzer{} spectroscopy and was simply not observed by \isofull{} (\iso).
Besides, this wavelength range is worthwhile to observe because, for instance, we could obtain the population of high-$J$ $\upsilon=0$ levels, which is usually assumed to follow $\upsilon>0$ levels \citep[e.g.~][]{Rho(2001)ApJ_547_885,Giannini(2006)A&A_459_821}, but have not been thoroughly checked yet.

Here we present the results of near-infrared spectral observations over $2.5-5.0$ \um{} for the shocked \Htwo{} gas in the SNR IC 443.
The observations were performed with the satellite \akari{} (section \ref{obs-red}).
We detected many \Htwo{} emission lines toward shocked molecular gas (section \ref{ana-res-line}), and found that the population of the shocked \Htwo{} gas cannot be described by any combination of \Htwo{} gas in Local Thermodynamic Equilibrium (LTE), which have been usually used for (section \ref{ana-res-level} and \ref{ana-res-cmp}).
Instead, we employed a non-LTE \Htwo{} gas model (section \ref{ana-res-plmod}), and interpreted the results in terms of a shock combination (C-type and J-type) together with the diversity in shock velocities (section \ref{dis-plmod}).
The observed background emission, attributed to diffuse warm \Htwo{} gas, was discussed as well (section \ref{dis-bg}).

\section{Observations and Data Reduction} \label{obs-red}
The spectral observations were performed with the InfraRed Camera \citep[IRC,][]{Onaka(2007)PASJ_59_S401} onboard the Japanese satellite \akari{} \citep{Murakami(2007)PASJ_59_S369}, on 2008-Sep-26th and 27th during the post-Helium phase.
During this phase, only near-infrared observations were possible, since the cryogenic cooling with liquid Helium had been run out.
We used the $5''\times48''$ slit and the grism, whose resolving power and wavelength coverage are $\Delta\lambda\sim0.03$ \um{} and $2.5-5.0$ \um, respectively.
The observation mode, called as Astronomical Observation Template (AOT), is IRCZ4 which is designed for general spectroscopic observations.
It has an imaging observation sandwiched by spectroscopic observations of four frames (cf.~\citealt{Onaka(2009)mana}).
Comparing this reference image to the 2MASS catalog \citep{Skrutskie(2006)AJ_131_1163}, we corrected the default astrometry of the slit, given from the satellite attitude information.

We observed four regions, three of which are the shocked CO clumps and the rest is the background.
Figure \ref{fig-slit} shows the four slit positions over the 2MASS $K_s$ RGB image \citep{Skrutskie(2006)AJ_131_1163}, which displays a diffuse `W' feature that traces the \HtwolineK{} 2.12 \um{} line emissions \citep{Rho(2001)ApJ_547_885}.
We name three on-source positions as `B', `C', and `G' after the names of the shocked CO clumps \citep{Denoyer(1979)ApJ_232_L165,Huang(1986)ApJ_302_L63}, and the background as `BG.'
Table \ref{tbl-obs} summarizes our observations---the region name, RA-Dec position, observation ID, and AOT.

The data were reduced through the official pipeline, supported by the \akari{} team \citep[cf.~][]{Onaka(2009)mana}.
We used the new spectral response curve for the post-Helium phase data\footnote{http://www.ir.isas.jaxa.jp/AKARI/Observation/DataReduction/IRC/SpecResponse\_091113/}, which shows a degraded sensitivity, $\sim$70\% of the Helium phase sensitivity.
Columns of the two dimensional spectral images are occasionally saturated by the very bright sources in the imaging area of the detector, which cause the column pull-down effect.
This effect was corrected by masking out the relevant columns.
Hot pixels of the detector were also masked out.
During the data reduction, no smoothing and tilt-correction were applied to the two dimensional spectral images.

In order to extract spectra, we chose certain sections along the slit length, then averaged the pixel values within those sections.
The extraction sections were carefully chosen for the \Htwo{} emission lines to be nearly uniform within the sections.
The sections extend 10 pixels for the clumps B and C, 14 pixels for the clump G, 25 pixels for BG, where one pixel corresponds to $1.46''$ \citep{Onaka(2007)PASJ_59_S401}; for the data set 1420806-002 of BG, we only chose 5 pixels as an extraction section to avoid abnormal single pixel peak.
Figure \ref{fig-spec} displays the extracted spectra at each region.
For clarity, the error bars were omitted and the spectrum of BG was enlarged by a factor of 20.
The statistical error bars can be seen in Figures \ref{fig-fit-B}-\ref{fig-fit-bg}.

For the error estimation, we included the systematic error caused by the calibration source type, in addition to the statistical error.
Our target is a diffuse source, hence the flux calibration referred to standard point sources are not suitable for our source, since the aperture loss and the slit loss would vary with the source type; the official pipeline uses the calibration from point sources.
We considered this type of systematic error and adopted 10\% of the signal intensity as the systematic error (private communication with the \akari{} helpdesk).
It was squarely summed to the statistical error of the line intensity, $\sqrt{\sigma_{st}^2+\sigma_{sys}^2}$, after measuring the line intensities (cf.~section \ref{ana-res}).

\section{Analysis and Results} \label{ana-res}
\subsection{Line Identification and Intensity Measurement} \label{ana-res-line}
Figure \ref{fig-spec} shows many emission lines.
Since our concerns are the \Htwo{} emission lines, we first compared the wavelength of the observed emission lines with those of \Htwo{}.
For easier identification of single and blended lines, we made template spectra of \Htwo{} gas in LTE at diverse temperature from 1000 K to 4000 K.
This temperature range was adopted because the shocked \Htwo{} gas usually shows such a range of excitation temperatures at the upper levels of \EvJ{} $\sim5\times10^3-2.5\times10^4$ K \citep[e.g.~][]{Rosenthal(2000)A&A_356_705,Rho(2001)ApJ_547_885,Giannini(2006)A&A_459_821}, which includes the upper levels of the \Htwo{} emission lines we detected.
In this way, we identified all the detected emission lines as \Htwo{} emission lines.

Some lines are blended with nearby lines, hence we could not tag them as a single line.
The uniquely identified lines and the blended lines are indicated by `$\mid$' and `+', respectively, in Figure \ref{fig-spec}, and their line identifications are listed in Table \ref{tbl-result}.
For a cross-check, this identification was compared with that of the shocked \Htwo{} gas observed in the Orion Molecular Cloud-1 \citep[OMC-1,][]{Rosenthal(2000)A&A_356_705}, and our identification turned out to be reliable.
Some contribution from Br$\beta$ 2.63 \um{} can be blended with the 2.63 \um{} blended line (cf.~Fig.~\ref{fig-spec}); however, we think it is unlikely since Br$\alpha$ 4.05 \um, which should be stronger than Br$\beta$, was not detected; the spontaneous transition probabilities of Br$\alpha$ and Br$\beta$ are $A_{54}=2.7\times10^6$ s$^{-1}$ and $A_{64}=7.7\times10^5$ s$^{-1}$, respectively.
The features seen at the edge of the band ($<2.6$ and $>4.9$ \um) were ignored, since they are likely to be inadequate to analyze.
The 2.56 \um{} and 4.95 \um{} features seem to be the $\upsilon=1-0$ Q(9) blended with nearby lines and the $\upsilon=1-1$ S(9), respectively.

The line intensities were measured by fitting their line profiles with a continuum plus Gaussian whose full-width-at-half-maximum (FWHM) is fixed to the spectral resolution of the IRC ($\Delta\lambda=0.03$ \um).
Adjacent lines whose profiles are overlapped with each other were fitted simultaneously.
As a baseline continuum for the fitting, we used a median-smoothed spectrum of each region; the kernel width ranging from 0.20 to 0.54 \um{} was carefully chosen to be wide enough to erase out the line emission features.
The feature near the edge of the spectrum (e.g.~2.56 \um{} feature in Figure \ref{fig-fit-B}) remains unchanged after the median filtering, since the filter only works on those pixels that are away from the edge by more than a half of the filter width.
The fitted profiles are displayed in Figures \ref{fig-fit-B}--\ref{fig-fit-bg}, and their measured intensities are listed in Table \ref{tbl-result}.
The intensities of blended lines are shown with the symbol `$<$', since we cannot determine the individual contribution of each line blended.
The intensities of weak lines, whose signal-to-noise ratio are lower than 3.0, are expressed as 90\% confidence upper limits.

\subsection{Reddening Correction and \Htwo{} Level Population} \label{ana-res-level}
Over $2.5-5.0$ \um{} wavelengths, the extinction optical depth becomes greater than one with the hydrogen nuclei column density, \defNH, of $\gtrsim10^{22}$ \Ncm{} \citep{Draine(2003)ARA&A_41_241}.
Since the observed regions are pervaded with dense molecular gas, the measured intensity should be corrected for the reddening by the intervening interstellar dust.
We exploited the extinction curve of ``Milky Way, $R_V=3.1$'' \citep{Weingartner(2001)ApJ_548_296,Draine(2003)ARA&A_41_241}, and adopted proper hydrogen nuclei column densities for each region.
\Av=13.5 was adopted for the clumps C as \cite{Neufeld(2008)ApJ_678_974} did based on the results of \cite{Richter(1995)ApJ_454_277}.
We adopted the same \Av{} for the clump B, since its extinction is known to be similar with that of the clump C \citep{Burton(1988)MNRAS_231_617}.
No extinction measurement exist toward BG; thus, we assumed it to be the \Av{} of the clump C, the nearest clump from BG. 
\Av=10.8 was adopted for the clump G; this was inferred from $A_{2.12}=1.3$, obtained by \cite{Richter(1995)ApJ_454_277}, employing the ``Milky Way, \Rv=3.1'' extinction curve.
The corresponding \NH{} was calculated with the equation \Av=\NH/($1.87\times10^{21}$ \Ncm) for \Rv=3.1 \citep{Bohlin(1978)ApJ_224_132}.

The \Htwo{} level population was derived from these reddening corrected intensities, assuming that the \Htwo{} emission lines are optically thin.
Since the infrared \Htwo{} emission is emanating from electric quadrupole transition, it is optically thin under a typical interstellar medium condition; for instance, the pure-rotational S(0) and S(1) lines become optically thick at the line center when \NHtwo{} $>10^{24}$ \Ncm{}, adopting the line width of 10 \kms.
We derived the reddening corrected level population of \Htwo{} gas from the following equation,
\begin{equation}
N_{rc}(\upsilon,J)=\frac{4\pi\lambda}{hc}\frac{I_{rc}(\upsilon,J\rightarrow\upsilon',J')}{A(\upsilon,J\rightarrow\upsilon',J')},
\end{equation}
where $I_{rc}(\upsilon,J\rightarrow\upsilon',J')$ and $A(\upsilon,J\rightarrow\upsilon',J')$ are the reddening corrected line intensity and the Einstein-A radiative transition probability of the transition from level $(\upsilon,J)$ to $(\upsilon',J')$, respectively.
The molecular data for \Htwo{} were obtained from the database provided by a simulation code, CLOUDY (version C08.00; \citealt{Ferland(1998)PASP_110_761}).
The results are listed in Table \ref{tbl-h2col} and their population diagrams are displayed in Figure \ref{fig-pop}.
$g_J$ is a weight factor which corresponds to $(2J+1)$ and $3(2J+1)$ for para (even $J$) and ortho (odd $J$) states, respectively; the population of LTE gas is appeared as a straight line in this diagram.

In Figure \ref{fig-pop}, the clumps B, C, and G show the population of $\upsilon=0,1,2$, while BG shows that of ($\upsilon,J$)=(0,11) only.
In the clumps B, C, and G, the $\upsilon=0$ population shows a similar shape and little zigzag pattern; when the ortho-to-para ratio is approaching to 3.0, the zigzag pattern disappears \citep[cf.~][]{Neufeld(2006)ApJ_649_816,Neufeld(2007)ApJ_664_890}.
The $\upsilon=1$ population, however, shows a similar shape and little zigzag pattern only in the clumps C and G; the clump B shows an evident zigzag pattern over the population of ($\upsilon,J$)=(1,1), (1,2), and (1,3).
Here we note that the $\upsilon=0$ and $\upsilon=1$ levels follow different branches.
This becomes clearer when plotted with the lower-$J$ $\upsilon=0$ population obtained from mid-infrared observations (see the following section).
BG show a much smaller but evident population of the ($\upsilon,J$)=(0,11) level.
It is about a factor of $12-38$ smaller than the clumps B, C, and G (cf.~Table \ref{tbl-h2col}).

\subsection{Comparison of \Htwo{} Populations with Previous Observations: the Clumps C and G} \label{ana-res-cmp}
As far as we know, the UKIRT CGS4 observation of \cite{Richter(1995)ApJ_454_277} is the only published near-infrared spectroscopic observation that targeted on the shocked \Htwo{} gas in IC 443; they observed the clumps C and G.
Besides, toward these two clumps, there are published results of mid-infrared spectroscopic observations for the shocked \Htwo{} gas, performed with \isofull{} \citep{Cesarsky(1999)A&A_348_945} and \spitzer{} \citep{Neufeld(2007)ApJ_664_890}.
For the clump B, mid-infrared spectroscopic observations were also performed with \spitzer{} \citep{Noriega-Crespo(2009)inproc}, however, they are under analysis and only covered three emission lines, $\upsilon=0-0$ S(0), S(1), and S(2).
Therefore, we concentrated on the data of the clumps C and G, and compared our \akari{} results with the ones from previous studies.

Before the comparison of the results, we compared the spectra extraction areas of \akari's and other's, to check whether we are comparing the same radiation sources.
The yellow circle and boxes in Figure \ref{fig-specext} indicate the extraction areas of the clumps C and G.
The details for the extraction area are given in the figure caption.
For the clump C, the centers of the extraction areas falls within $\sim20''$.
Especially, that of \spitzer{} data \citep{Neufeld(2007)ApJ_664_890}---the center of a Gaussian taper---well falls into the extraction area of the \akari{} data; hence, it is likely that the \Htwo{} emissions, observed by \akari{} and \spitzer{}, represent almost the same gas, unless the average physical property drastically varies over a few arcsec scale ($<0.07$ pc).
For the clump G, the centers of the extraction areas falls within about $30''$.
In this case, the extraction areas of \akari{} and \iso{} \citep{Cesarsky(1999)A&A_348_945} data overlap about 40 \% with a $\sim10''$ ($\sim0.07$ pc) separation.
Thus, as in the clump C, the \Htwo{} emissions, observed by \akari{} and \iso{}, likely represent almost the same gas, unless the average physical property drastically varies over a $\sim0.07$ pc scale.

Prior to the comparison of level population from different observations, we here note that the same extinction-correction curve and \Av{} values, used in section \ref{ana-res-level}, were used for the consistent comparison.
We first compared our results with those of mid-infrared observations \citep{Cesarsky(1999)A&A_348_945,Neufeld(2007)ApJ_664_890}.
The \emph{top} panels of Figure \ref{fig-popall} shows the comparison.
We adopted 15\% systematic calibration error, which is a dominant component, for both clumps, C and G \citep{Cesarsky(1999)A&A_348_945,Neufeld(2007)ApJ_664_890}.
We here note that, in both clumps, $\upsilon=0$ levels are more populated than $\upsilon=1$ levels at a fixed level energy with a 3-$\sigma$ significance at least.
This means that the \Htwo{} level population cannot be reproduced by any combination of \Htwo{} gas in LTE, which is usually adopted for the description of shocked \Htwo{} gas (e.g.~\citealt{Rho(2001)ApJ_547_885,Giannini(2006)A&A_459_821}).
This invalidity was previously pointed out in the study on the shocked \Htwo{} gas of the supernova remnant HB 21 \citep{Shinn(2010)AdSpR_45_445}.

Secondly, we compared our results with those of ground near-infrared observations by \cite{Richter(1995)ApJ_454_277}.
The extinction-corrected level populations were derived from their ``averaged'' intensities of ``position 1'' and ``position 3,'' each for the clumps C and G, respectively.
The comparisons are plotted in the \emph{middle} panels of Figure \ref{fig-popall}.
As the figures show, there is an inconsistency between the same ($\upsilon,J$) levels of two observations.
We attribute this inconsistency to the flux calibration difference; the $\upsilon=0,1$ levels show almost constant, vertical gaps in the population diagram, about a factor of $3-4$ in the column density.
Our \akari{} calibration is likely to be correct, because the \akari{} $\upsilon=0$ level populations are seamlessly merged with those obtained from previous mid-infrared observations (cf. the \emph{top} panels of Fig.~\ref{fig-popall}).
However, we cannot rule out that the inconsistency is caused by the difference of observed regions, since the average properties of the shocked \Htwo{} gas may change over $\sim30''\sim0.2$ pc scale (see Fig.~\ref{fig-specext}).

Lastly, we note the importance of the space observations, fully covering $2.5-5.0$ \um, for the study of shocked \Htwo{} gas.
With ground observations, the gap between $\upsilon=0$ and 1 levels is hardly inspectable, since only a few $\upsilon=0$ levels can be observable at largely separated \EvJ{} (see the \emph{grey} plots in \emph{middle} panels of Figure \ref{fig-popall}).
Hence, when using the results of ground observations, we are likely to think that a combination of \Htwo{} gas in LTE looks viable to reproduce the level populations up to \EvJ{} $\sim25,000$ K (see the \emph{bottom} panels of Figure \ref{fig-popall}).
However, such a gap can be immediately inspectable from the space observations, fully covering $2.5-5.0$ \um, since $\upsilon=0,1$ levels are covered continuously; the \akari{} near-infrared observation is a good example.

\subsection{Power-law Thermal Admixture Model of \Htwo{} Gas} \label{ana-res-plmod}
As seen in the previous section, the combination of \Htwo{} gas in LTE cannot reproduce the observed level population of the shocked \Htwo{} gas in the clump C and G, obtained from \akari{} and previous mid-infrared observations.
Therefore, we applied the power-law thermal admixture model of \Htwo{} gas, which successfully reproduced the level population of shocked \Htwo{} gas before \citep{Neufeld(2008)ApJ_678_974,Shinn(2009)ApJ_693_1883,Neufeld(2009)ApJ_706_170,Shinn(2010)AdSpR_45_445,Lee(2010)ApJ_709_L74,Takami(2010)ApJ_720_155,Yuan(2011)ApJ_726_76}.
The model configuration was the same with that used in the study of the supernova remnant HB 21 \citep{Shinn(2009)ApJ_693_1883,Shinn(2010)AdSpR_45_445} and the young stellar object (YSO) L 1251A \citep{Lee(2010)ApJ_709_L74}, except the inclusion of the H atom as an additional collider.
Our \akari{} spectra include several ro-vibrational transition lines (e.g.~$\upsilon=1\rightarrow0, 2\rightarrow1$), which are sensitive to the collision with H atoms \citep{Neufeld(2008)ApJ_678_974,Shinn(2009)ApJ_693_1883,Shinn(2010)AdSpR_45_445}.
Thus, we updated the previous model to reflect the collisions with H atoms.

The \Htwo{} column density was calculated from the following equation,
\begin{eqnarray}
&dN=a T^{-b} dT, \\
&\textrm{where, }a=\frac{\textrm{\scriptsize N}(H_2; T>100 \textrm{ \scriptsize K})(b-1)}{T_{min}^{1-b}-T_{max}^{1-b}} \nonumber \\
&T_{min}=\textrm{100 K}, \, T_{max}=\textrm{4,000 K}& \nonumber
\end{eqnarray}
$a$ and $b$ are constants, and $\textrm{N}(H_2; T>100 \textrm{ K})$ is a total column density of molecular hydrogen warmer than 100 K.
At each temperature, the statistical equilibrium was assumed, and the collisional partners were \Htwo, He, and H.
The collisional deexciation rates were obtained from \cite{LeBourlot(1999)MNRAS_305_802} for \Htwo{} and He, and from \cite{Wrathmall(2007)MNRAS_382_133} for H.
For the \Htwo{} collider, newer rates were calculated by \cite{Lee(2008)ApJ_689_1105}; however, their results are similar with \citeauthor{LeBourlot(1999)MNRAS_305_802}'s over $100-6000$ K and only include for those levels of $J_{up}\leq8$.
Hence, we kept using \citeauthor{LeBourlot(1999)MNRAS_305_802}'s.
The collisional excitation rates were calculated from the detailed balance relation.
The \Htwo{} density, \nHtwo, and the relative abundance of H atom to \Htwo, $X_H\equiv\textrm{log}\left[\frac{n(\textrm{\tiny H I})}{n(\textrm{\tiny H}_2)}\right]$, were set as free parameters, while \nHe{} was assumed to be $0.2\times$\nHtwo.
The ortho-to-para ratio was set to 3.0, since the level populations show little zigzag pattern (cf.~the \emph{top} panels of Fig.~\ref{fig-popall}).

First, we tried to fit with a single \nHtwo{} value, but we failed.
No single \nHtwo{} model could successfully reproduce the population of high energy-levels.
This result reaffirms the previously-noted tendency that the single \nHtwo{} model does not reproduce the level population over \EvJ{} $=0-25,000$ K (cf.~section \ref{dis-plmod-cmp}).
Neither the addition of another $b$ was successful.
Instead, we added an additional \nHtwo{} component, and then we obtained successful results.
The fitting results are displayed in Figure \ref{fig-mfit}.
For the clump G, the fewer data points from mid-infrared observations caused weaker constraints for the fitting parameters than the clump C; therefore, we fixed the $X_H$ value as $-1.7$ in view of the fitting results for the clump C.
Then, we scanned the $\chi^2$ space for fitting parameters and the results are displayed in Figure \ref{fig-mconf}; the step size for the scan was 0.1 for all parameters displayed in Figure \ref{fig-mconf}.
The fit parameters with their 90\% confidence intervals, together with the reduced chi-square values, are listed in Table \ref{tbl-mfit}.

As seen from Figure \ref{fig-mfit}, the reddening-corrected level populations are well reproduced by the power-law thermal admixture model with two different \nHtwo{}s: one low \nHtwo{} $\sim10^3-10^4$ \ncm{} and the other high \nHtwo{} $\sim10^5-10^6$ \ncm{} (cf.~Table \ref{tbl-mfit}).
The \emph{lower and higher} density \Htwo{} gases mainly contribute to the \emph{lower and higher} upper-energy levels, respectively.
This indicates that some model parameters may not be uniquely determined, since only a small portion of the modeled population is constrained by the observed population.
This point is more discussed in section \ref{dis-plmod-cmp}.
The two kinds of \Htwo{} gases share a common power-law index $b$; it is 1.6 and 3.5 for the clumps C and G, respectively.
The column density ratios of the lower to higher density gas are not much different; they are 40 and 13 for the clumps C and G, respectively (see section \ref{dis-plmod-cmp} for more notes about the column densities, however).
We obtained $X_H=-1.7$, which corresponds to \nHone/\nHtwo=0.02; this value is similar with the one where the collision with H atoms starts to dominate the collision with \Htwo{} for the rovibrational transition lines, as mentioned by \cite{Neufeld(2008)ApJ_678_974} and \cite{Shinn(2009)ApJ_693_1883,Shinn(2010)AdSpR_45_445}, and smaller than those obtained for protostellar outflows of LDN 1157, \nHone/\nHtwo{} $=0.1-0.3$ \citep{Nisini(2010)ApJ_724_69}.

The $\chi^2$ contour for the column densities show a rough correlation for both clumps, C and G (cf.~Fig.~\ref{fig-mconf}).
This seems to be caused by the fact as follows.
If both column densities increase or decrease together, then the $\chi^2$ can be decreased by adjusting the power-law index $b$ shared by the lower and higher density \Htwo{} gases.
However, if one column density increases and the other decreases, or vice versa, then the $\chi^2$ cannot be decreased by the same way as before, since the total \Htwo{} population changes its shape, which is related with the power-law index $b$.
The variation of \nHtwo{} and \NHtwo{} are not effective as of $b$ in adjusting the fitting, as the confidence intervals indicate (Table \ref{tbl-mfit}).
For the clump C, almost no correlation is shown between the densities and between $b$ and $X_H$.
For the clump G, the densities show a complex $\chi^2$ surface.
It may be caused by the absence of the lowest level data, ($\upsilon,J$)=(0,2) and (0,3), which imposes a weaker constraint for the fitting.

\section{Discussion} \label{dis}
\subsection{Shocked \Htwo{} Gas Described by the Power-law Thermal Admixture Model} \label{dis-plmod}
\subsubsection{Comparison Between Model Parameters from Previous and Our Studies} \label{dis-plmod-cmp}
The first attempt to describe the level population of shocked \Htwo{} gas with the power-law mixture of thermal \Htwo{} gas was by \cite{Brand(1988)ApJ_334_L103} (see also \citealt{Burton(1989)inproca}).
They assumed that the \Htwo{} gas, shocked by J-type shocks \citep{Draine(1993)ARA&A_31_373}, is in LTE and the postshock temperature profile is determined through the \Htwo{} radiative cooling, which is proportional to $T^{4.7}$.
In this way, they modeled the level population of thermally mixed \Htwo{} gas at the postshock region, which is equivalent to the power-law thermal admixture of \Htwo{} gas in LTE with the power-law index $b=4.7$ (cf. section \ref{ana-res-plmod}).
They applied this model to the observational data of OMC-1, and found that it is successful in describing the available data at that time.
However, the omission of important coolants---like CO, OH, H$_2$O, and grain---at dense environments as OMC-1 and of the magnetic field make the model assumption doubtful \citep{Chang(1991)ApJ_378_202,Draine(1993)ARA&A_31_373}.

Then, \cite{Oliva(1990)A&A_240_453} first tried to use the method of power-law thermal mixing as a phenomenological description tool for the \Htwo{} level population, with no background physics; they simply mixed \Htwo{} gas in LTE according to the power-law distribution, $dN(T)\sim T^{-b}dT$.
In this approach, they found that the \Htwo{} population of the SNR RCW 103, obtained from near-infrared observations, can be described with the models whose power-law indices are between $b=3.8$ and $b=4.7$.
This method was extended to the non-LTE case by \cite{Neufeld(2008)ApJ_678_974}.
Applying to the \spitzer{} IRAC data, they found that the shocked \Htwo{} gas of the supernova remnant IC 443 can be described with the power-law index $b$ of $\sim3-6$ and the molecular hydrogen density of \nHtwo{} $\sim10^6-10^7$ \ncm.
Afterwards, the same method was applied to the shocked \Htwo{} of the SNR HB 21 \citep{Shinn(2009)ApJ_693_1883,Shinn(2010)AdSpR_45_445} and of the outflows from YSOs \citep{Neufeld(2009)ApJ_706_170,Shinn(2010)AdSpR_45_445,Lee(2010)ApJ_709_L74,Takami(2010)ApJ_720_155,Nisini(2010)ApJ_724_69,Yuan(2011)ApJ_726_76}.
All of the non-LTE application results are summarized in Table \ref{tbl-nh2b}; we excluded the work of \cite{Nisini(2010)ApJ_724_69}, because they varied $T_{min}$ for the model application.

Table \ref{tbl-nh2b} shows that \nHtwo{} and $b$ vary with the estimated levels ($\upsilon=0,J$).
\nHtwo{} and $b$ are both tend to be smaller when the model applied to lower $J$ levels; \nHtwo{} shows a more drastic variation than $b$.
This tendency was already noted in \cite{Shinn(2010)AdSpR_45_445} through the model application to the data of the SNR HB 21 and OMC-1; now, the tendency is strengthened by additional data from other SNRs and outflows of YSOs.
This indicates that the level population of shocked \Htwo{} gas is not described by the power-law admixture model with \emph{single} \nHtwo{} and $b$.
Moreover, it suggests we should analyze the level population over as many levels as we can get; if not, we may have a limited picture on the shocked \Htwo{} gas.

In the point of extending the data coverage, our observations are important because we extended the $\upsilon=0$ level population obtained from previous mid-infrared studies, from $E(\upsilon,J)\lesssim7,000$ K up to $\lesssim22,000$ K (see Fig.~\ref{fig-popall}).
We could reproduce the estimated \Htwo{} population with two \nHtwo{} and one $b$ after including the H atom as an additional collision partner (Table \ref{tbl-mfit}).
The derived two \nHtwo{}s, \nHtwo=$10^{2.8-3.8}$ \ncm{} and \nHtwo=$10^{5.4-5.8}$ \ncm, fall into the range previously obtained, \nHtwo=$10^{2.7-7.0}$ \ncm{} (Table \ref{tbl-nh2b}).
As distinct from the previous applications (Table \ref{tbl-nh2b}), we included H atoms in the model, which efficiently excite high-$J$ $\upsilon=0$ levels; hence, we obtained smaller \nHtwo{} for the higher density component than we did without H atoms, by a factor of 0.8 dex and 1.7 dex for the clump C and G, respectively.

As mentioned in section \ref{ana-res-plmod}, the lower and higher density component mainly contribute to the lower and higher energy levels, respectively (Fig.~\ref{fig-mfit}).
This clearly show why we get different \nHtwo{}s depending on which levels are used for the estimation (cf. Table \ref{tbl-nh2b}).
This density property also indicates that there may be a non-uniqueness in the model parameter, because the observed population constrains only small portion of the modeled population (cf.~section \ref{ana-res-plmod}).
Indeed, the $T_{max}$ of the lower density gas and the $T_{min}$ of the higher density gas can be lower and higher, respectively, since small variations of those values may have a negligible effect on the fitting results.
We checked these possibilities by varying $T_{max}$ and $T_{min}$, with the obtained fitting parameters (Table \ref{tbl-mfit}).
In the case of clump C, we found that the $T_{max}$ of the lower density gas cannot be decreased from the initial model setting 4000 K, while the $T_{min}$ of the higher density gas can be relaxed up to 1000 K.
This result means that a small amount of warm, high density \Htwo{} gas can explain the high energy-level population, i.e., $N$(\Htwo; $T>1000$ K)$\sim4\times10^{18}$ \Ncm.
In the similar sense, the $X_H$ in the lower density gas can be lower, even down to the hydrogen-free case, \nHone/\nHtwo=0, since the excitation of the low-$J$ $\upsilon=0$ levels is dominantly determined by the collisions with \Htwo{} rather than \astH{I}.

\subsubsection{Nature of the Shocks} \label{dis-plmod-shock}
There have been many observational studies trying to identify the shock type at the clump C and G at diverse wavelengths---millimeter, sub-millimeter, and infrared.
Much of the observational results were compared with several J- and C-type shocks by \cite{Snell(2005)ApJ_620_758}, and they concluded that a combination of shocks (dissociative and non-dissociative) is required to explain all the observational results.
Later, \spitzer{} mid-infrared spectral observations showed the emission lines likely from such a combination of dissociative (J-type) and non-dissociative (C-type) shocks at the clump C \citep{Neufeld(2007)ApJ_664_890}.
Besides, previous studies on the shocked \Htwo{} gas of protostellar outflows which covered a similar \EvJ{} range with ours ($\sim0-25,000$ K) showed that a mixture of C- and J-type properties are required to explain the \Htwo{} level population \citep{Flower(2003)MNRAS_341_70,Giannini(2006)A&A_459_821,Gusdorf(2008)A&A_490_695}.

Under the shock-combination preference, the immediate question is which type of shocks the two component \Htwo{} gas originate from.
We first considered the lower density gas which occupies most mass of the shocked \Htwo{} gas (Fig.~\ref{fig-mfit} and Table \ref{tbl-mfit}).
If the lower density gas originates from J-type shocks, the recombinational lines of \astH{I} are also expected.
We checked the relative intensity of Br$\alpha$ 4.05 \um{} line and \Htwo{} $\upsilon=0-0$ S(3) line from a theoretical model \citep{Hollenbach(1989)ApJ_342_306}.
The ratio of Br$\alpha$/[\Htwo{} $\upsilon=0-0$ S(3)] is between $\sim1-10$ over the shock velocity $50-100$ \kms{} and preshock density $10^3-10^6$ \ncm.

The \Htwo{} density obtained from the model fitting likely represent the postshock density, since the \Htwo{} emissions are mainly emanating from the reformed \Htwo{} gas in J-type shocks \citep{Hollenbach(1989)ApJ_342_306}.
The typical J-type shock velocity is around 100 \kms{} under a general interstellar medium condition \citep{Draine(1993)ARA&A_31_373}, hence the density of the molecular reformation regions at far down stream would be $\gtrsim100$ times higher than the preshock density; this density ratio can be seen from many numerical studies \cite[e.g.~][]{Allen(2008)ApJS_178_20,Flower(2010)MNRAS_406_1745}.
Therefore, the preshock density should be \nHtwo{} $\lesssim10^{1.8}$ \ncm{} for the lower-density gas to be from J-type shocks.
This preshock density range is somewhat lower than considered in the above theoretical model ($10^3-10^6$ \ncm), but the similar ratio Br$\alpha$/[\Htwo{} $\upsilon=0-0$ S(3)] would be maintained since the postshock structure is insensitive to the preshock density in J-shocks.
Therefore, if the lower density gas is from J-shocks, then there should be Br$\alpha$ of $\sim2\times10^{-3}-2\times10^{-2}$ \luerg{}, based on the observed intensity of \Htwo{} $\upsilon=0-0$ S(3) \citep{Neufeld(2007)ApJ_664_890}.
However, no Br$\alpha$ line of such a high intensity was detected (Fig.~\ref{fig-spec}).

In addition, theoretically, a J-shock produce more abundant high-T ($\sim10^3$ K) \Htwo{} gas than a C-shock \citep{Wilgenbus(2000)A&A_356_1010,Flower(2010)MNRAS_406_1745}.
If the lower density gas is from J-shocks, the \Htwo{} population diagram must show a much up-turn curvature than we obtained (Fig.~\ref{fig-mfit}).
Overall, the lower density gas is not likely from J-type shocks.

We then checked whether the lower density gas originate from C-type shocks.
The postshock gas behind a planar C-type shock can be approximated as an isothermal gas whose density \nHtwo{} and temperature $T_s$ are as follows \citep{Neufeld(2006)ApJ_649_816},
\begin{eqnarray}
\textrm{\nHtwo}=1.5\,n_0 \\
T_s=375\,b_B^{-0.36}\,\upsilon_{s6}^{1.35} \textrm{ K,}
\end{eqnarray}
where $10\,\upsilon_{s6}$ \kms{} is the shock velocity and $b_B(n_0/\textrm{\ncm})^{1/2}$ $\mu$G is the assumed preshock magnetic field for a shock propagating in material of preshock \Htwo{} density $n_0$.
We here note the above approximation was tested over $n_0=10^4-10^6$ \ncm{} \citep{Neufeld(2006)ApJ_649_816}.
From this approximation, we can say the lower density gas is from C-shocks propagating into a preshock medium \nHtwo{} $\sim10^{2.6-3.6}$ \ncm.
Interestingly, this density is similar with the typical density of molecular clouds ($\gtrsim10^3$ \ncm), where the hydrogen dominantly exists in the molecular form \Htwo{} \citep{Snow(2006)ARA&A_44_367}.
Such a similarity gives a natural way to explain the mass dominance of the lower density gas (Fig.~\ref{fig-mfit}), since \Htwo{} survives C-type shocks.
Therefore, C-type shocks are more suitable to explain the lower density gas than J-type shocks.

If the lower density gas is from C-type shocks, the power-law index $b$ may reflect the multiple shocks of different velocities.
According to the approximation with $b_B=1$, it ranges from 3 to 58 \kms, which corresponds to $\sim100-4000$ K gas.
This velocity range is theoretically expected for a C-type shock whose preshock density is $\sim10^3$ \ncm{} and $b_B=1$ \citep{LeBourlot(2002)MNRAS_332_985}.
Also, such a diversity in the shock velocity is consistent with the results of \cite{Hewitt(2009)ApJ_694_1266} who found two C-shocks are preferred to explain the excitation of shocked \Htwo{} gas around SNRs over the range \EvJ{} $\sim0-8,000$ K.
However, we here note that this shock velocity range depends on the model parameters we fixed ($T_{min},T_{max}$); therefore, the shock velocity range mentioned above ($3-58$ \kms) is not observationally determined one.
Instead, it should be understood that multiple shocks of a few to a few tens of kilometer per second are required to explain the level population of the lower density gas.

The diversity in the shock velocity may be a natural result during the shock-cloud interaction, since the shock driving pressure would be higher at the head than the side.
This kind of property can be inferred from a numerical simulation for radiative clouds \citep{Nakamura(2006)ApJS_164_477}, which showed that the velocity dispersion of the shocked cloud is larger along the shock-normal (the blast wave direction; $z$-axis in \citealt{Nakamura(2006)ApJS_164_477}) than the shock-tangential ($\varpi$-axis in \citealt{Nakamura(2006)ApJS_164_477}); this velocity dispersion does not exactly corresponds to the dispersion of the shock velocity in the cloud, however, it shows the tendency which direction the shock would propagate faster and cause the larger velocity dispersion.

We assigned the same \astH{I} abundance both for the lower and higher density gas, and obtained $X_H=-1.7$, equivalent to \nHone/\nH{} $=0.01$ and \nHone/\nHtwo{} $=0.02$ (Table \ref{tbl-mfit}).
This value is higher for typical molecular clouds \citep[\nH{} $\gtrsim10^4$ \ncm, ][]{Draine(1983)ApJ_264_485}.
However, as mentioned in section \ref{dis-plmod-cmp}, we can achieve the population of the lower density gas even with zero \astH{I} abundance, since the population is dominantly determined by the collisions with \Htwo{} rather than \astH{I}.
Considering this fact, the C-shock interpretation for the lower density gas has no contradiction with the \astH{I} abundance.

Now we move to discuss which type of shock the higher density gas originate from, considering the \astH{I} abundance first.
The \astH{I} abundance, $X_H$, is mainly determined by the vertical gap between $\upsilon=0$ and $\upsilon>0$ levels in the population diagram, and the gap is dominantly described by the higher density gas (see Fig.~\ref{fig-mfit}).
The obtained $X_H=-1.7$, equivalent to \nHone/\nH{} $=0.01$ and \nHone/\nHtwo{} $=0.02$, is expected for a typical diffuse cloud (\nH{} $\sim10^2$ \ncm), but high for a typical molecular cloud (\nH{} $\gtrsim10^4$ \ncm) which has $X_H\lesssim-4.0$ \citep{Draine(1983)ApJ_264_485}.
Therefore, the obtained $X_H$ must be from one of the following cases: (1) non-dissociative (C-type) shocks propagating into molecular clouds of the high $X_H$ value; (2) non-dissociative (C-type) shocks propagating into diffuse clouds; (3) dissociative (J-type) shocks propagating into molecular clouds.

We checked the first case.
In view of the calculations of \cite{Solomon(1971)ApJ_165_41}, \cite{Draine(1983)ApJ_264_485}, and \cite{Goldsmith(2005)ApJ_622_938}, the total ionization rate for \Htwo{} must be $\zeta\sim10^{-13}-10^{-12}$ s$^{-1}$ to make the obtained abundance in the higher density \Htwo{} gas of \nHtwo{} $\sim10^{5.0}$ \ncm.
However, the measured rate is much smaller than the required: $\zeta\sim5.5\times10^{-16}$ s$^{-1}$ \citep{Hewitt(2009)ApJ_706_L270} and $\zeta\sim2.6\times10^{-15}$ s$^{-1}$ \citep{Indriolo(2010)ApJ_724_1357}.
The second case is negative either.
It was shown that the postshock gas behind a planar C-type shock can be approximated with a isothermal gas whose density is equal to $\sim1.5\,n_{preshock}$ \citep{Neufeld(2006)ApJ_649_816}.
If the shock is C-type, the obtained density of \nHtwo{} $=10^{5.4-5.8}$ \ncm{} is contradict with the requirement that the preshock medium must be a diffuse cloud (\nH{} $\sim10^2$ \ncm).
In the third case, however, the high $X_H$ of $-1.7$ is expected under appropriate conditions; such an $X_H$ can be achieved at the $T\sim1000-4000$ K region where the \Htwo{} emission is mainly emanating from reformations \citep{Hollenbach(1989)ApJ_342_306,Flower(2010)MNRAS_406_1745}.

If the higher density gas is from J-type shocks where the \Htwo{} emission dominantly comes from the postshock reformation regions, the obtained model parameters for the higher density gas (Table \ref{tbl-mfit}) can be regarded to represent \emph{characteristic} physical parameters of the reformation regions.
Based on this, the preshock density should be a factor of $\lesssim100$ smaller than the obtained density of \nHtwo{} $=10^{5.4-5.8}$ \ncm{} since the typical velocity of J-shocks is around 100 \kms.
Also, there should exist shocked \astH{I} gas, the raw material for the reformation of \Htwo.
The shocked \astH{I} gas has already been observed along the ``W'' ridge (Fig.~\ref{fig-slit}) we are studying \citep{Braun(1986)A&A_164_193,Lee(2008)AJ_135_796}.
\cite{Lee(2007)phdth} measured the column densities of the shocked \astH{I} gas at the clump C and G, which are about $\sim3\times10^{21}$ \Ncm.
If we assume that the shocked \astH{I} spreads over along our line-of-sight as long as the width of the ``W'' ridge ($\sim2'\sim1$ pc), then the mean \astH{I} density of $\sim10^3$ \ncm{} is obtained, which is smaller than the postshock \Htwo{} density \nHtwo{} $=10^{5.4-5.8}$ \ncm.
These \nHone{} and \nHtwo{} are consistent with the J-shock interpretation, since the \astH{I} recombines before the \Htwo{} and the postshock density increases along the distance from the shockfront.
We also checked the line profile of \astH{I} 21cm emission \citep{Lee(2007)phdth} and \HtwolineK{} emission \citep{Rosado(2007)AJ_133_89} at the clump C, and found that they both show the shocked component at the similar velocity $\sim-30$ \kms.
Therefore, the J-shock interpretation for the higher density gas seems plausible again.

We then considered the meaning of the power-law index $b$ in the higher density gas.
As mentioned in section \ref{dis-plmod-cmp}, $T_{min}$ of the higher density gas can be increased from 100 K to 1000 K without making any difference on the final level population of \EvJ{} $\gtrsim10^4$ K levels.
This means that the gas in $T=1000-4000$ K is necessary to describe the observed population of \EvJ{} $\gtrsim10^4$ K levels.
We thus checked whether the $b$ variation makes any significant change to the level population of \EvJ{} $\gtrsim10^4$ K, summed over $T=1000-4000$ K; the obtained \nHtwo{} and $X_H$ was fixed while the $b$ value was varied.
We found that the $b$ variation makes a negligible effect over $b=1.0-4.0$.
This is because the population of \EvJ{} $\gtrsim10^4$ K levels has an almost same slope over the temperature integration range ($1000-4000$ K) in the power-law thermal model.
Therefore, the $b$ value has almost no practical meaning for the higher density gas.

From all the above discussion, the lower and higher density gases are likely from the C-type and J-type shocks, respectively.
This conclusion is consistent with the conclusion from previous studies which claimed a mixture of dissociative and non-dissociative shocks for the description of the observed results \citep{Snell(2005)ApJ_620_758,Neufeld(2007)ApJ_664_890}.
If we interpret our results based on the hierarchical picture of molecular clouds \citep{Bergin(2007)ARA&A_45_339,Williams(2000)2000prpl.conf__97}, the C-type shocks propagate into ``clumps,'' while the J-type shocks propagate into ``clouds'' (interclump media); the typical densities of these two constituents (``clouds'' and ``clumps'') are $50-500$ \ncm{} and $10^3-10^4$ \ncm, respectively.
This interpretation is consistent with the requirement that the ram pressure ($\sim\rho v_{s}^{2}$) should be similar between C-type and J-type shocks.
In the above discussion, we claimed that C-type shocks propagate into a preshock medium of $\sim10^3$ \ncm{} with a shock velocity of a few 10 \kms, while J-type shocks propagate into a preshock medium of $<10^3$ \ncm{} with a shock velocity of $\sim100$ \kms.

Before closing this section, we note that the \HtwolineK{} is dominantly originating from J-type shocks rather than C-type shocks, according to our results.
Interestingly, it was observed that the [\astSi{II}] 34.8 \um{} emission well follows the \HtwolineK{} emission at the clump C \citep{Richter(1995)ApJ_454_277,Neufeld(2007)ApJ_664_890}; the [\astSi{II}] 34.8 \um{} emission is an efficient cooling line at $T\lesssim5000$ K where the \Htwo{} reformation also happens \citep{Hollenbach(1989)ApJ_342_306}.
The intensity of [\astSi{II}] and \HtwolineK{} are $0.97$ and $1.8$ in $10^{-4}$ \luerg, respectively (\citealt{Neufeld(2007)ApJ_664_890}, Table \ref{tbl-h2col}).
The ratio [\astSi{II}]/\HtwolineK{} is 0.54, which is significantly lower than the theoretical expectation of a 100 \kms{} J-shock into $10^3$ \ncm{} gas, $\sim5.0$ \citep{Hollenbach(1989)ApJ_342_306}.
In spite of this difference, the J-shock origin of \HtwolineK{} emission is worth to check further whether it is general in shocked regions, considering that the shocked \Htwo{} gas usually shows a similar shape in the level population over \EvJ{}$=0-25,000$ K (cf.~section \ref{dis-plmod-coher}).

\subsubsection{Embedded Coherence in the Obtained Model Parameters} \label{dis-plmod-coher}
Table \ref{tbl-mfit} shows some coherences of the obtained model parameters for the two-component gases, although the clump G has no data for the population of \EvJ=(0,2) and (0,3) which are important in the determination of $b$ and density.
For example, the density ratio and the column density ratio of the two-component gases are not much different at the each clump, C and G.
This may originate from some common properties of the preshock medium.
One step forward, \cite{Richter(1995)ApJ_454_277,Richter(1995)ApJ_449_L83} already noted that the similarity between the \Htwo{} level population of SNRs and outflows of YSOs.

We can study such similarities with the two-\nHtwo{} power-law thermal admixture model by comparing the model parameters, and may lead out the properties of the preshock medium.
Especially, the power-law index $b$ would give information about how shock velocity is diverse at the shock-cloud interaction regions, which may be related with the geometrical structure of the clouds.
The coming James Webb Space Telescope which can perform spectral observations over the $\sim1-30$ \um{} wavelength range would provide an excellent opportunity for such studies.

\subsection{Warm Diffuse Background \Htwo{} Gas} \label{dis-bg}
At the background region (BG in Fig.~\ref{fig-slit}), we detected the \Htwo{} $\upsilon=0-0$ S(9) line (Fig.~\ref{fig-fit-bg}).
The extinction corrected column density gives log[N(\Htwo; $\upsilon,J$=0,11)]=$15.08\pm0.11$ \Ncm{} (Table \ref{tbl-h2col}).
This emission may be related with the \emph{diffuse background \Htwo{} gas}, previously observed towards SNRs \citep{Neufeld(2007)ApJ_664_890,Hewitt(2009)ApJ_694_1266}.
In order to check whether the obtained ($\upsilon,J$=0,11) population is plausible for such an diffuse background \Htwo{} gas, the population was compared with those obtained towards other Galactic lines of sight.
Since the relative shape of the level population determines the physical condition of the \Htwo{} gas, we must compare such an relative shape, thus we need the population of other levels in addition to ($\upsilon,J$=0,11) level.

For this, we employed the ($\upsilon,J$=0,2) population of the diffuse background \Htwo{} gas, previously observed towards the clump C of IC 443 \citep{Neufeld(2007)ApJ_664_890}, as the ($\upsilon,J$=0,2) level population of our BG region (Fig.~\ref{fig-slit}).
We adopted $I_{diff}=1.5\times10^{-6}$ \luerg{} as the intensity for the \Htwo{} $\upsilon=0-0$ S(0) line \cite[cf.~Fig.~19 of][]{Neufeld(2007)ApJ_664_890}, and corrected the extinction with $A_V=13.5$ employing the ``Milky Way, \Rv=3.1'' extinction curve \citep{Weingartner(2001)ApJ_548_296,Draine(2003)ARA&A_41_241}.
The population is log[N(\Htwo; $\upsilon,J$=0,2)]=$19.05\pm0.07$ \Ncm.
After all, the weighted populations of the ($\upsilon,J$)=(0,2) and (0,11) levels for the BG region are log[N($\upsilon,J$)/g$_J$]=18.3 and 13.2, respectively.

We compared the populations of these two levels with those of ($\upsilon,J$)=($0,\le8$) obtained toward the ``translucent lines of sight'' to the Galactic background stars, from the far-ultraviolet observations of \Htwo{} absorption lines \citep{Jensen(2010)ApJ_711_1236}.
Since the population of ($\upsilon,J$)=($0,11$) is absent in the results of \cite{Jensen(2010)ApJ_711_1236}, the relative curvature in the population diagram was compared, by scaling up or down the populations of ($\upsilon,J$)=(0,2) and (0,11) levels.
From the comparison, we found that our populations make a more upturn curvature than expected from the excitation temperature of $J\ge2$ levels, $\sim200-550$ K \citep{Jensen(2010)ApJ_711_1236}.
This means the existence of \Htwo{} gas with higher temperature than 550 K, which is consistent with the suggestion of \cite{Jensen(2010)ApJ_711_1236}: more than two temperature components are probably necessary to describe the population.

\cite{Jensen(2010)ApJ_711_1236} tried to fit the population obtained toward two sample targets, with a model for photodissociation regions; however, the fits were poor.
Such a mismatch was also recognized by \cite{Gry(2002)A&A_391_675} and \cite{Falgarone(2005)A&A_433_997}, and they suggested that another collisional excitation mechanisms are required to explain the observed population, such as magnetohydrodynamic shocks or intermittent dissipation of turbulence.
The diffuse background \Htwo{} gas towards the SNR IC 443 may originate from these mechanisms.

\section{Conclusions} \label{concl}
\begin{figure}
\center{
\includegraphics[scale=0.5,angle=270]{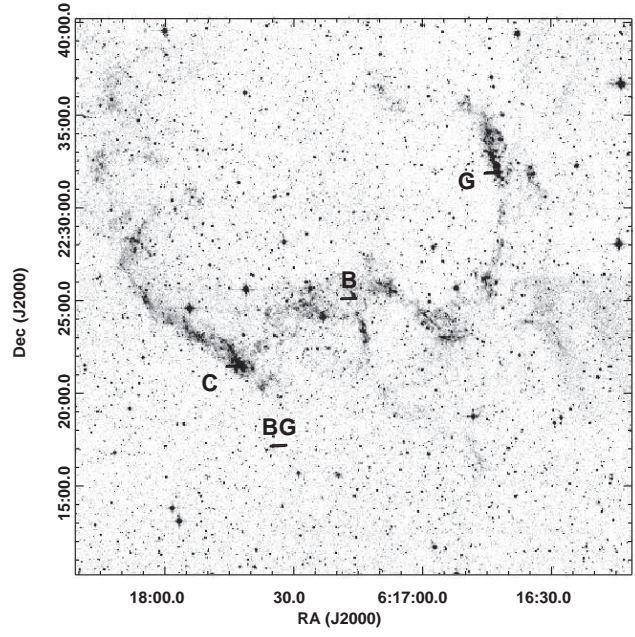}
}
\caption{The observed slit positions. Slit positions are indicated on the 2MASS $K_s$ image of the SNR IC 443 \citep{Skrutskie(2006)AJ_131_1163}. Slit positions are named after their representative clump names---`B', `C', and `G.' `BG' means the background. The red `W' ridge is dominated by \Htwo{} emissions.} \label{fig-slit}
\end{figure}

\begin{figure}
\center{
\includegraphics[scale=0.8]{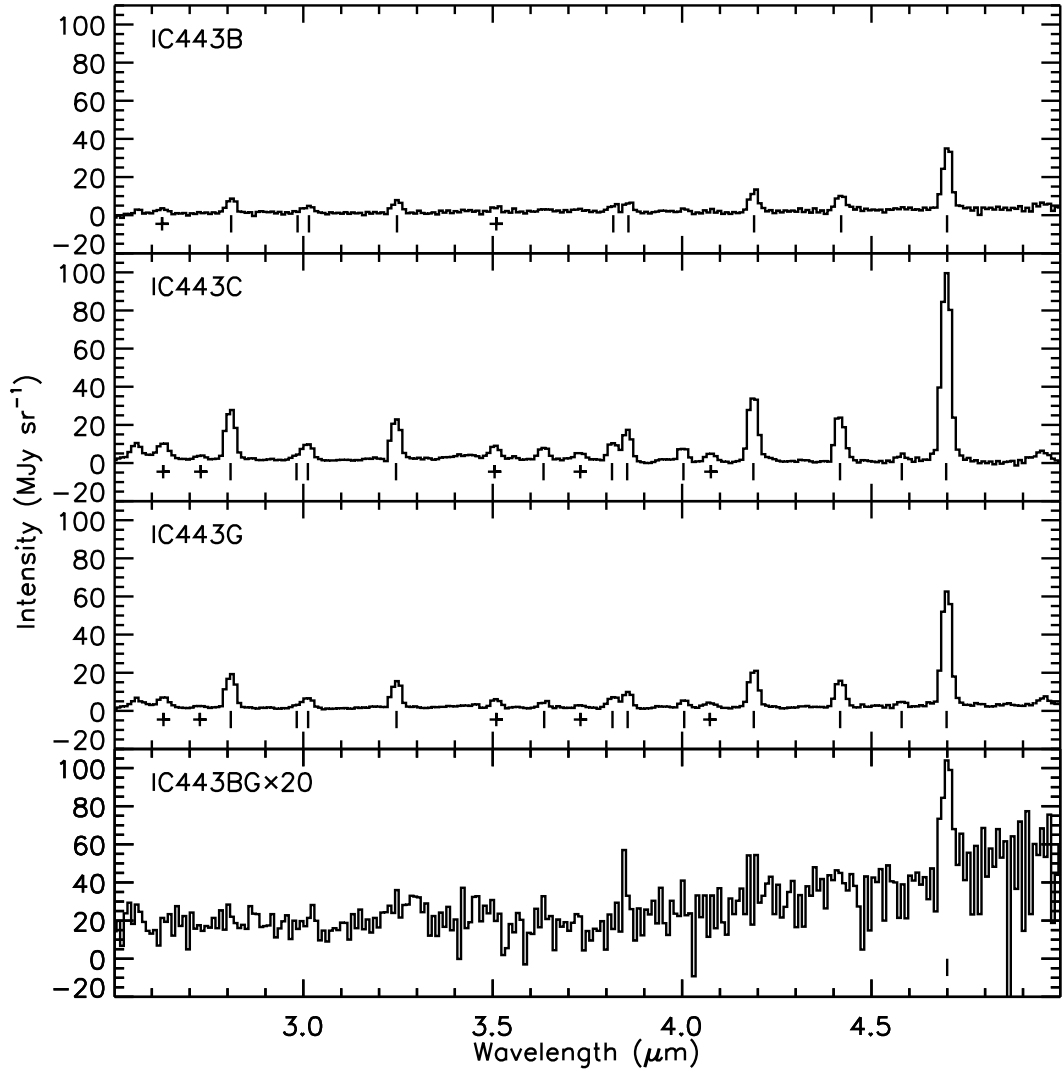}
}
\caption{The \akari{} IRC near-infrared spectra. Singular and blended lines are indicated with vertical bars and crosses, respectively. These lines are all the \Htwo{} emission lines (cf. Table~\ref{tbl-result}). In the case of `BG', a twenty-times enlarged spectrum is plotted for the identification of the $\upsilon=0-0$ S(9) 4.69 $\mu$m line.} \label{fig-spec}
\end{figure}

\begin{figure}
\center{
\includegraphics[scale=0.33]{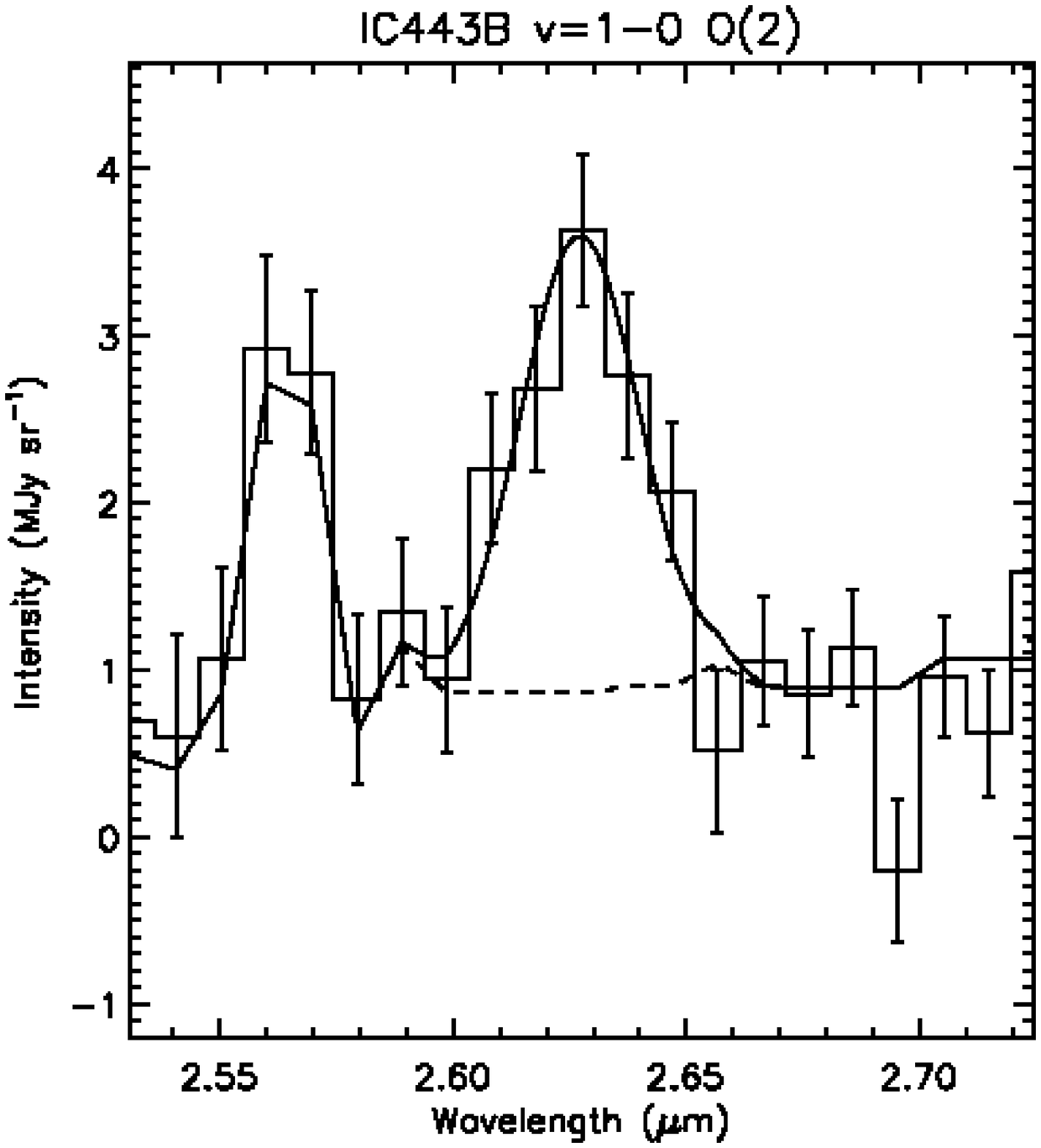}
\includegraphics[scale=0.33]{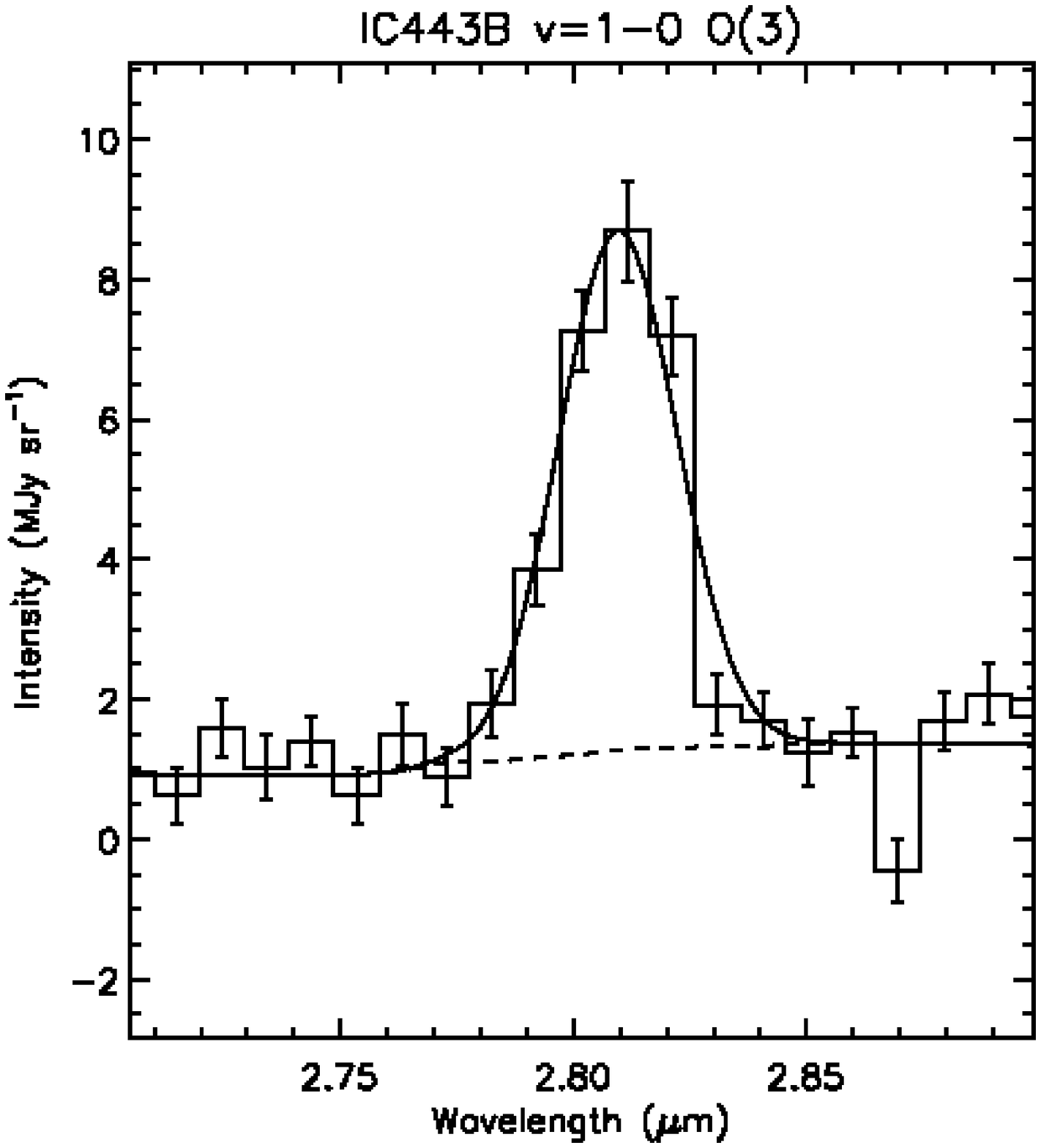}
\includegraphics[scale=0.33]{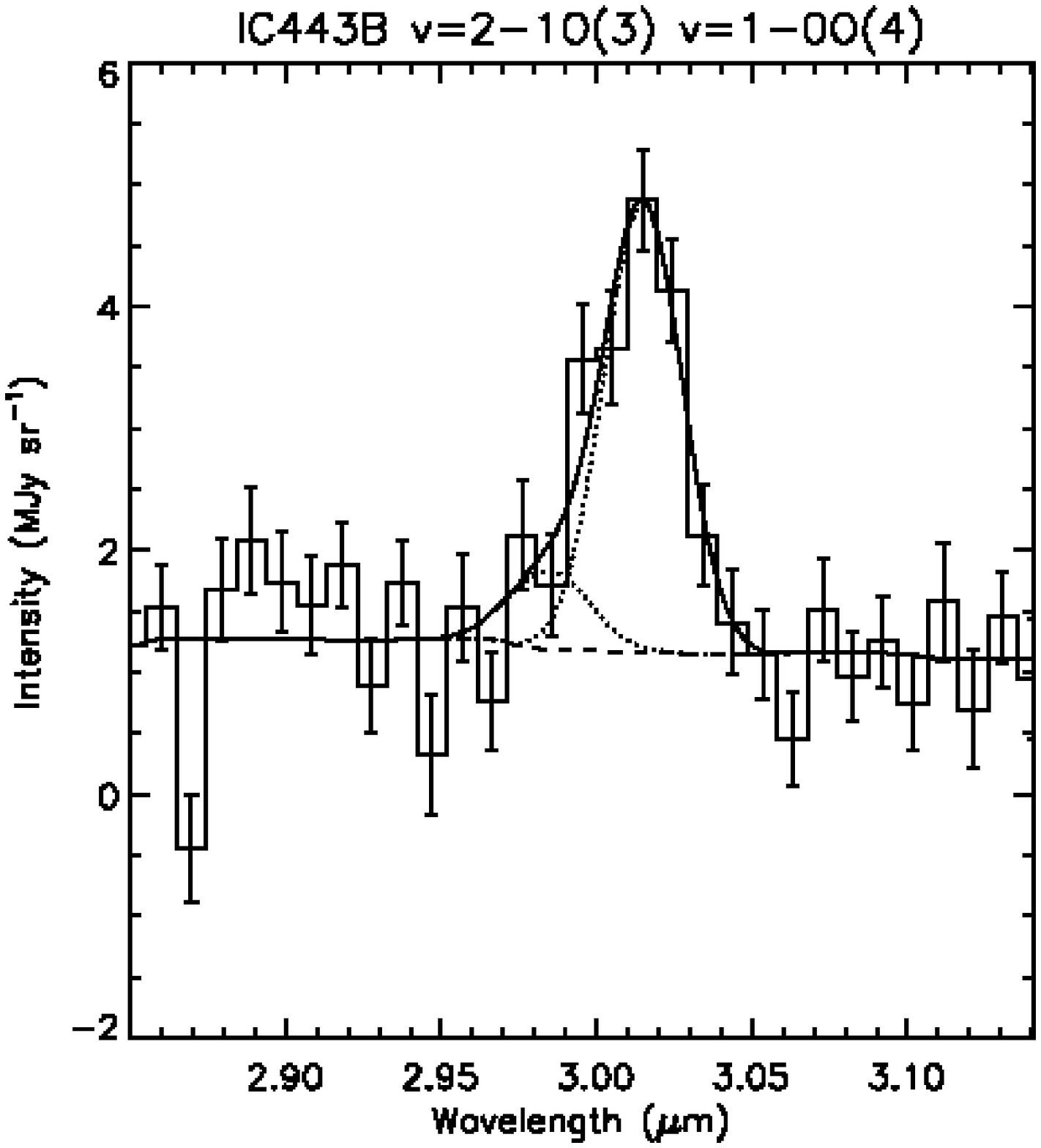} \\
\includegraphics[scale=0.33]{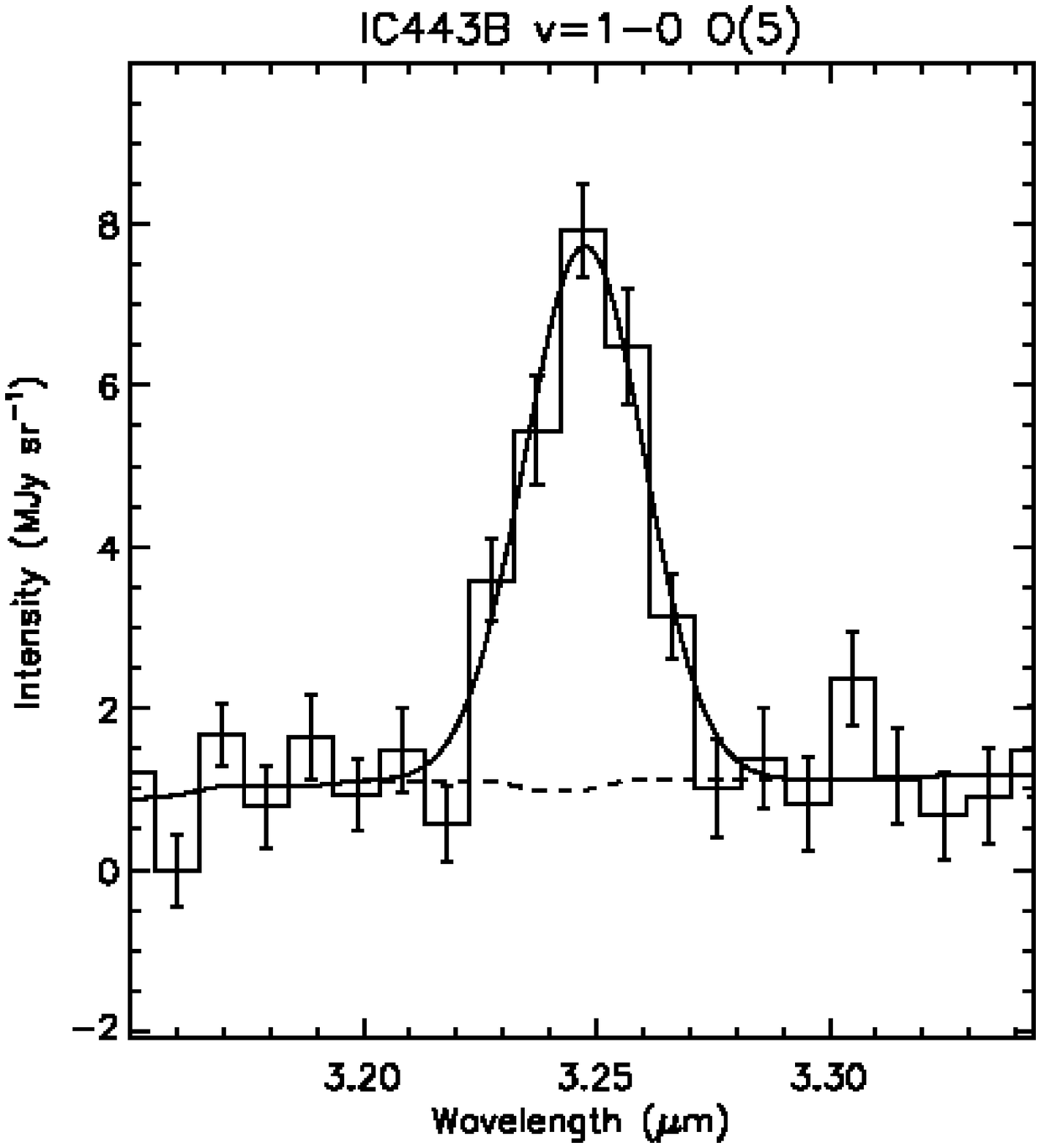}
\includegraphics[scale=0.33]{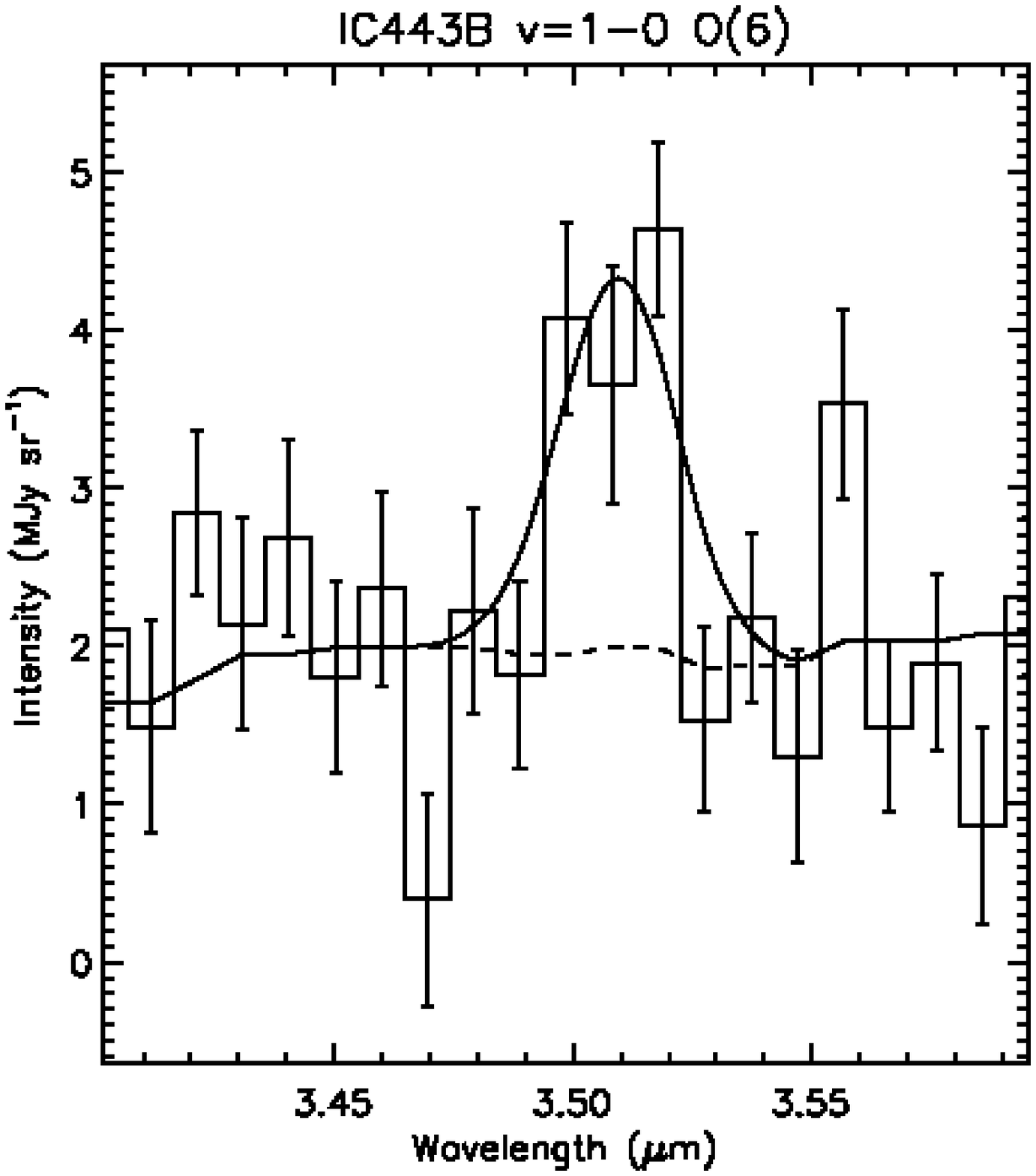}
\includegraphics[scale=0.33]{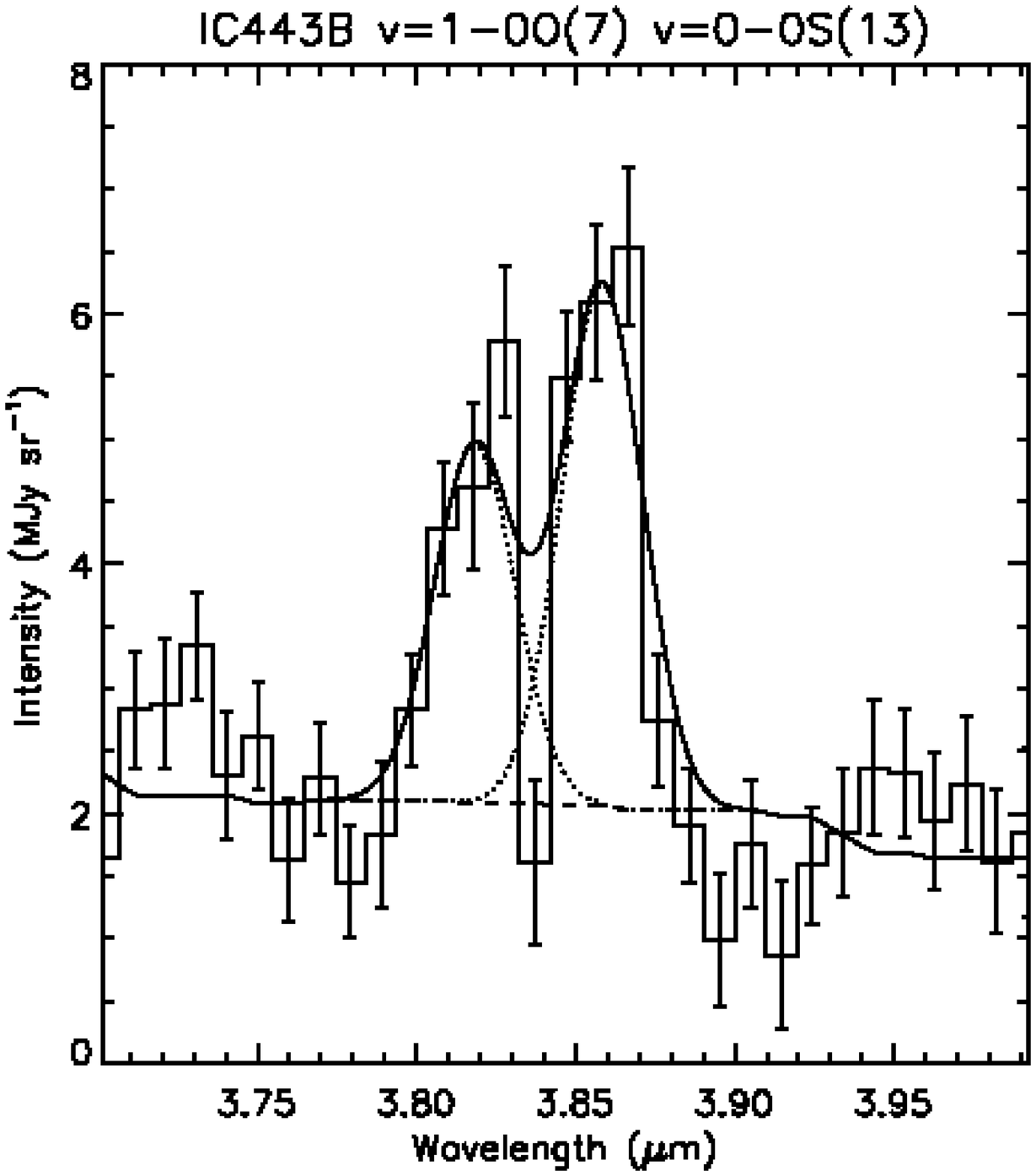} \\
\includegraphics[scale=0.33]{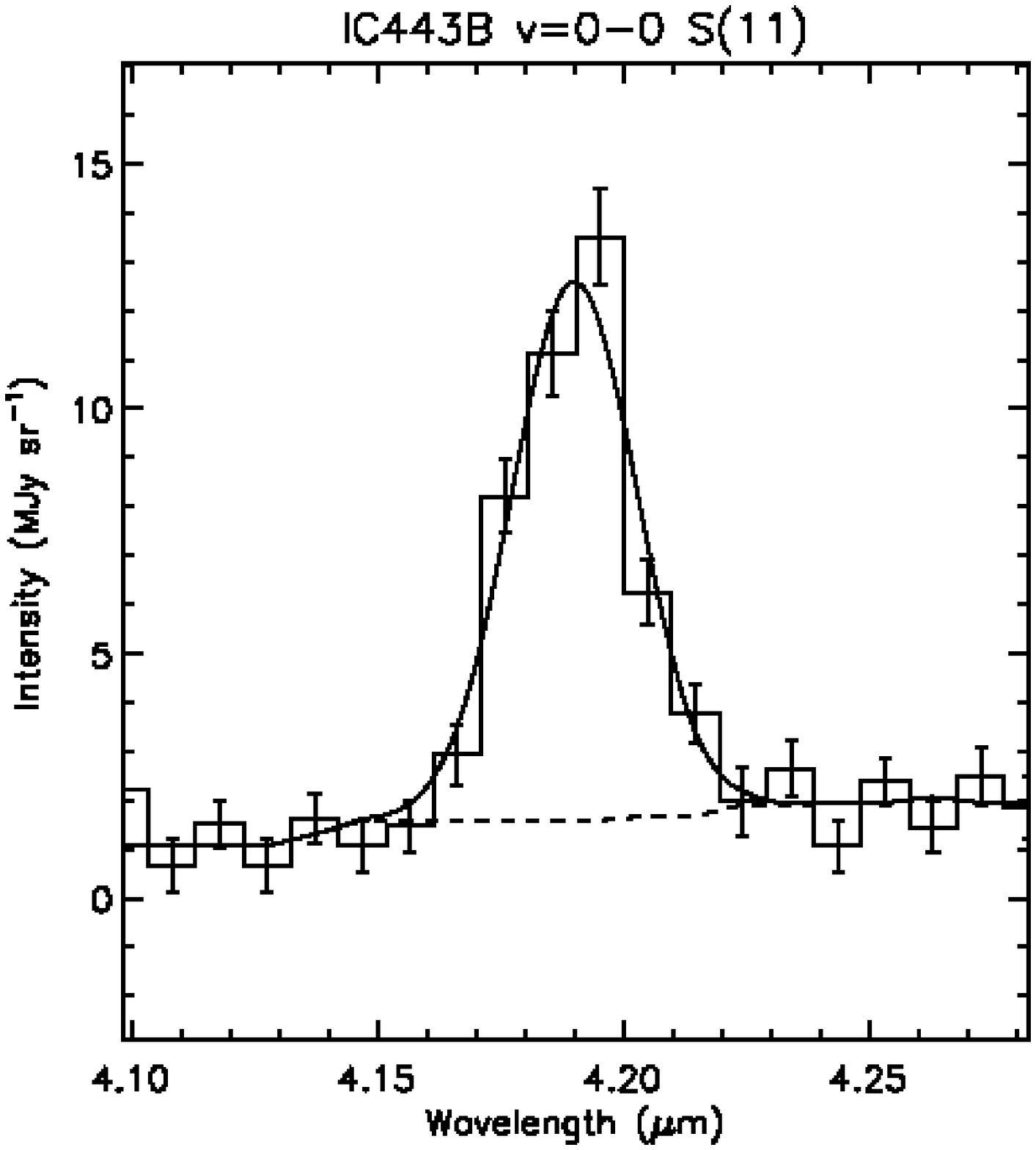}
\includegraphics[scale=0.33]{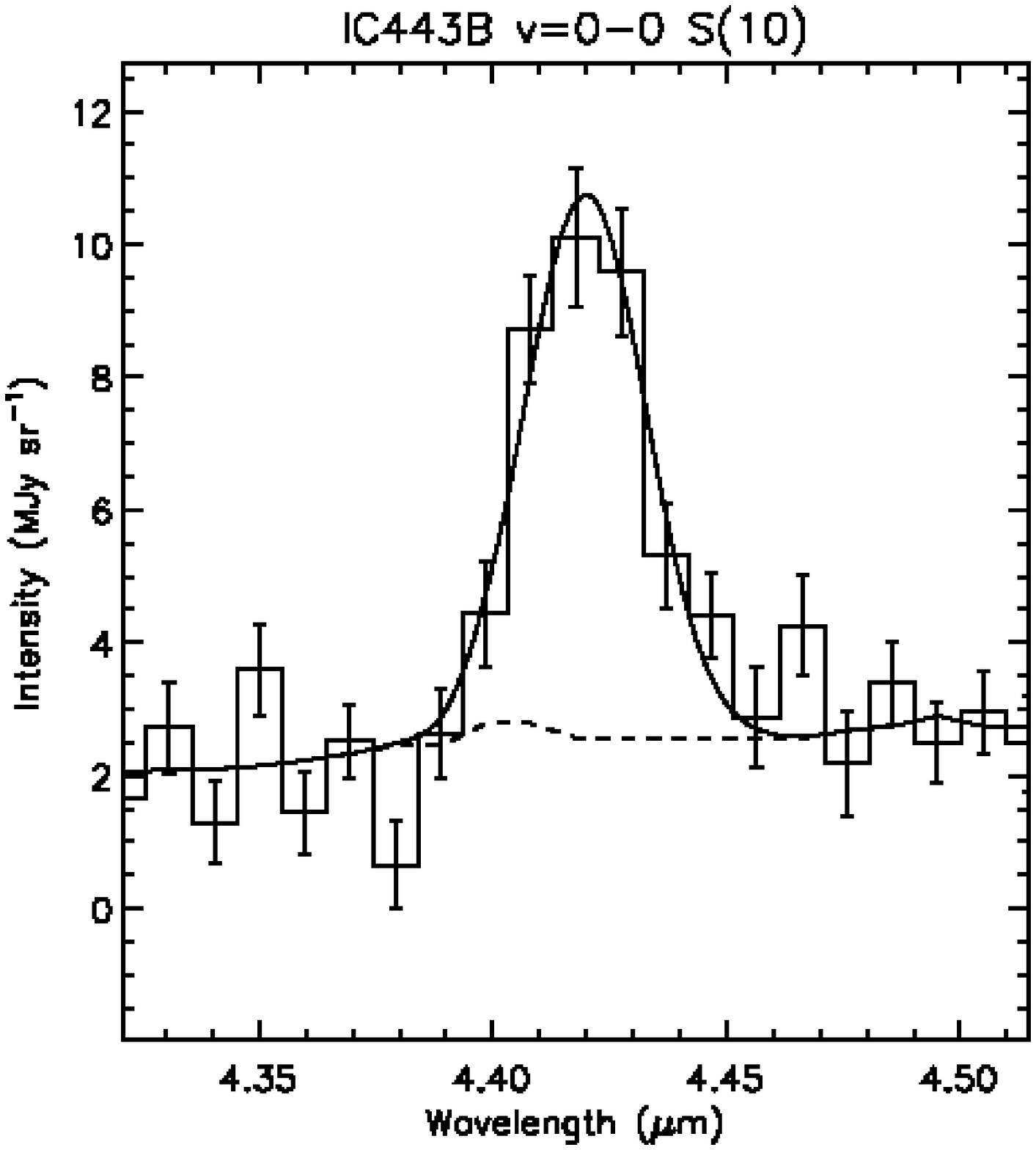}
\includegraphics[scale=0.33]{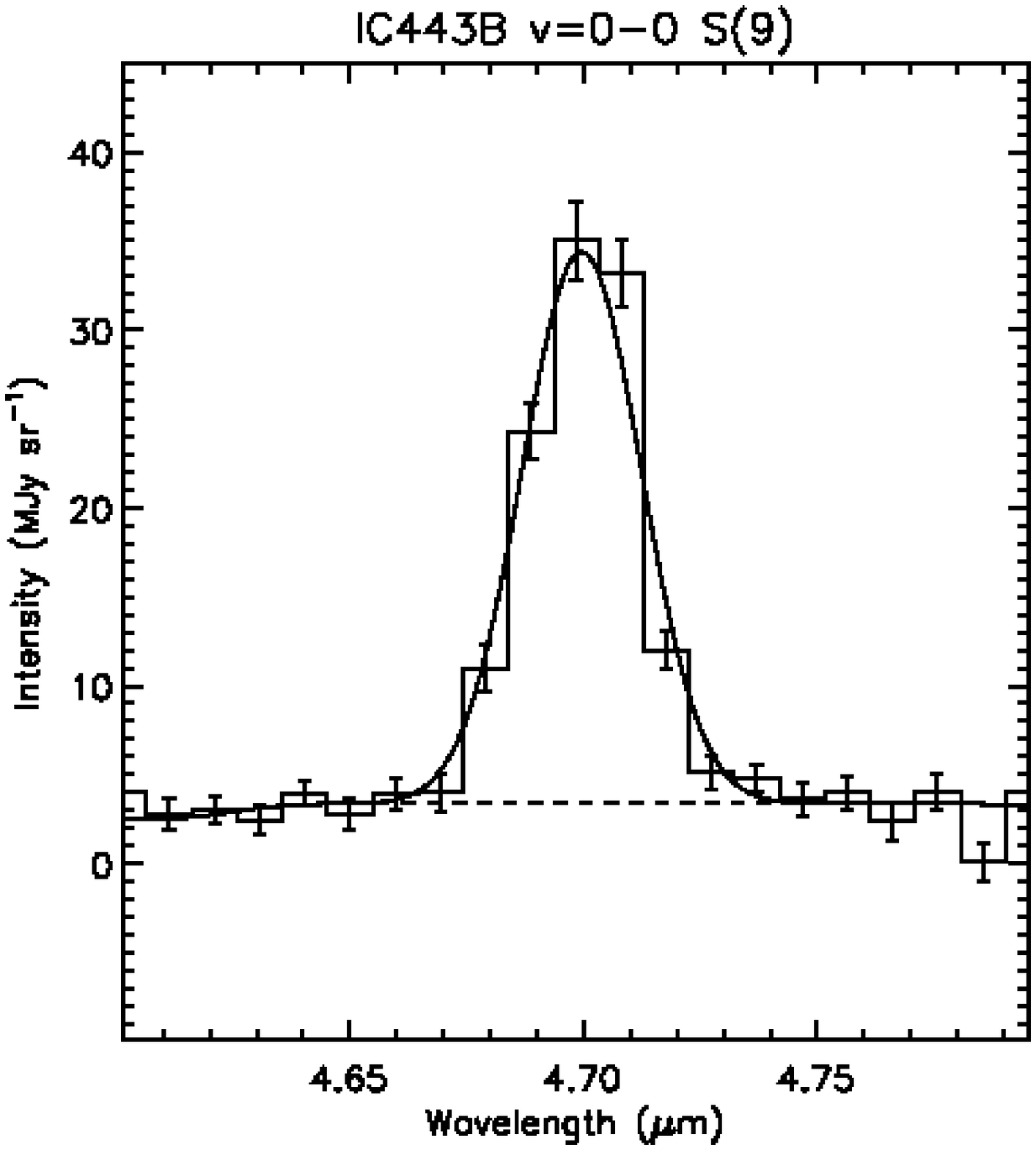}
}
\caption{Fitting for the \Htwo{} emission lines, observed toward the clump B. The \emph{solid} and \emph{dashed} lines indicate the continuum+line and continuum, respectively. The blended line components are indicated by the \emph{dotted} lines, if any. The feature around 2.56 \um{} is ignored since it is near the end of the band coverage.} \label{fig-fit-B}
\end{figure}

We present near-infrared ($2.5-5.0$ \um) spectra toward three shocked \Htwo{} clumps (B, C, and G) of the SNR IC 443 and one background (BG) region (cf.~Fig.~\ref{fig-slit}).
The observations were performed with the satellite \akari{} during its post-Helium phase.
Only \Htwo{} emission lines were detected toward all four directions.
We measured the line intensities and, after the reddening-correction, obtained the relevant level populations.
The level populations were compared with the ones from previous near- and mid-infrared observations.
\akari{} level populations are well fitted with those from previous mid-infrared space observations, while there is a systematic difference with those from previous near-infrared ground observations.
We attributed this difference to the calibration error of previous near-infrared observations, although it may be caused by the different position of observed regions.

With the \akari{} near-infrared observations, we could extend the level population of shocked \Htwo{} gas in the clump C and G obtained from mid-infrared observations, from $\sim7000$ K to $\sim22000$ K continuously.
From these combined level populations, we found that the $\upsilon=0$ levels are more populated than the $\upsilon>0$ levels at a fixed level energy, which means the population cannot be reproduced with any combination of \Htwo{} gas in LTE, usually used for the description of shocked \Htwo{} gas.
The populations were well described two-\nHtwo{} power-law thermal admixture model, including the H atom as a collisional partner in addition to \Htwo{} and He.
The model parameters are two number densities \nHtwo, two column densities \NHtwo, one power-law index $b$, and the \astH{I} to \Htwo{} abundance $X_H$ (cf.~Table \ref{tbl-mfit}).

We attributed the lower (\nHtwo{} $=10^{2.8-3.8}$ \ncm) and higher (\nHtwo{} $=10^{5.4-5.8}$ \ncm) density gases to the shocked \Htwo{} gas behind C-type and J-type shocks, respectively, based on several arguments such as the preshock density, the amount of shocked gas, the obtained high \astH{I} abundance, the shape of level populations, the line profile, etc.
This interpretation is consistent with the hierarchical picture of molecular clouds whose constituents are classified into ``clouds,'' ``clumps,'' and ``cores'' \citep{Bergin(2007)ARA&A_45_339,Williams(2000)2000prpl.conf__97}.
The C-type shocks are likely propagating into the ``clumps'' ($\sim10^3$ \ncm), while the J-type shocks are propagating into the ``clouds'' (interclump media, $<10^3$ \ncm).
The power-law index $b$, mainly determined by the lower density gas, is attributed to the diversity in the shock velocity propagating into the clouds.
Such a velocity diversity may be a natural result during the shock-cloud interaction, since the shock driving pressure would be higher at the head than the side.
According to our results, the \HtwolineK{} emission is mainly from J-type shocks propagating into interclump media.
In addition, our power-law thermal admixture model would be useful to perform statistical studies on the shocked \Htwo{} gas observed around other SNRs and YSOs, which shows coherent excitation characteristics.

\Htwo{} emission was also detected at the BG region, and we attributed it to the diffuse \Htwo{} gas, pervaded in the Milky Way.
This gas may be excited by collisional processes, like shocks or turbulence dissipation, in addition to ultraviolet photon pumping.

\acknowledgments
This work is based on observations with \akari, a JAXA project with the participation of ESA. 
The authors thank all the members of the \akari{} project.
J. H. S. is grateful to Jae-Joon Lee for the useful discussion on the two-density interpretation.
This research has made use of SAOImage DS9, developed by Smithsonian Astrophysical Observatory \citep{Joye(2003)inproc}.

{\it Facilities:} \facility{Akari}

\bibliographystyle{G:/Work/Publication/bibtex/astronat/apj/apj}
\bibliography{G:/Work/Publication/bibtex/paper,G:/Work/Publication/bibtex/book,G:/Work/Publication/bibtex/manual}

\clearpage

\clearpage

\clearpage

\clearpage
\begin{figure}
\center{
\includegraphics[scale=0.33]{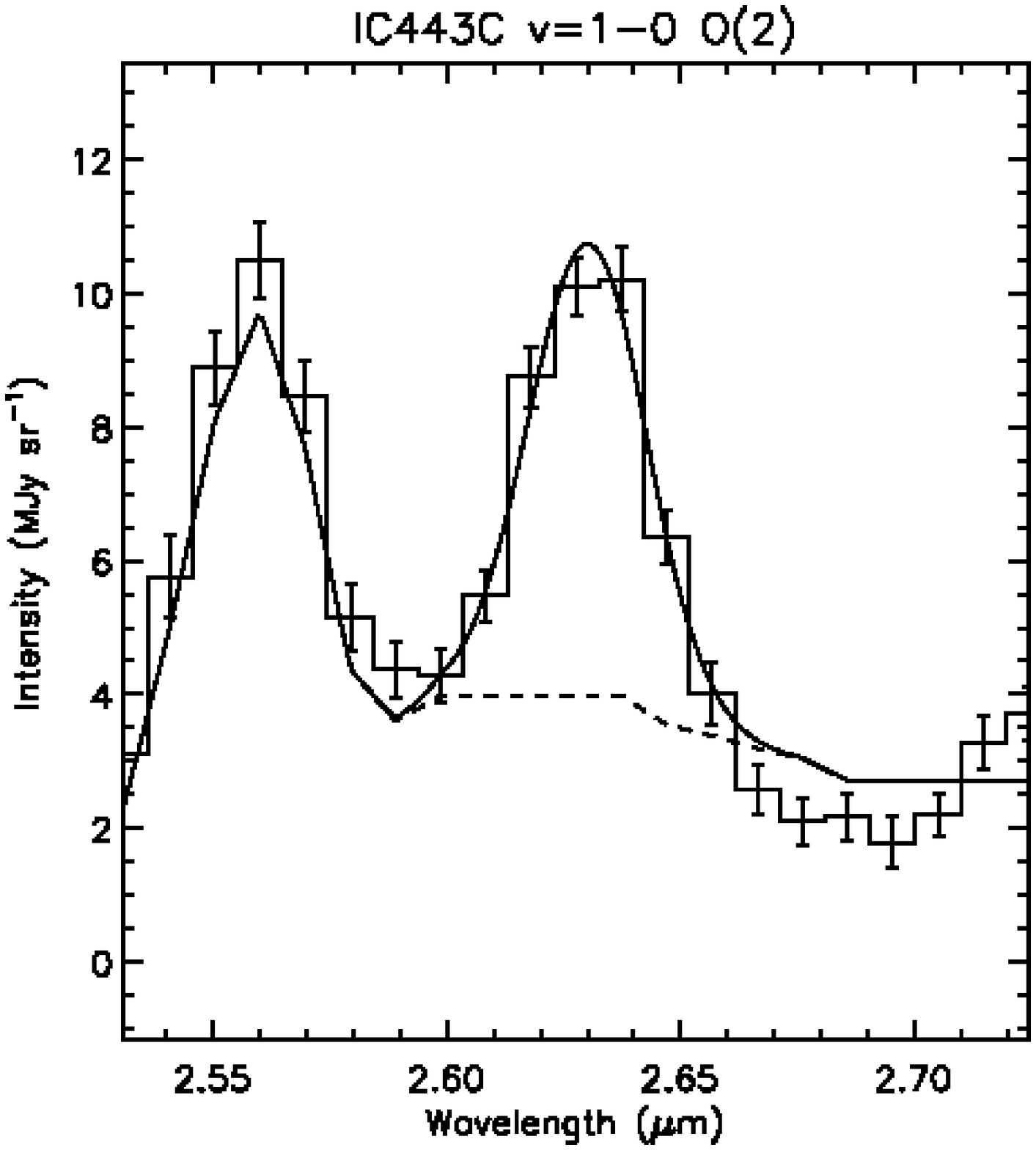}
\includegraphics[scale=0.33]{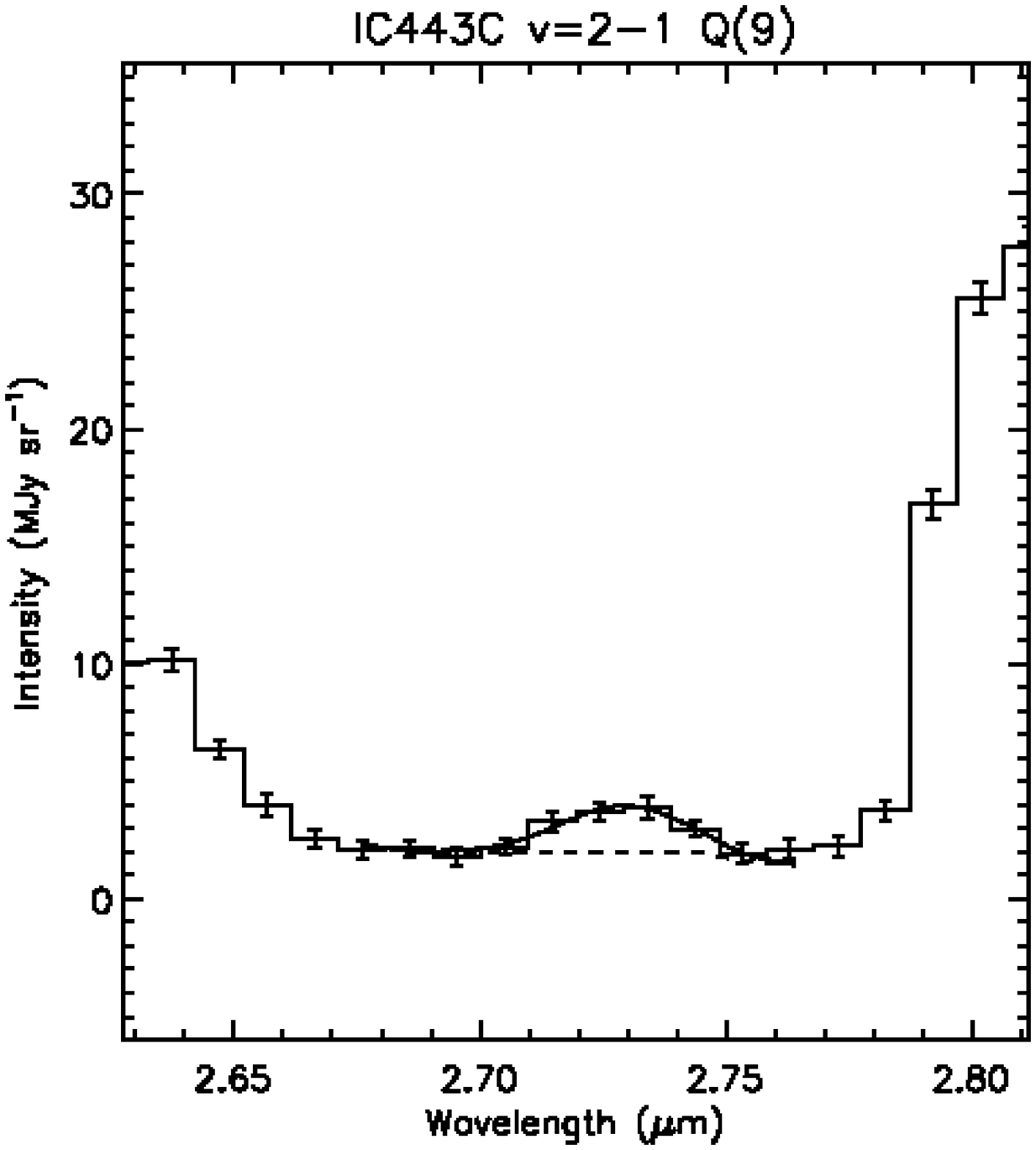}
\includegraphics[scale=0.33]{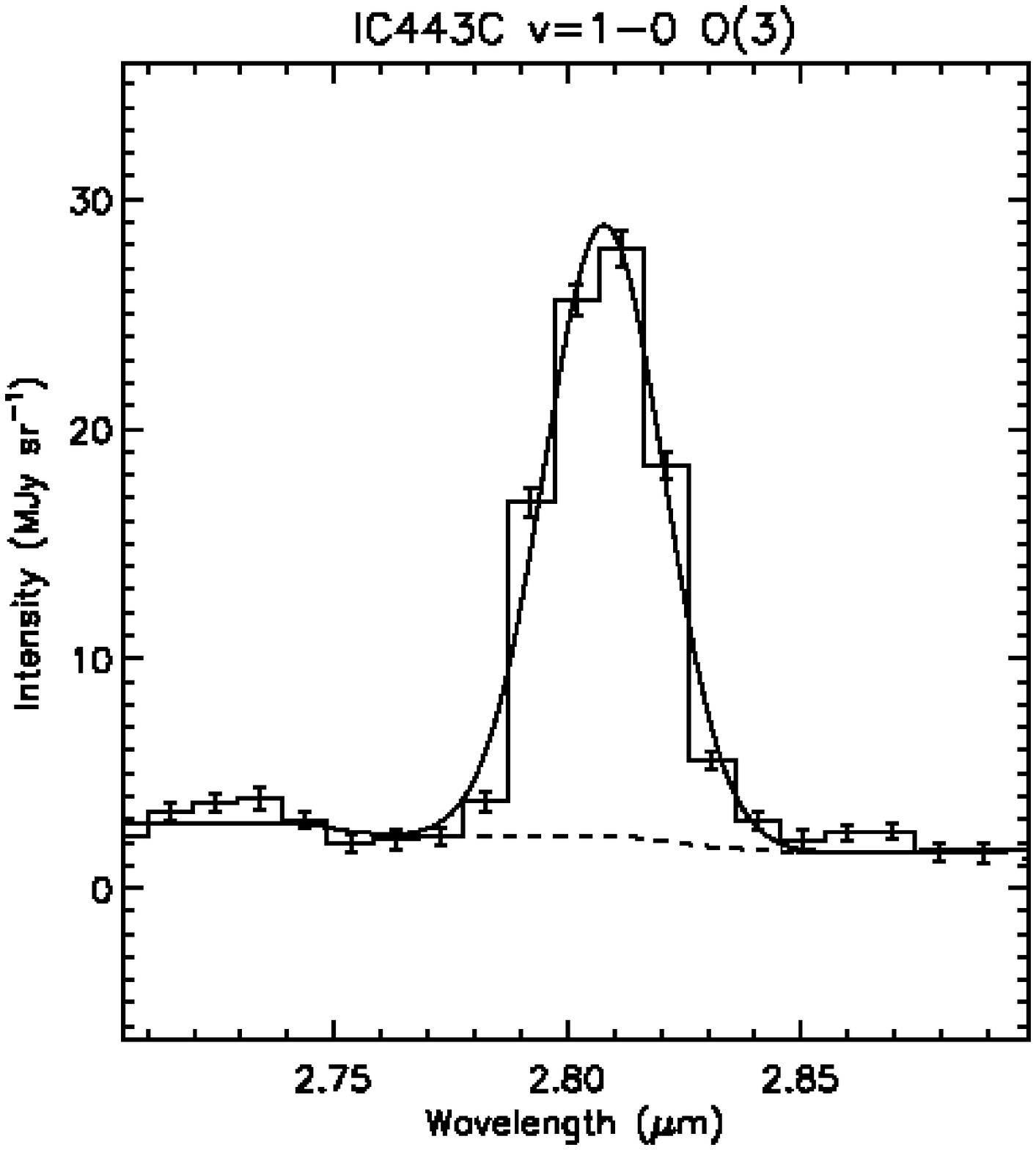} \\
\includegraphics[scale=0.33]{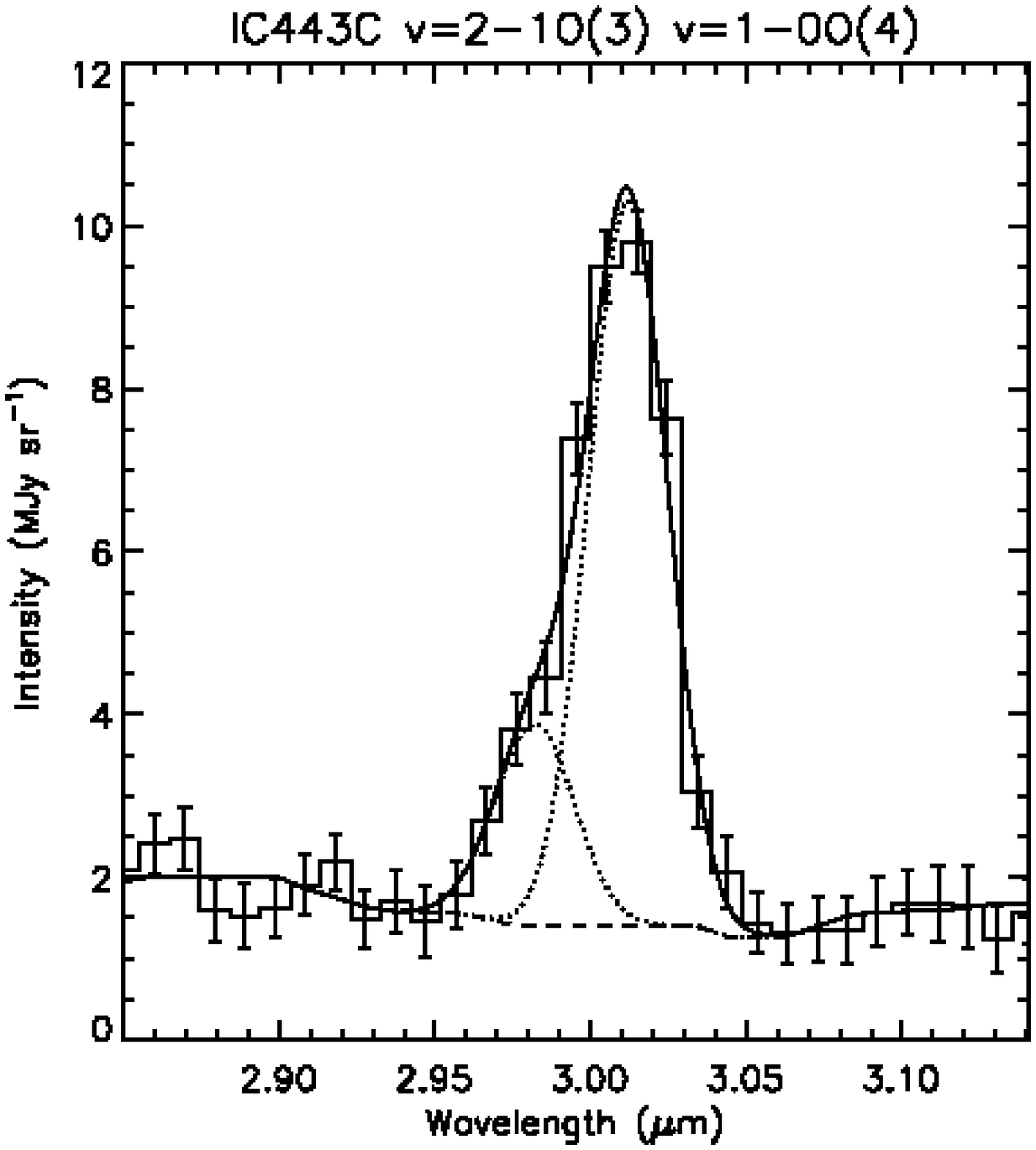}
\includegraphics[scale=0.33]{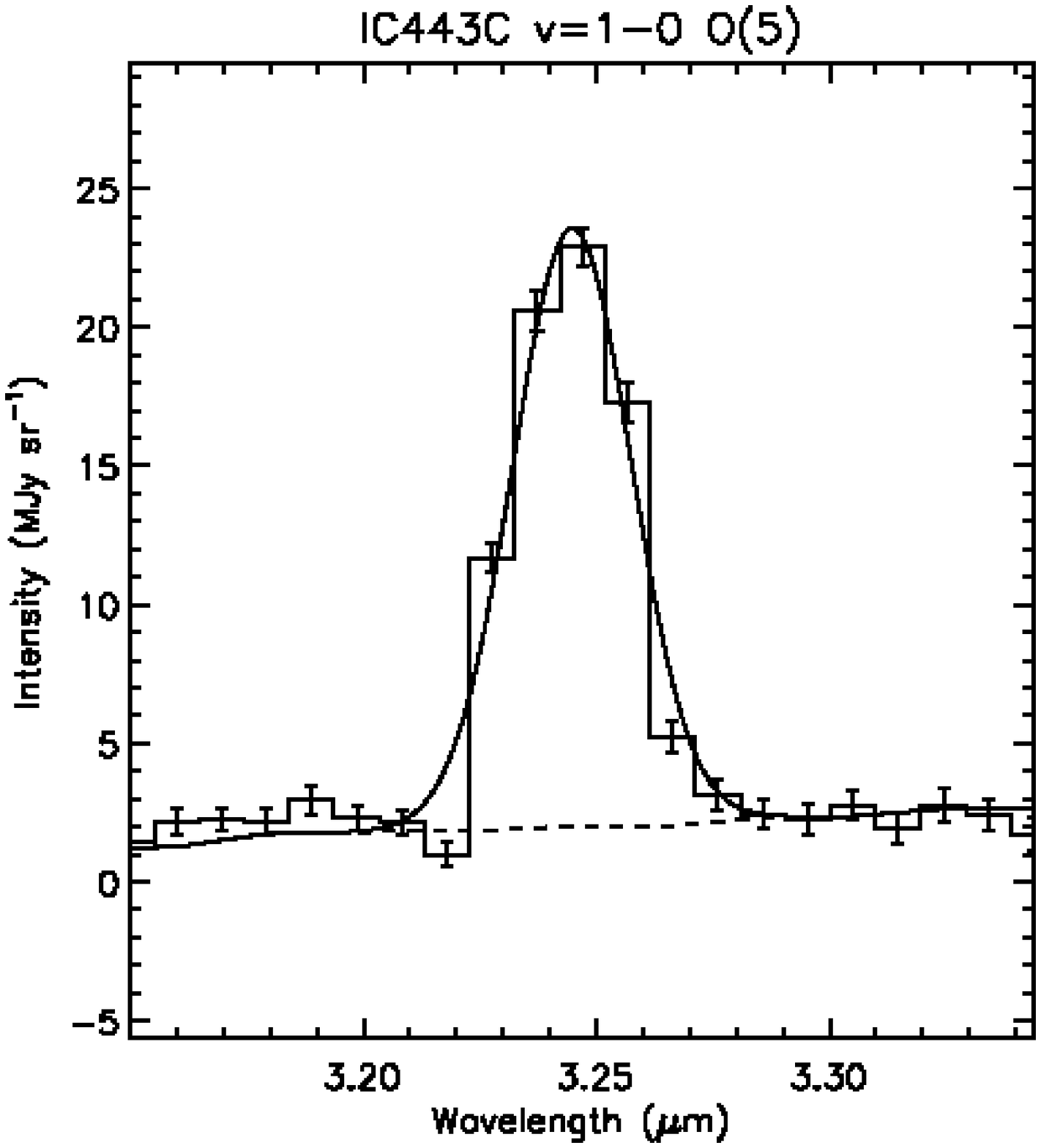}
\includegraphics[scale=0.33]{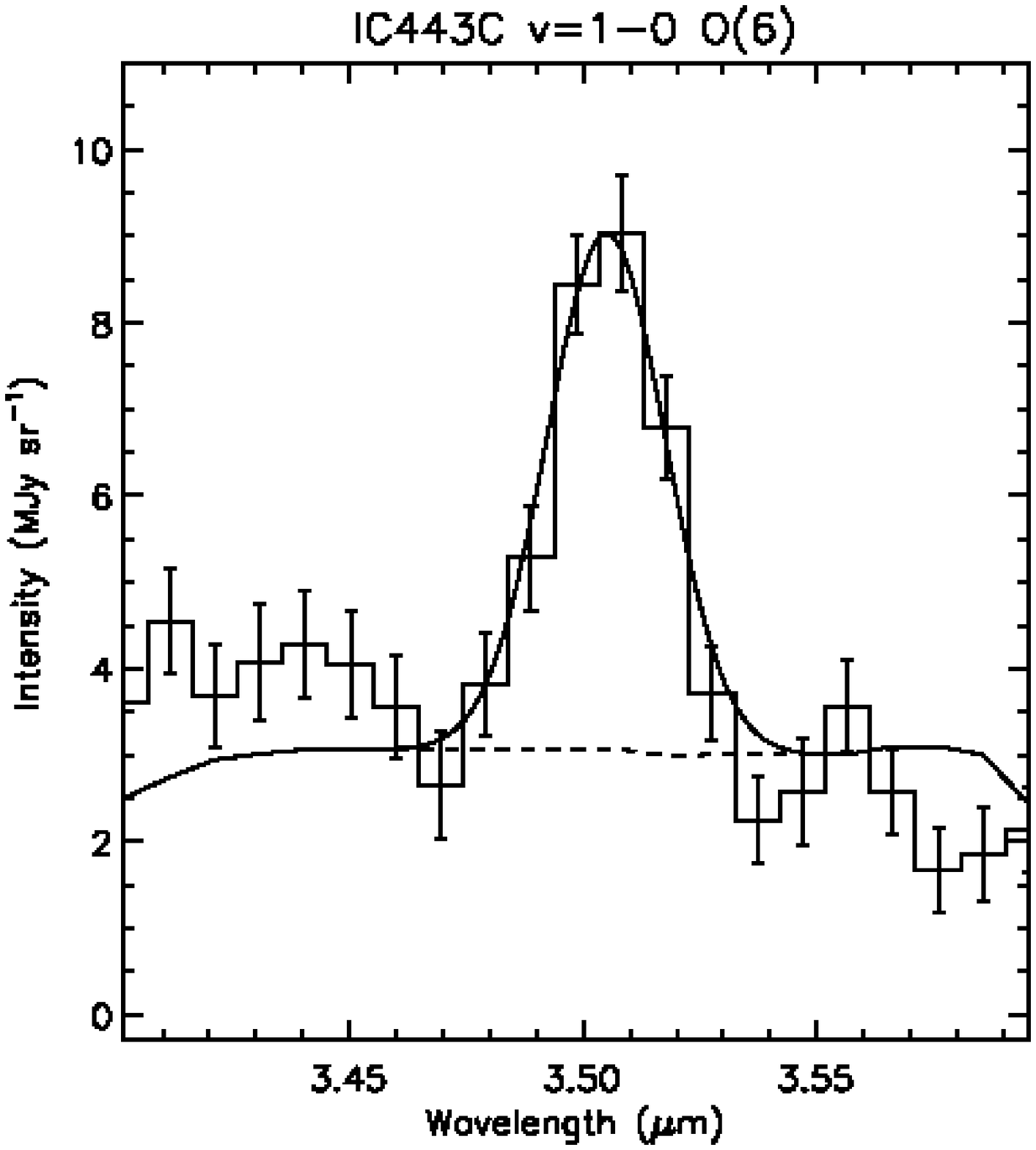} \\
\includegraphics[scale=0.33]{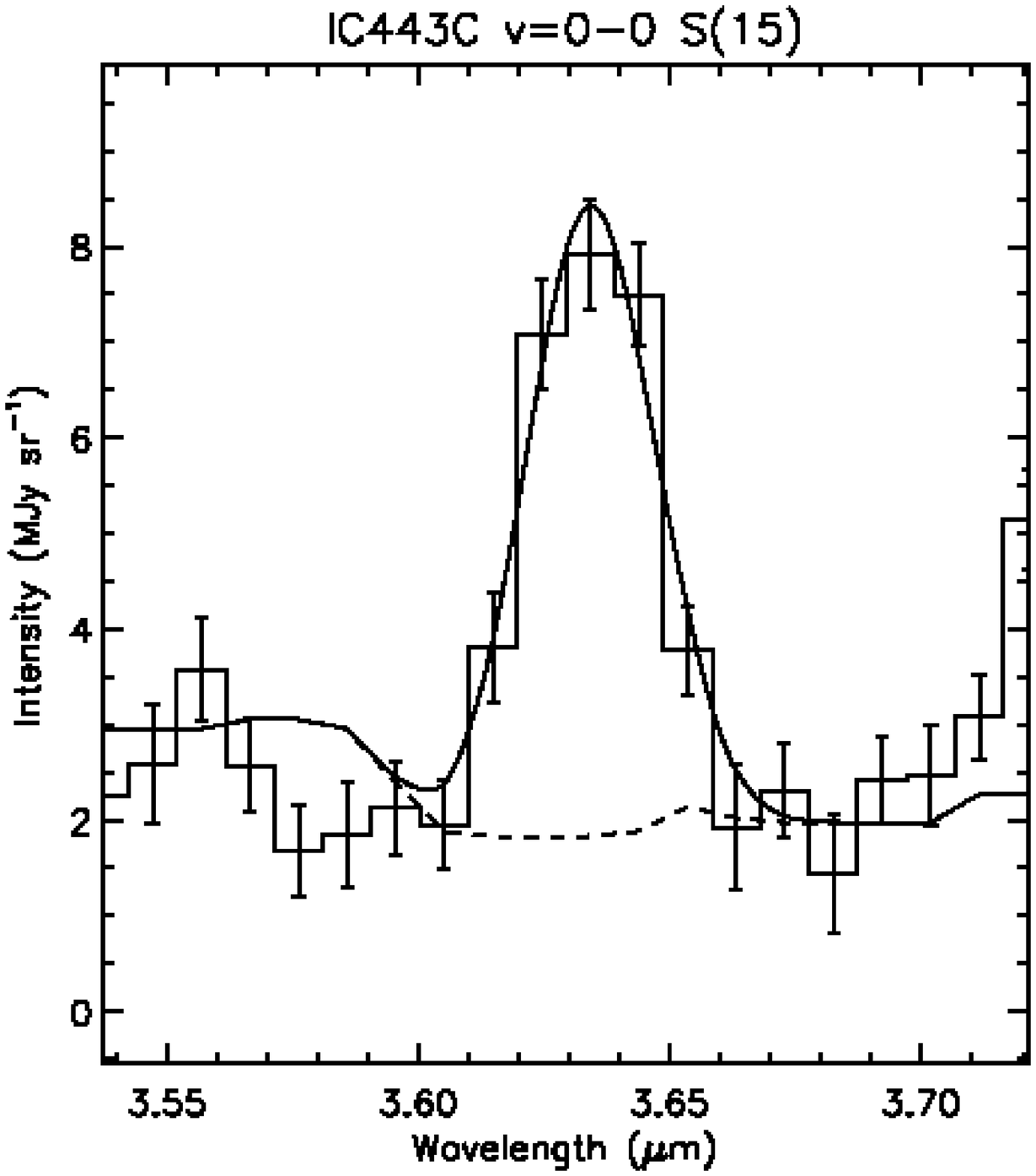}
\includegraphics[scale=0.33]{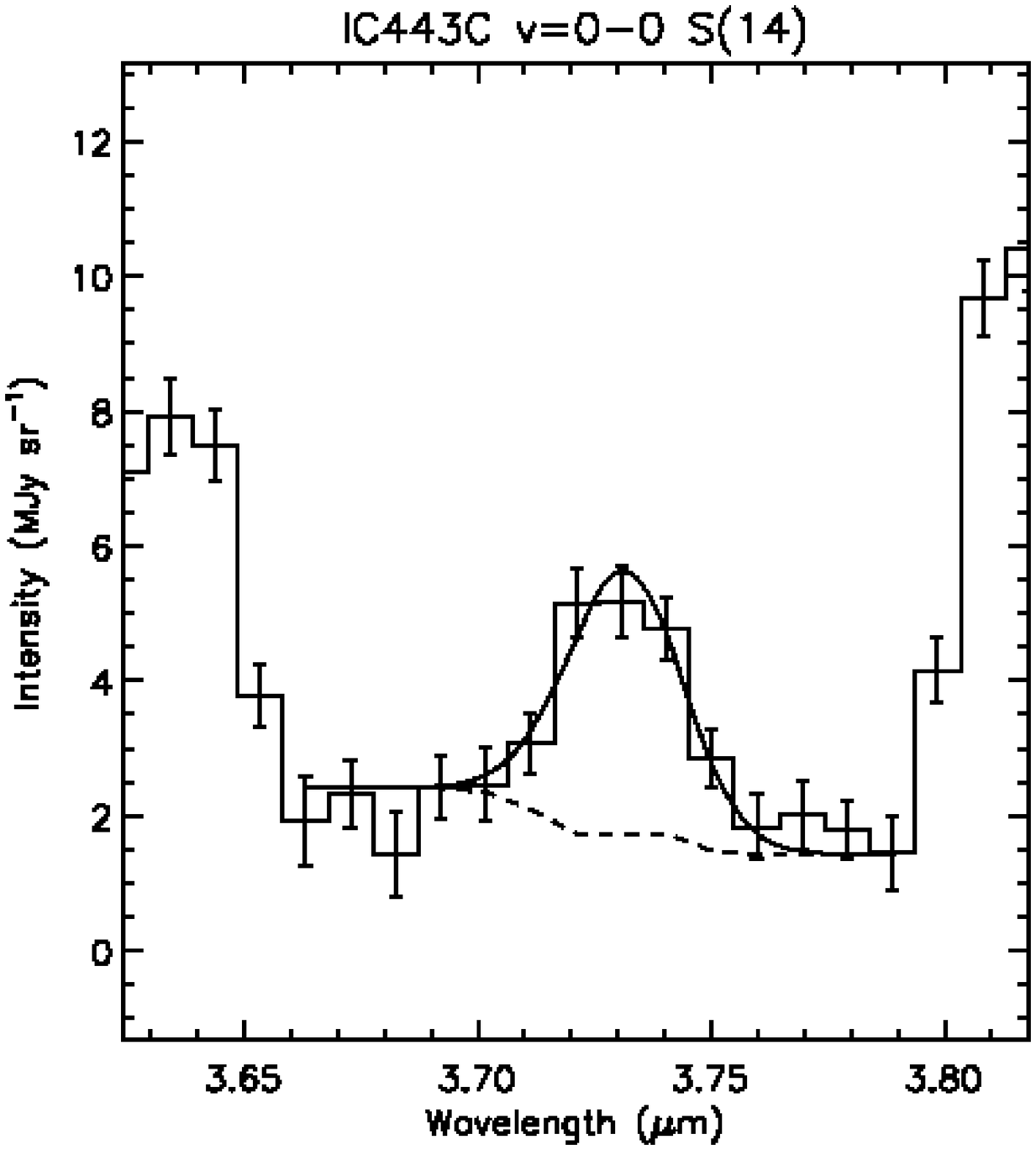}
\includegraphics[scale=0.33]{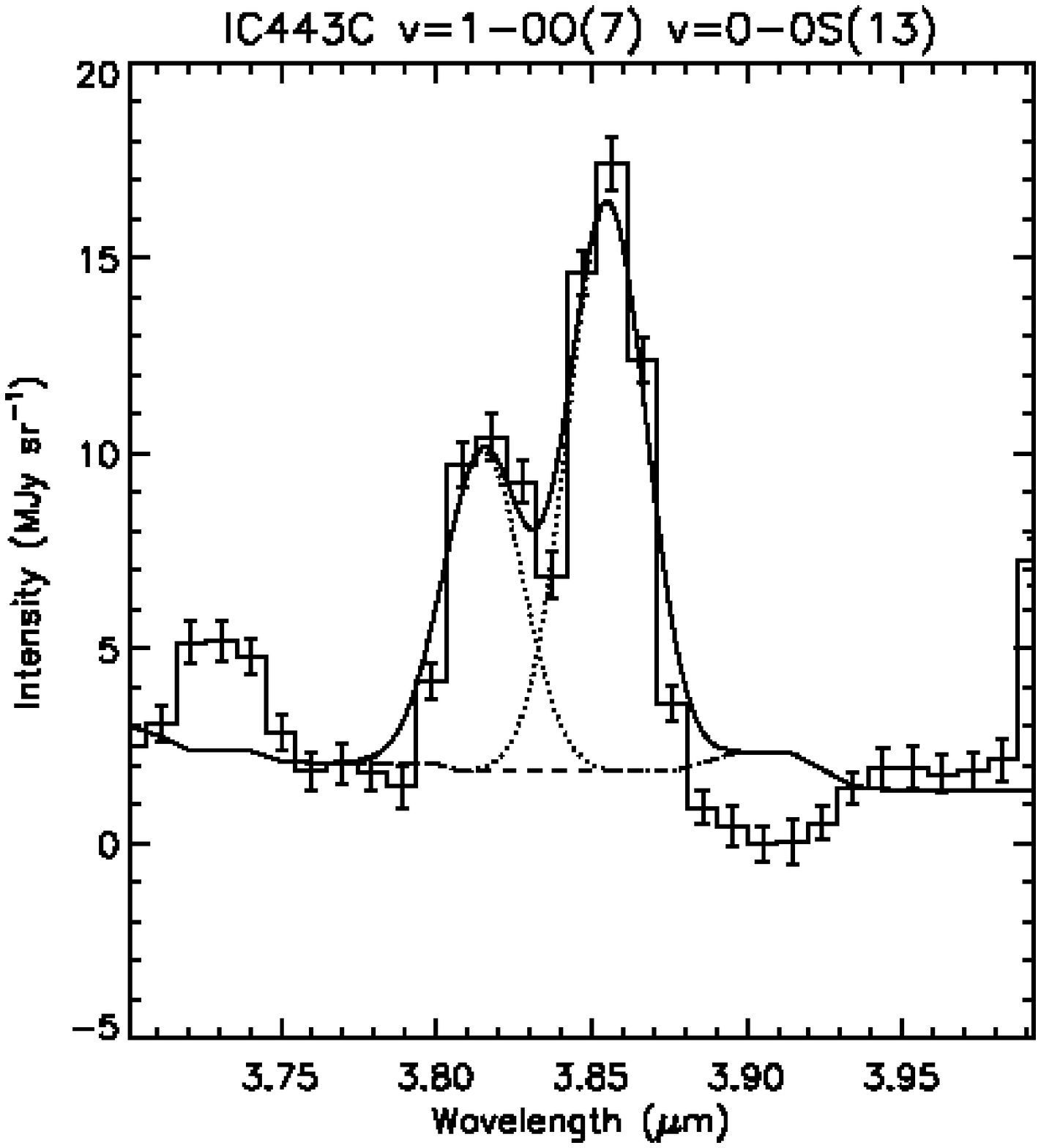} \\
}
\caption{Fitting for the \Htwo{} emission lines, observed toward the clump C. The rest is the same as Figure \ref{fig-fit-B}.} \label{fig-fit-C}
\end{figure}

\clearpage
\begin{figure}
\figurenum{4}
\center{
\includegraphics[scale=0.33]{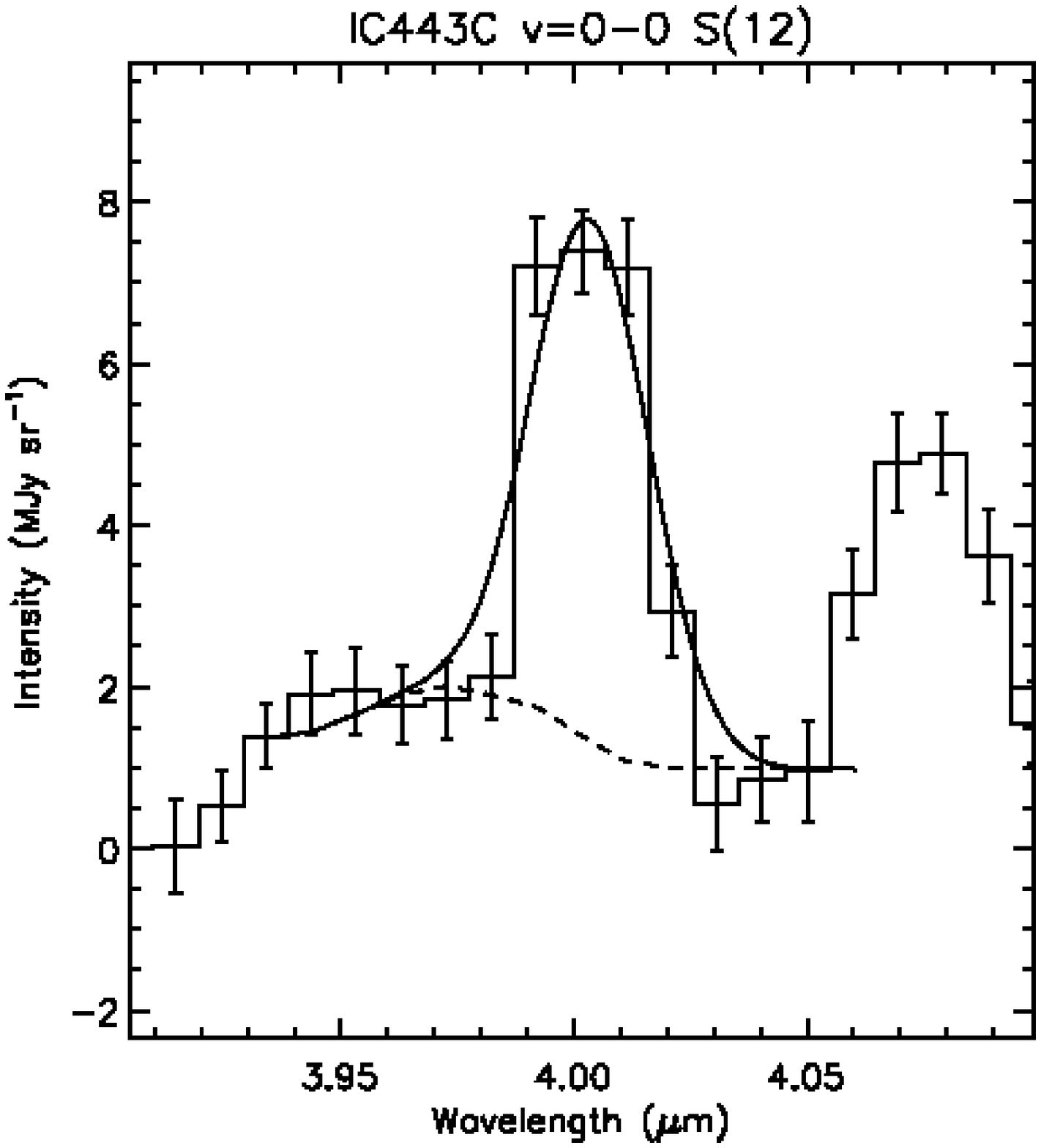}
\includegraphics[scale=0.33]{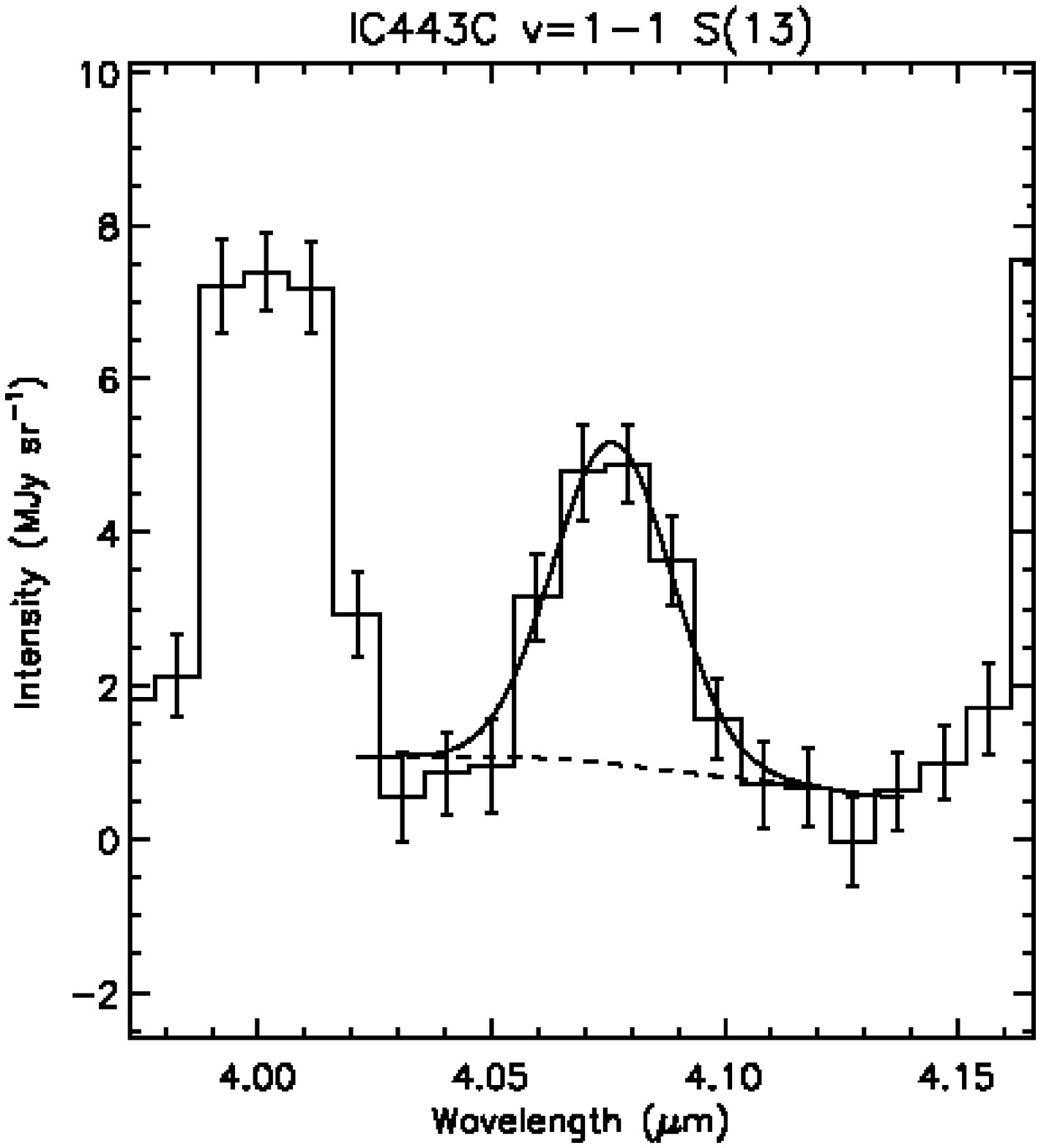}
\includegraphics[scale=0.33]{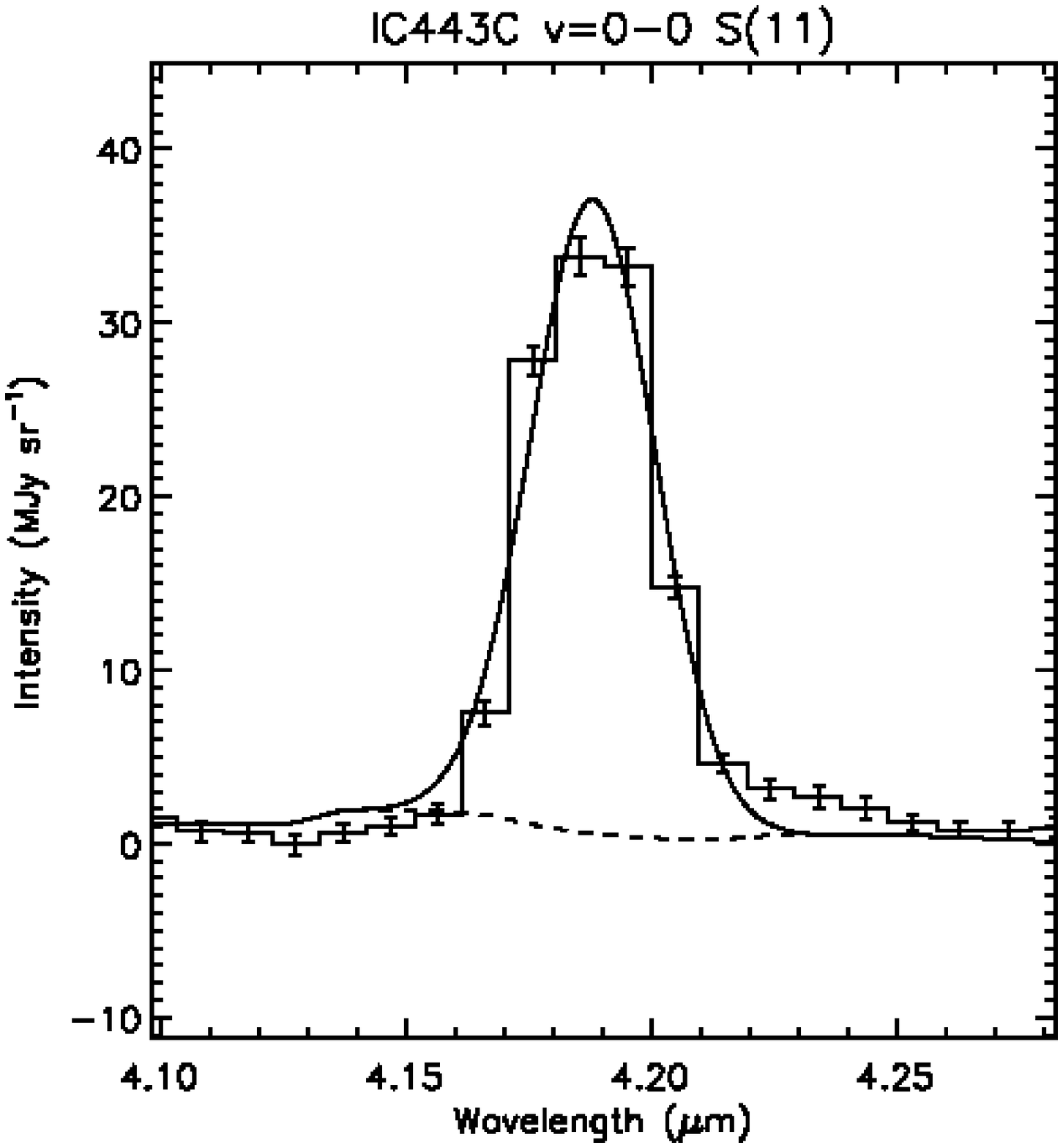} \\
\includegraphics[scale=0.33]{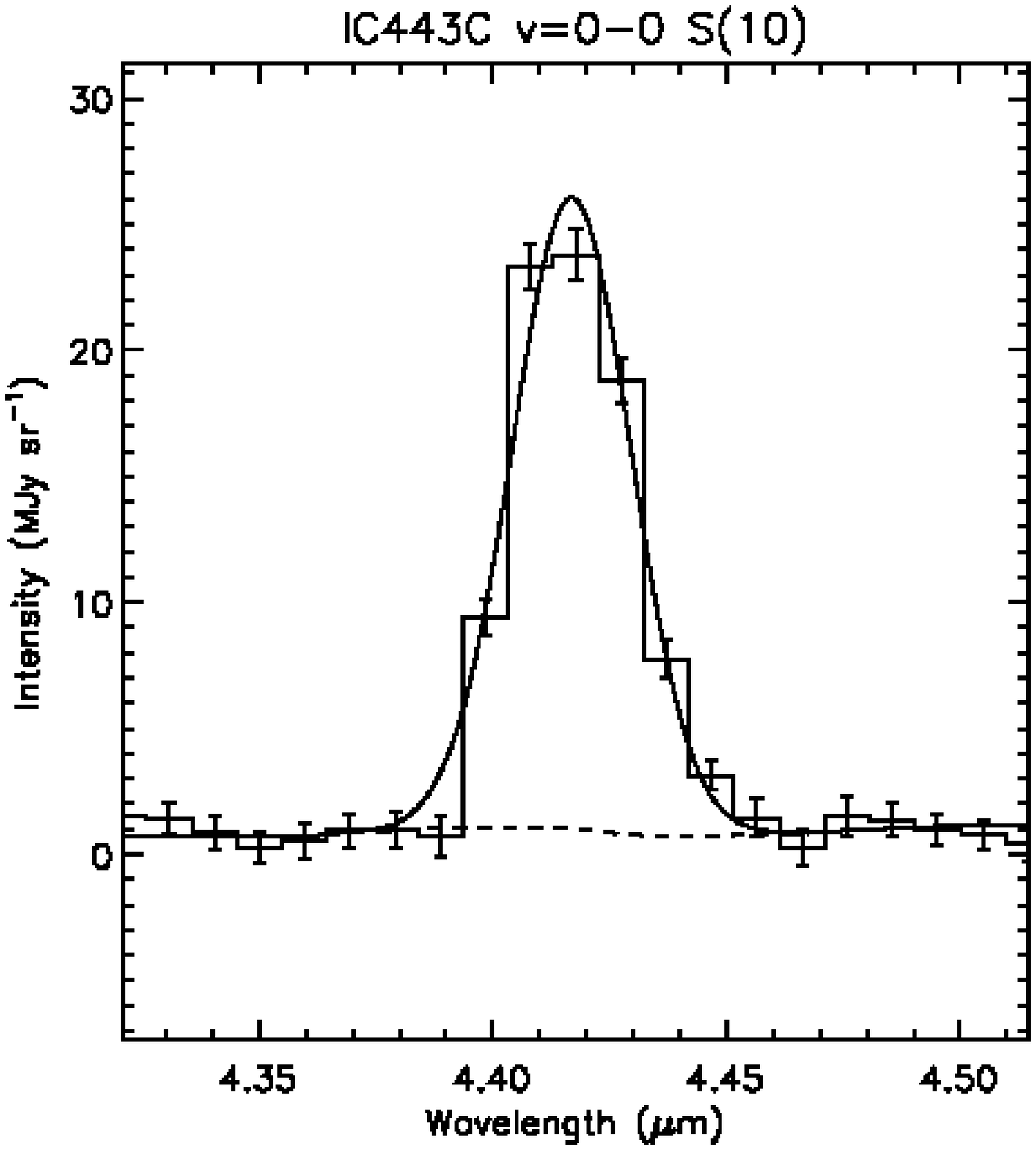}
\includegraphics[scale=0.33]{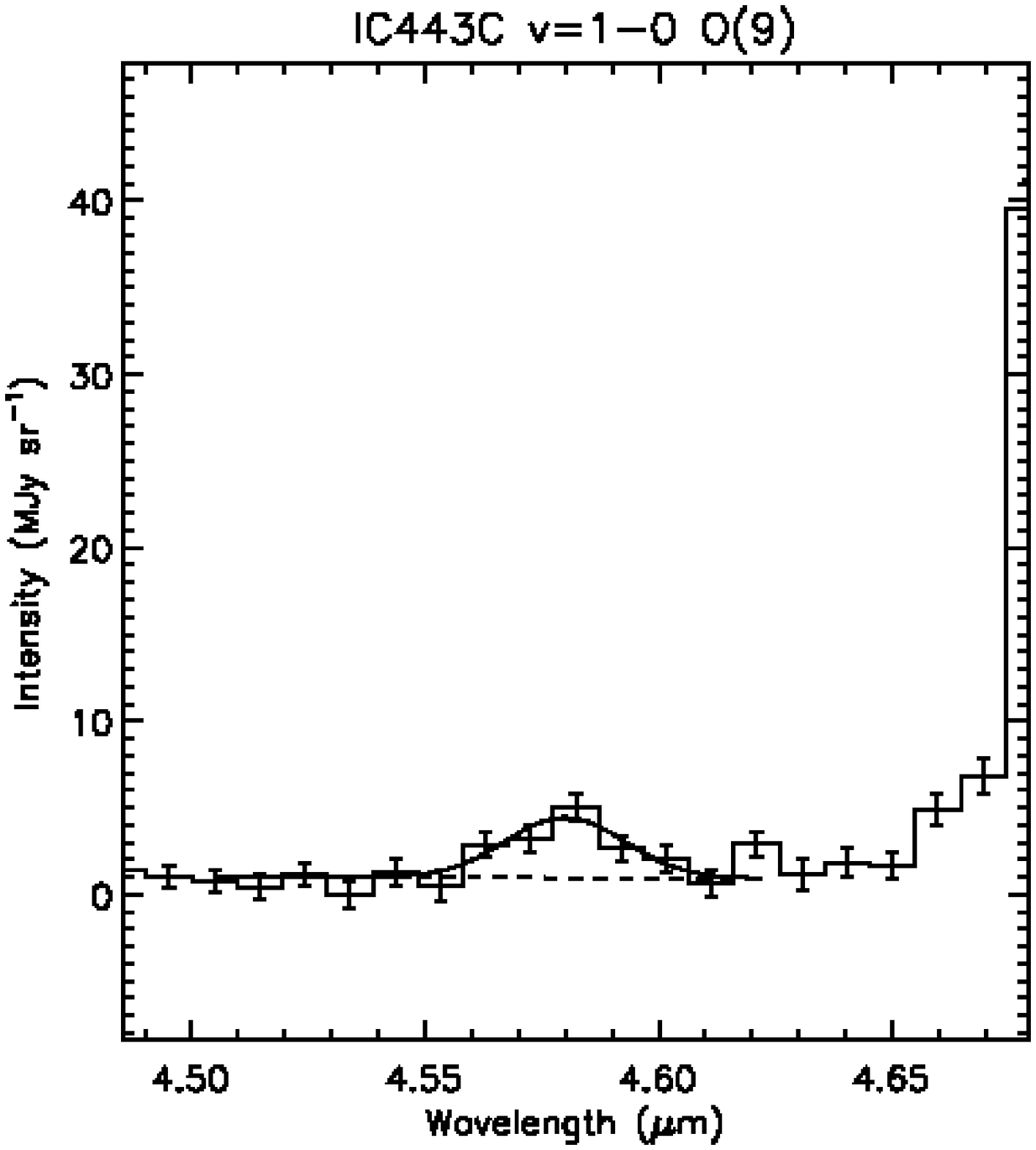}
\includegraphics[scale=0.33]{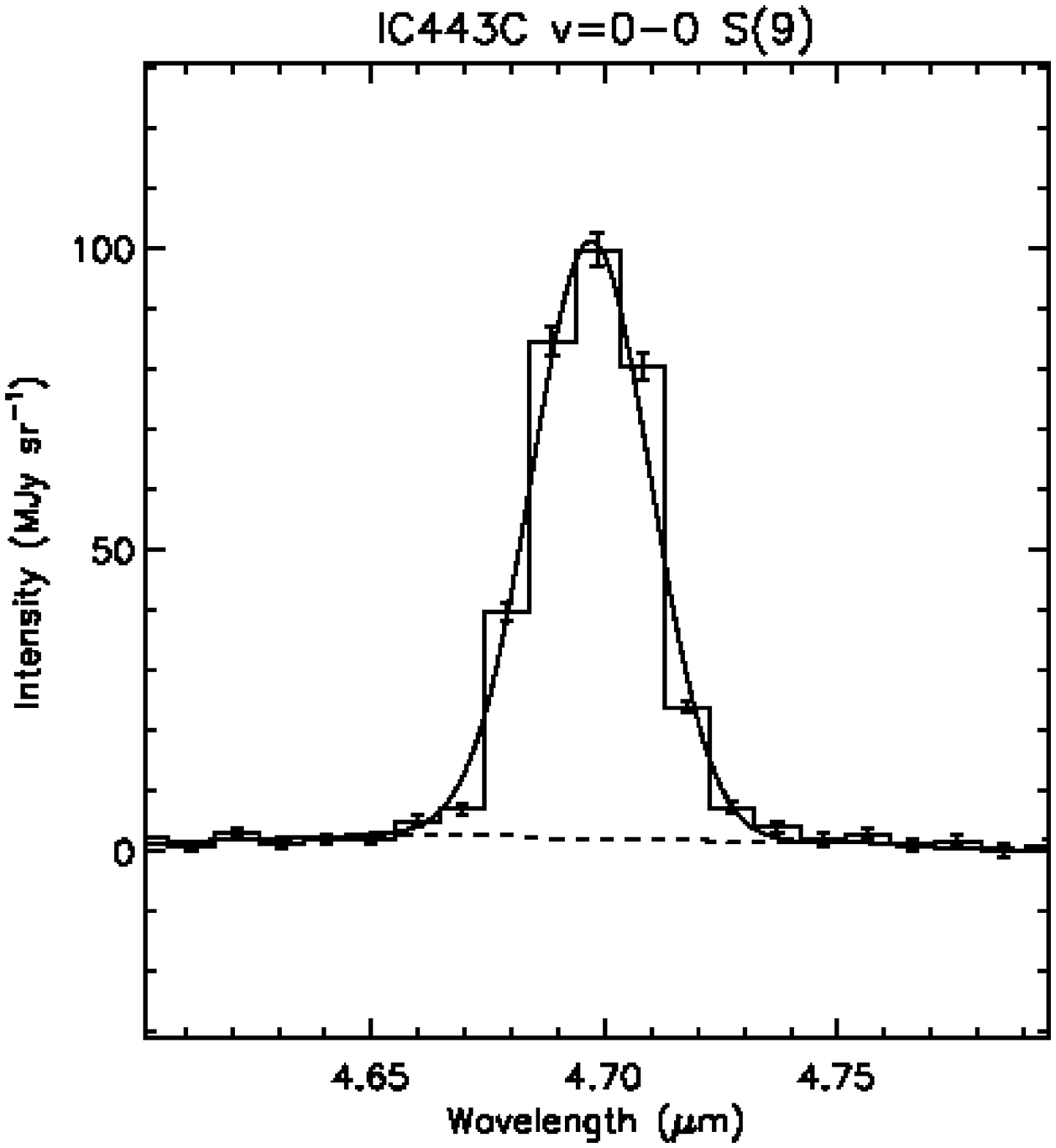}
}
\caption{Continued.}
\end{figure}

\clearpage
\begin{figure}
\center{
\includegraphics[scale=0.33]{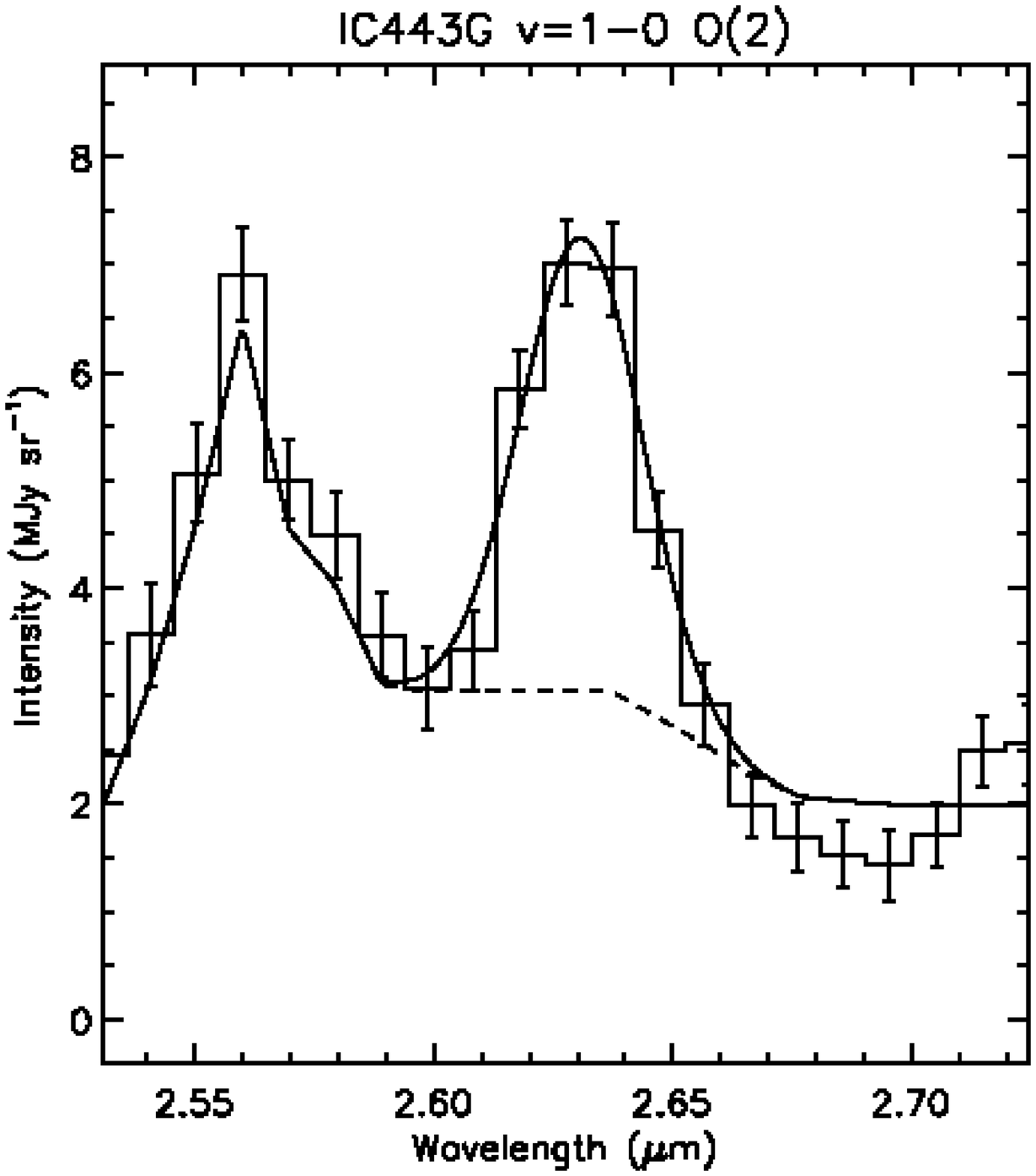}
\includegraphics[scale=0.33]{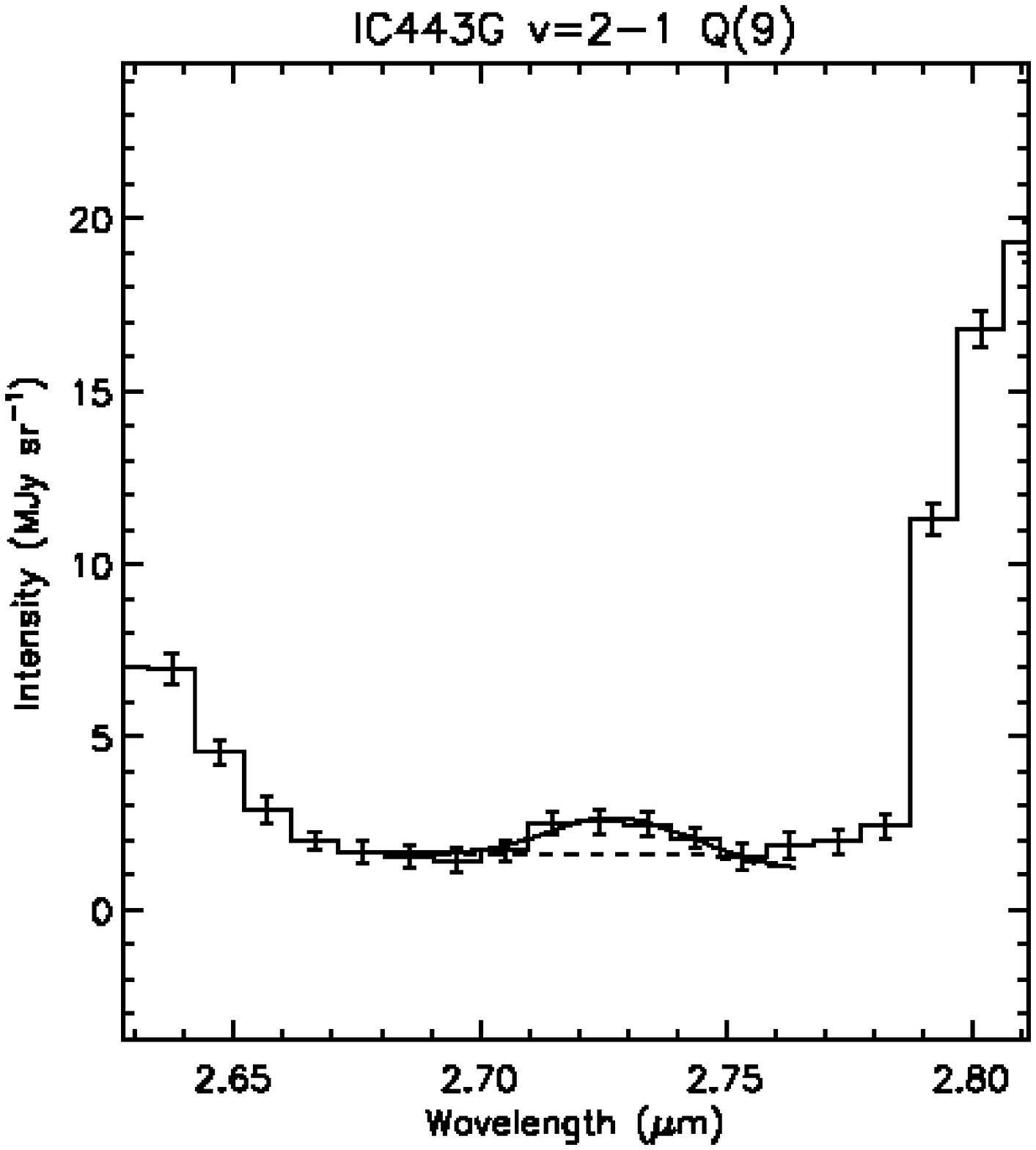}
\includegraphics[scale=0.33]{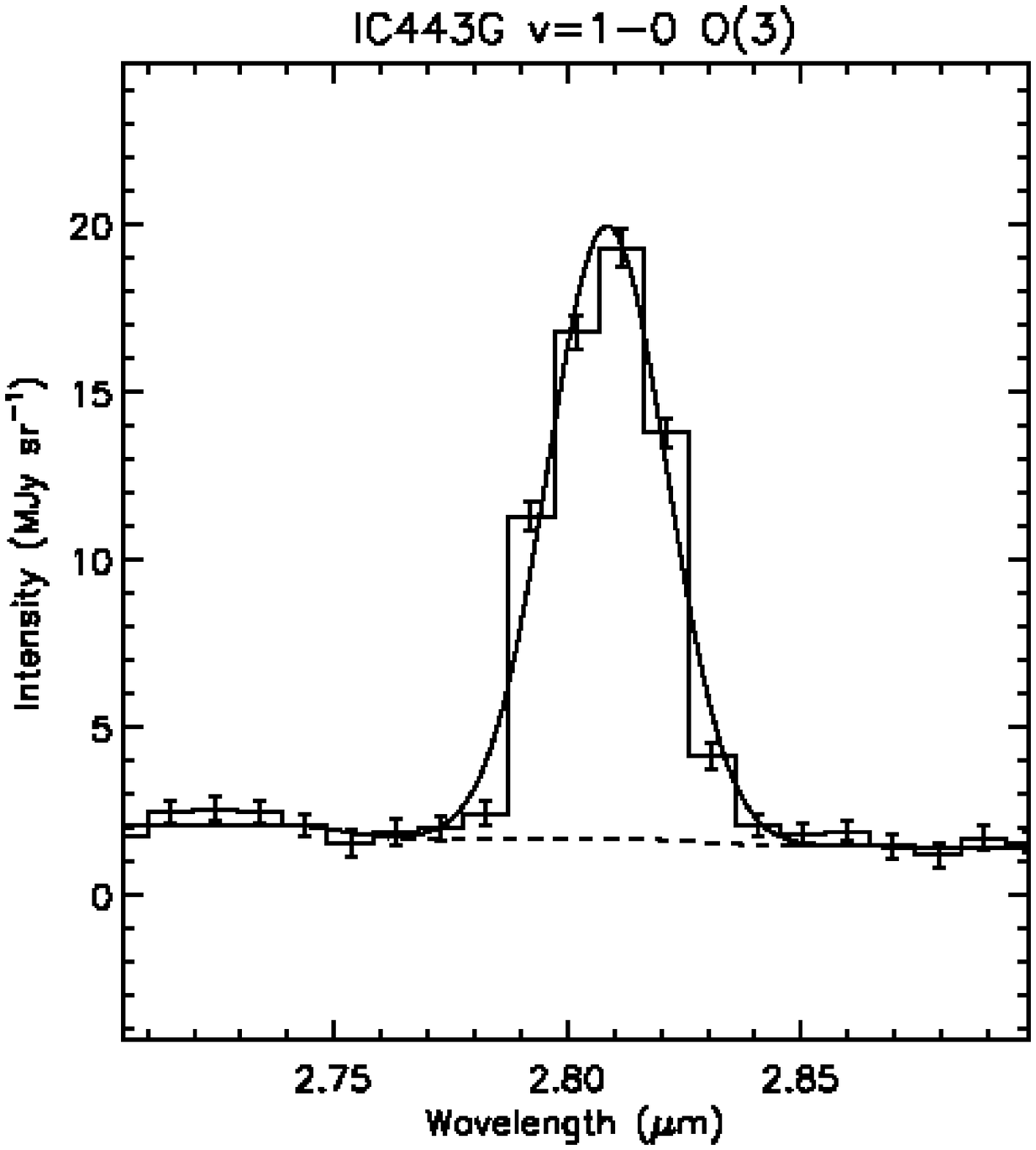} \\
\includegraphics[scale=0.33]{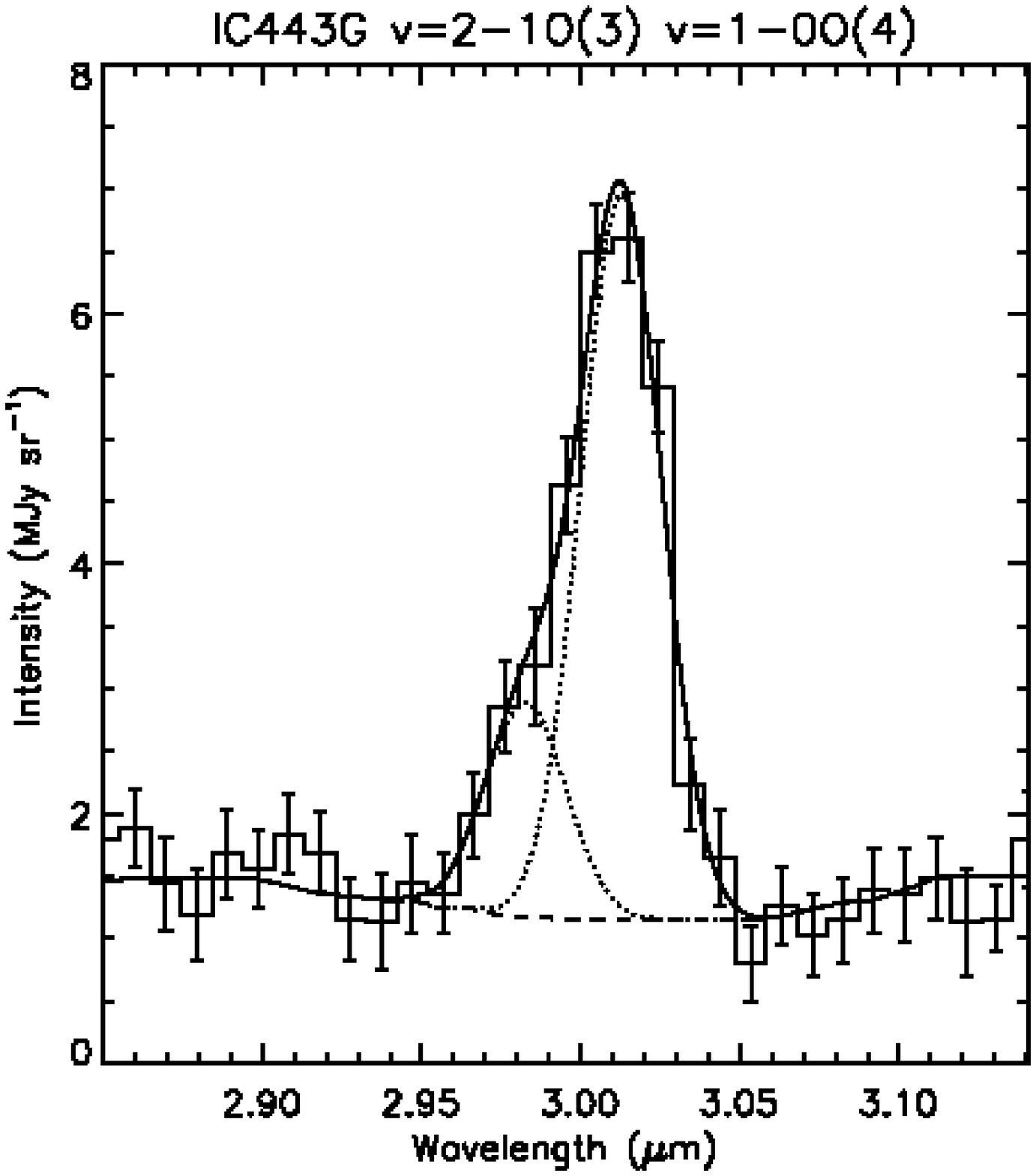}
\includegraphics[scale=0.33]{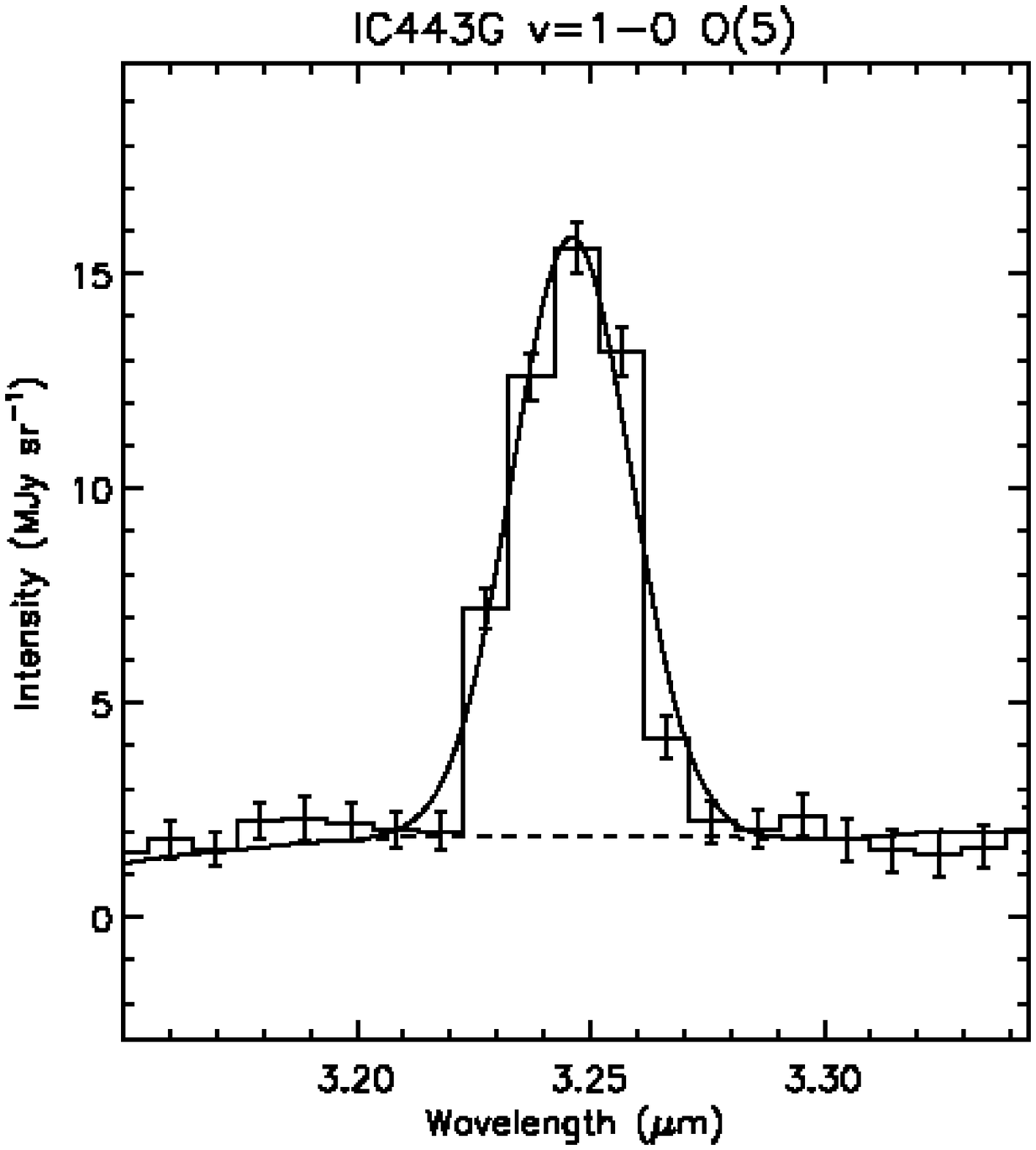}
\includegraphics[scale=0.33]{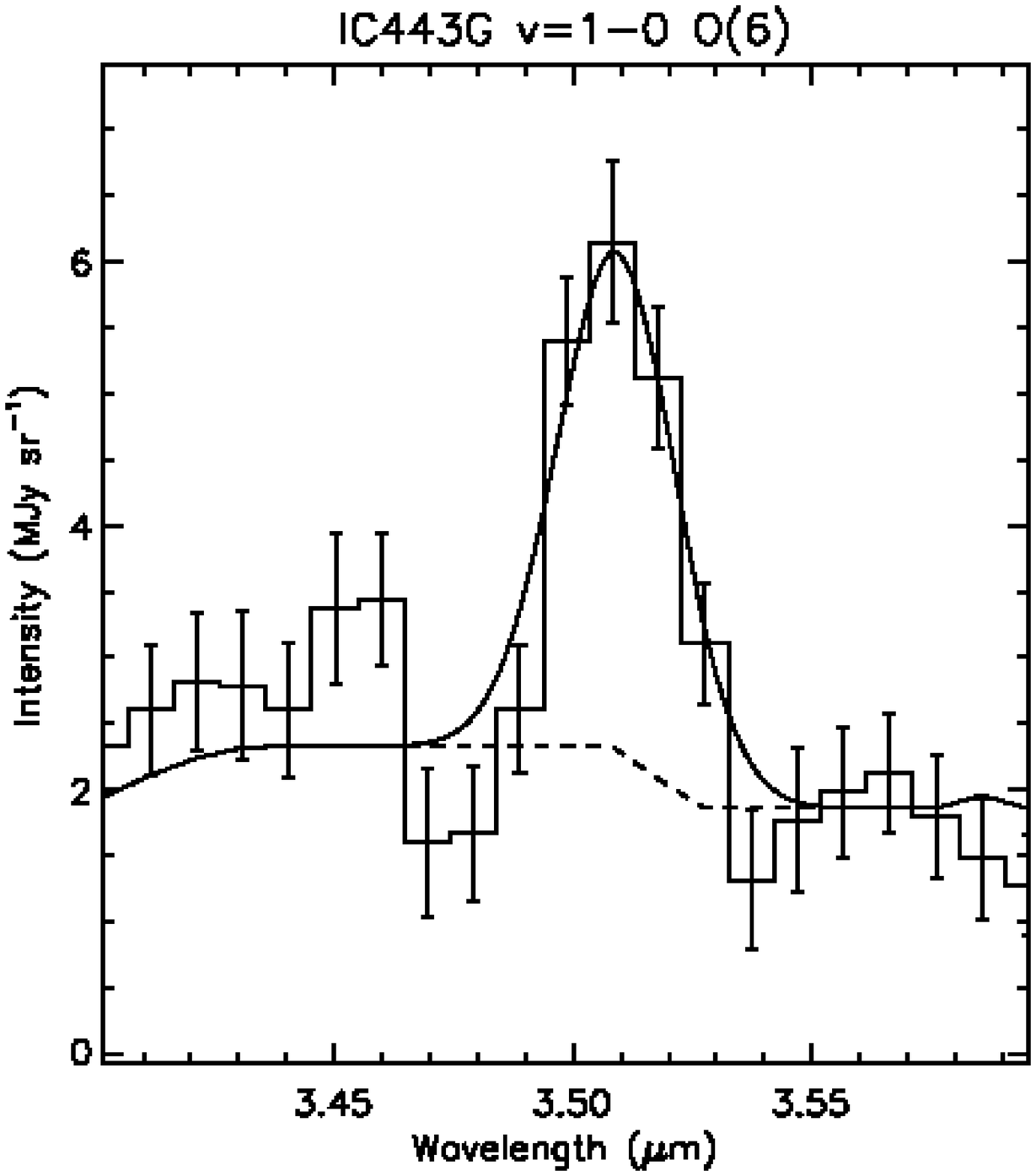} \\
\includegraphics[scale=0.33]{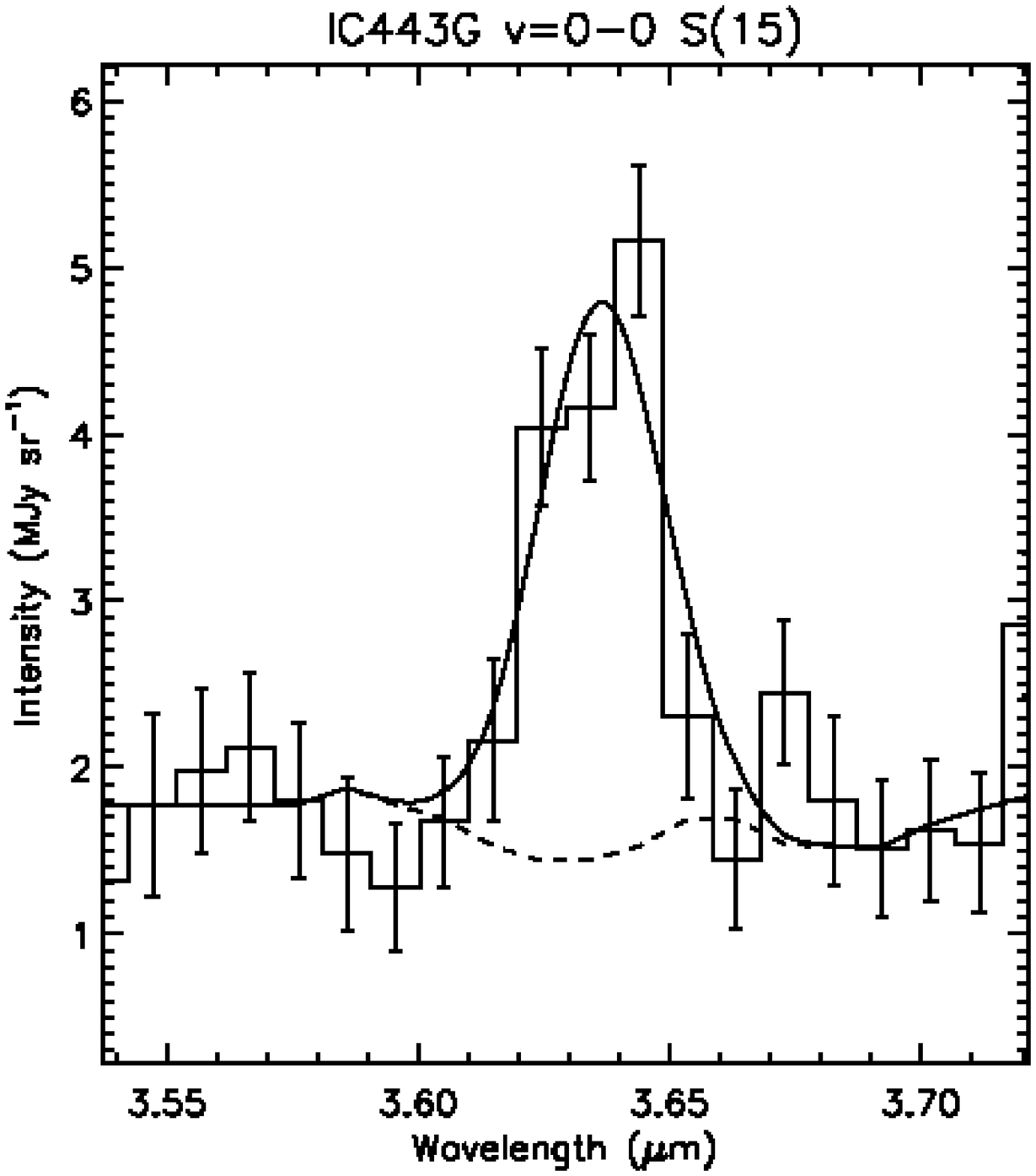}
\includegraphics[scale=0.33]{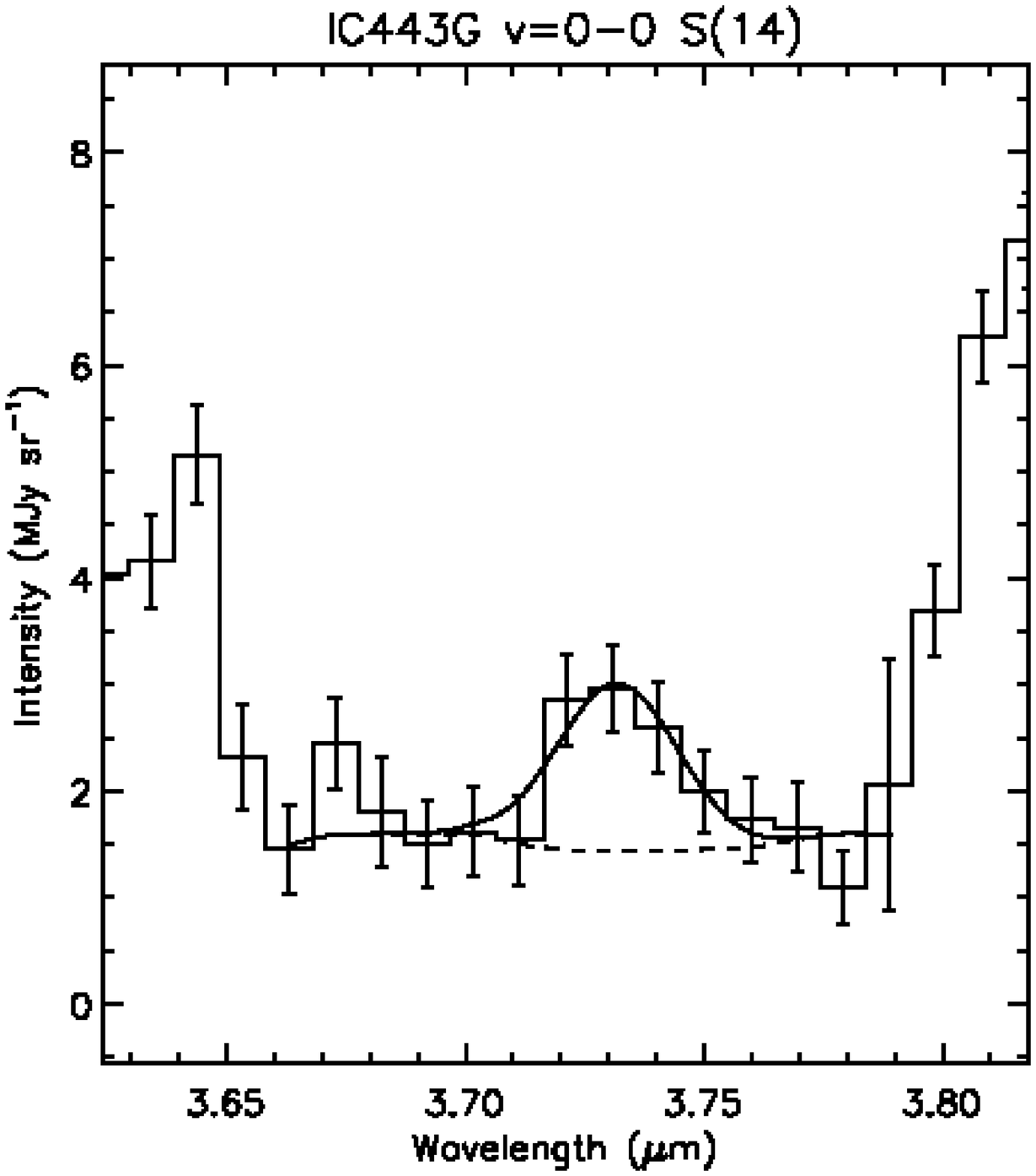}
\includegraphics[scale=0.33]{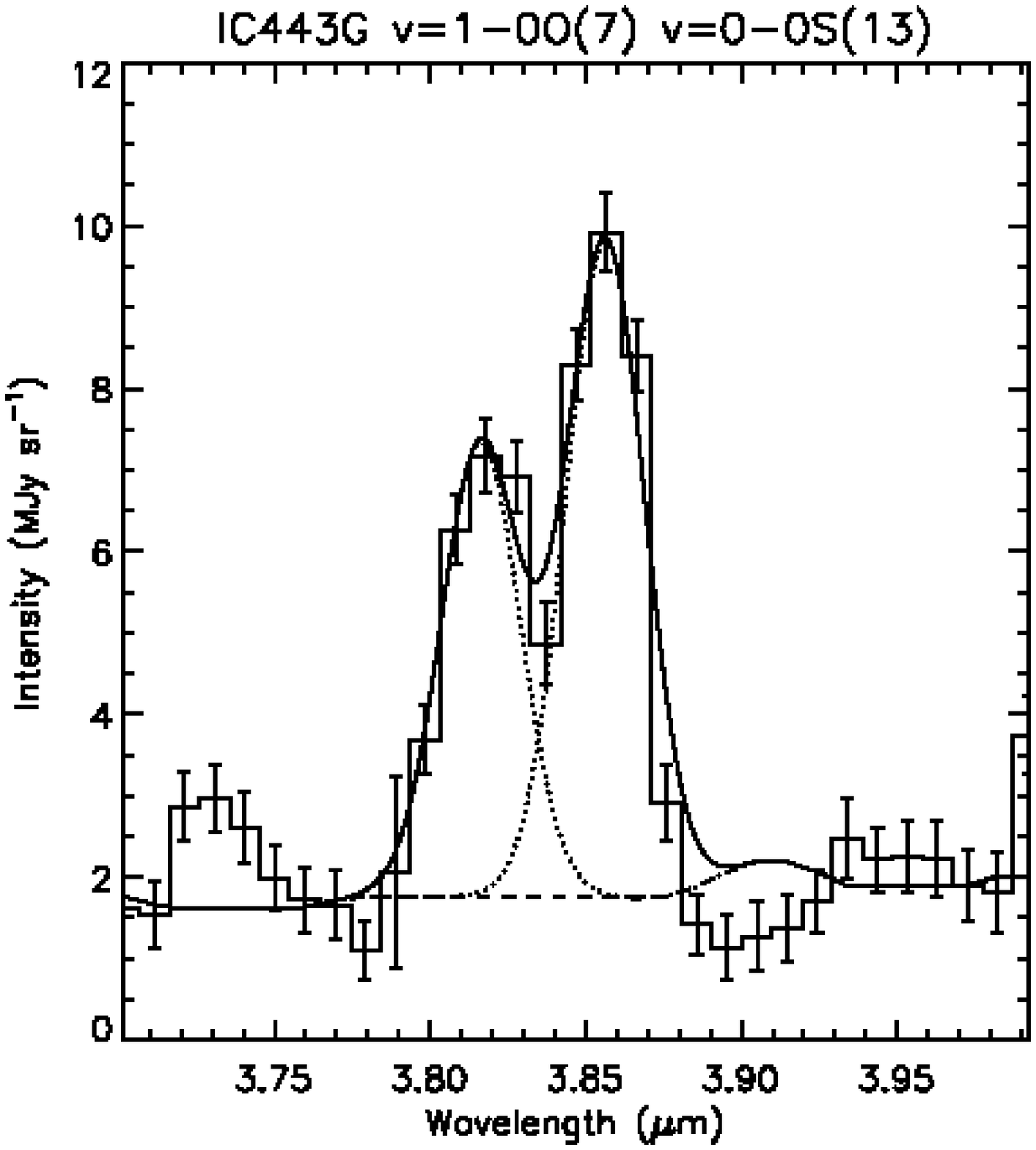} \\
}
\caption{Fitting for the \Htwo{} emission lines, observed toward the clump G. The rest is the same as Figure \ref{fig-fit-B}.} \label{fig-fit-G}
\end{figure}

\clearpage
\begin{figure}
\figurenum{5}
\center{
\includegraphics[scale=0.33]{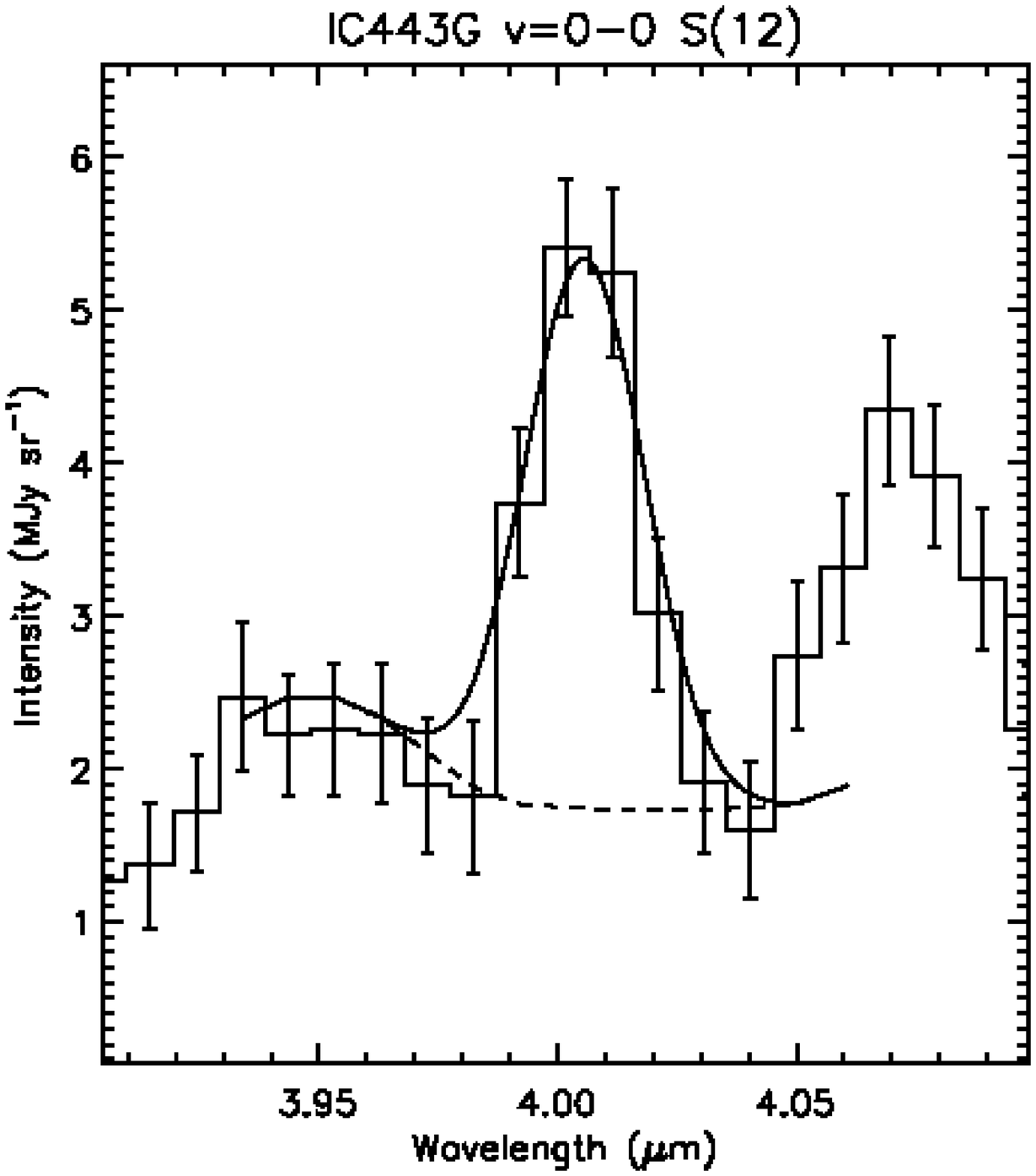}
\includegraphics[scale=0.33]{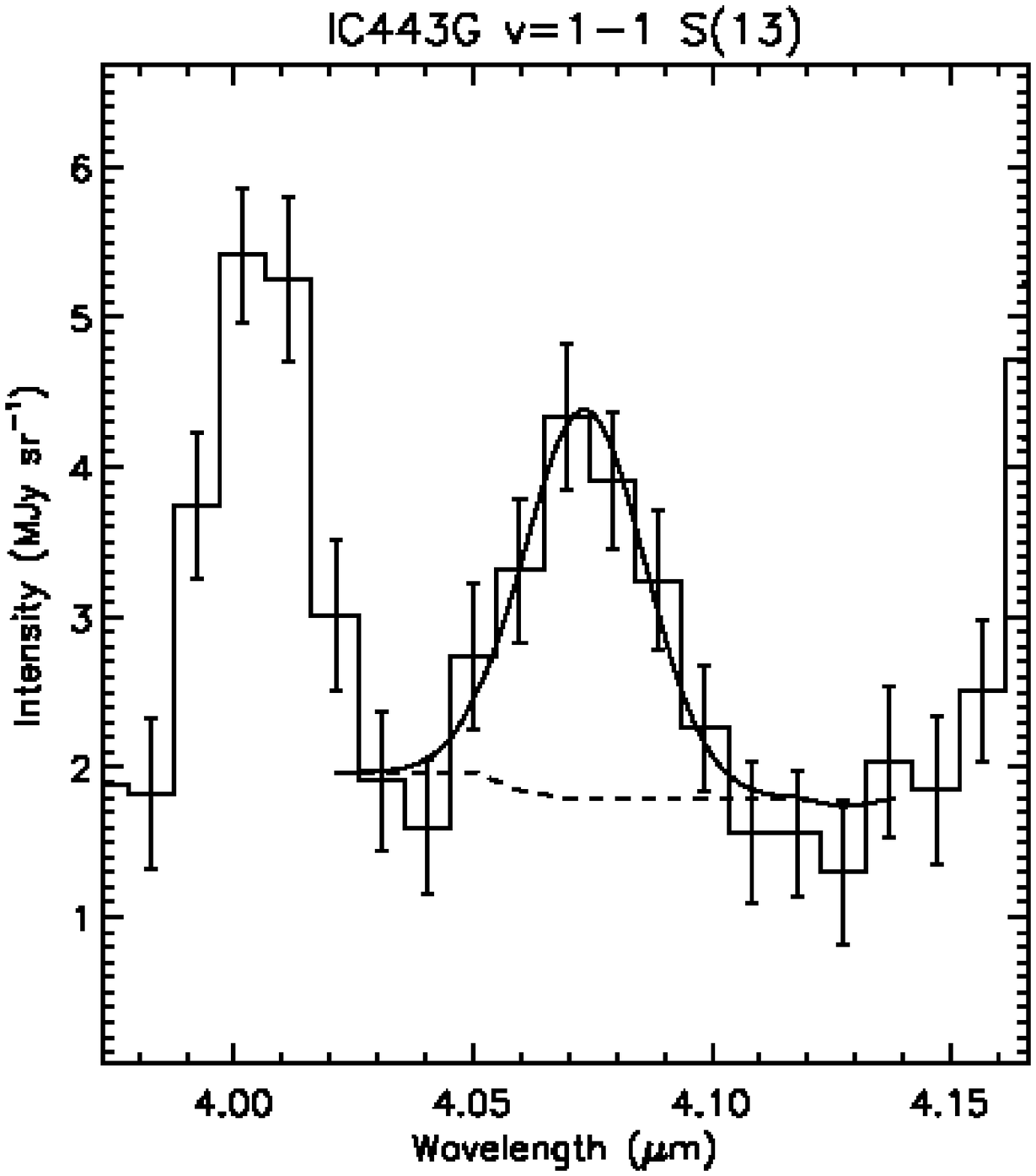}
\includegraphics[scale=0.33]{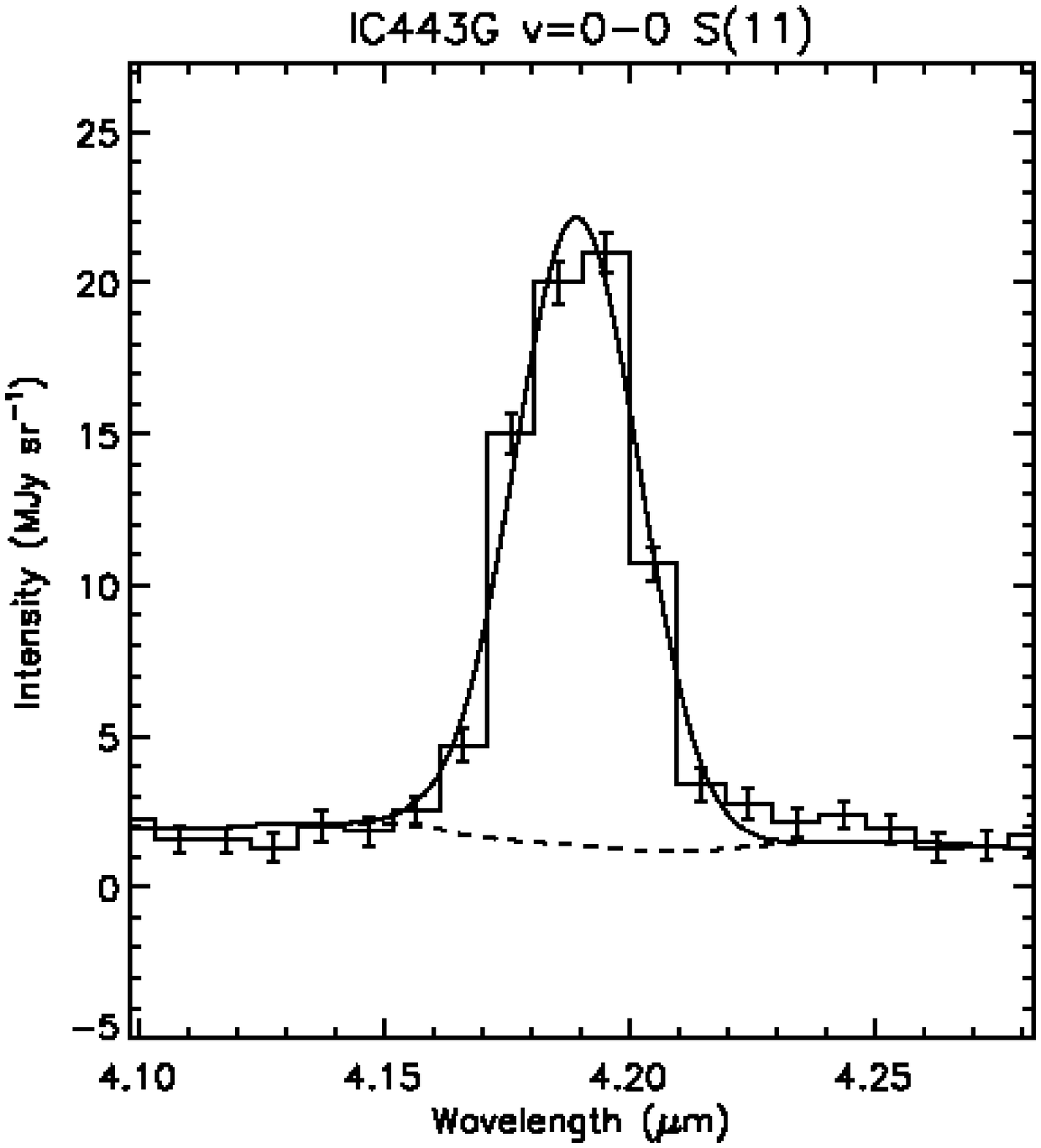} \\
\includegraphics[scale=0.33]{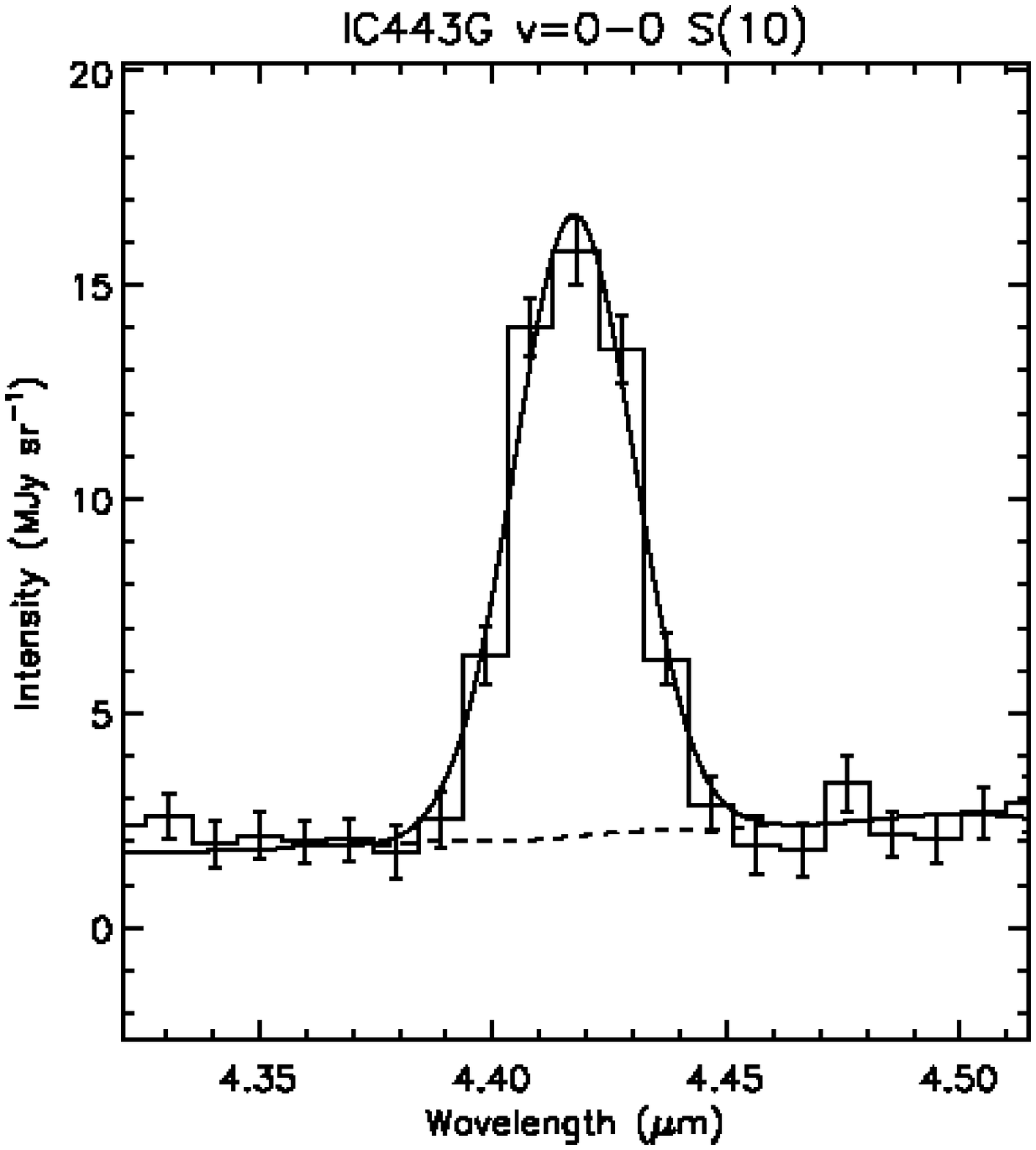}
\includegraphics[scale=0.33]{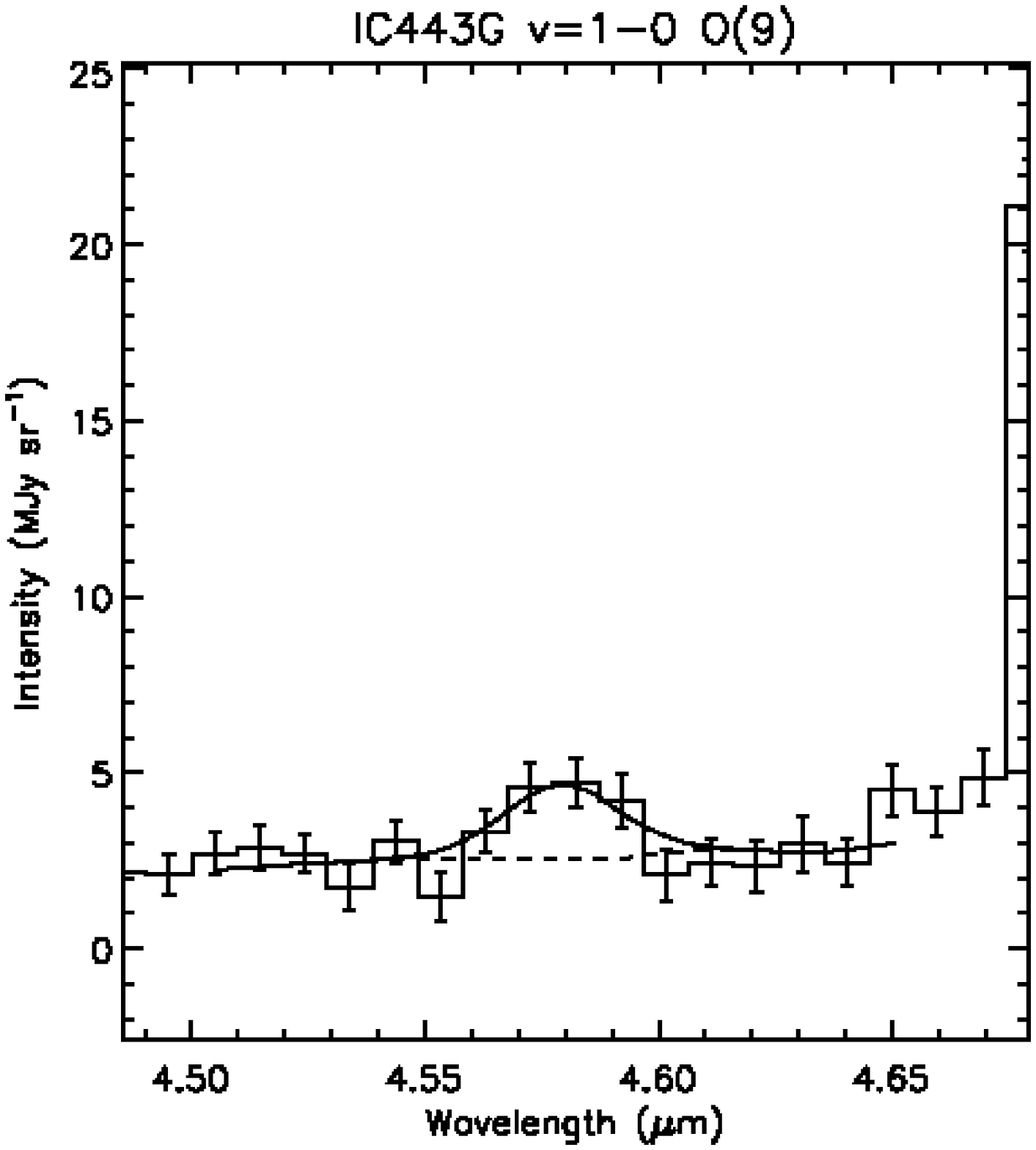}
\includegraphics[scale=0.33]{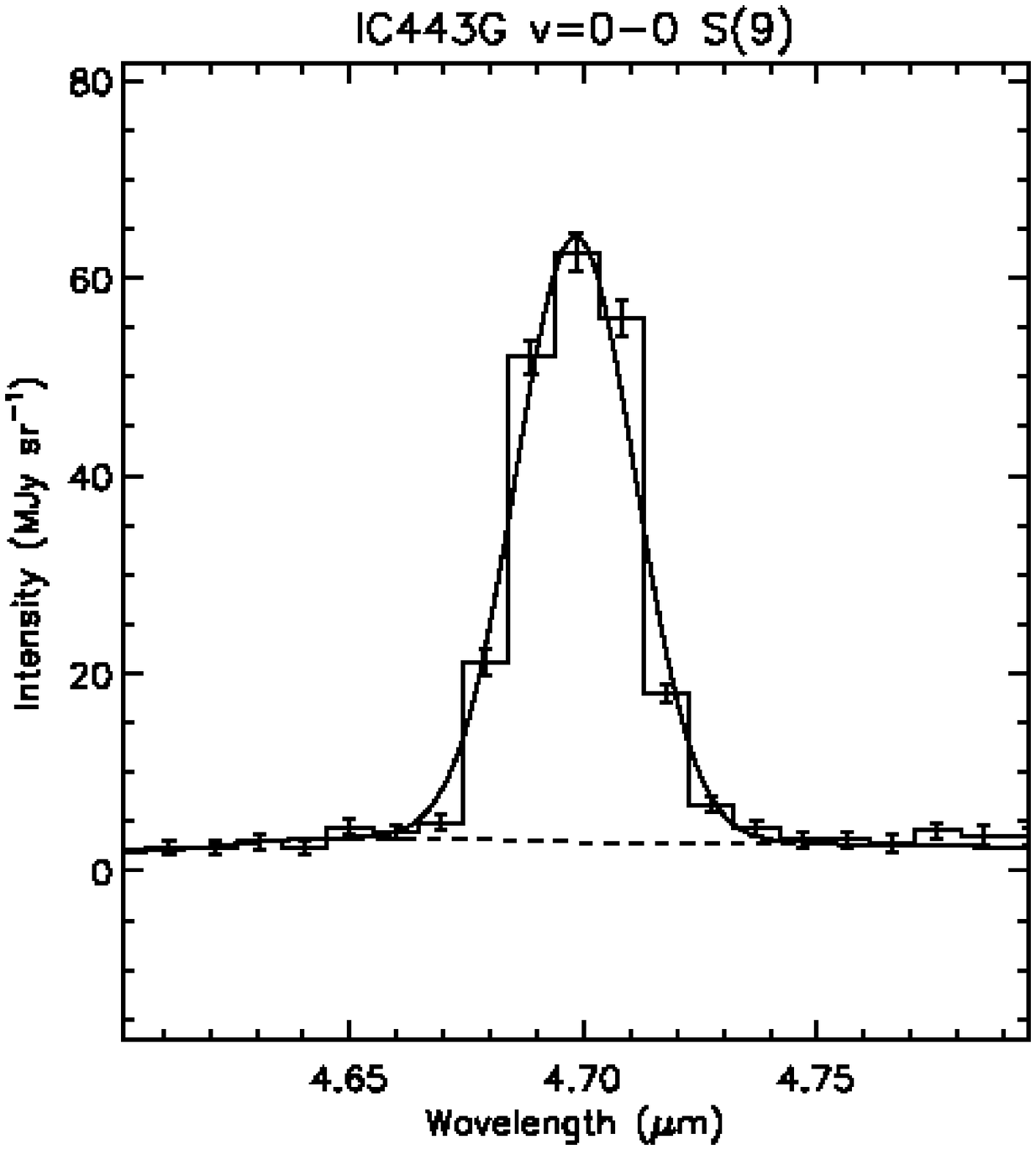}
}
\caption{Continued.}
\end{figure}

\clearpage
\begin{figure}
\center{
\includegraphics[scale=0.33]{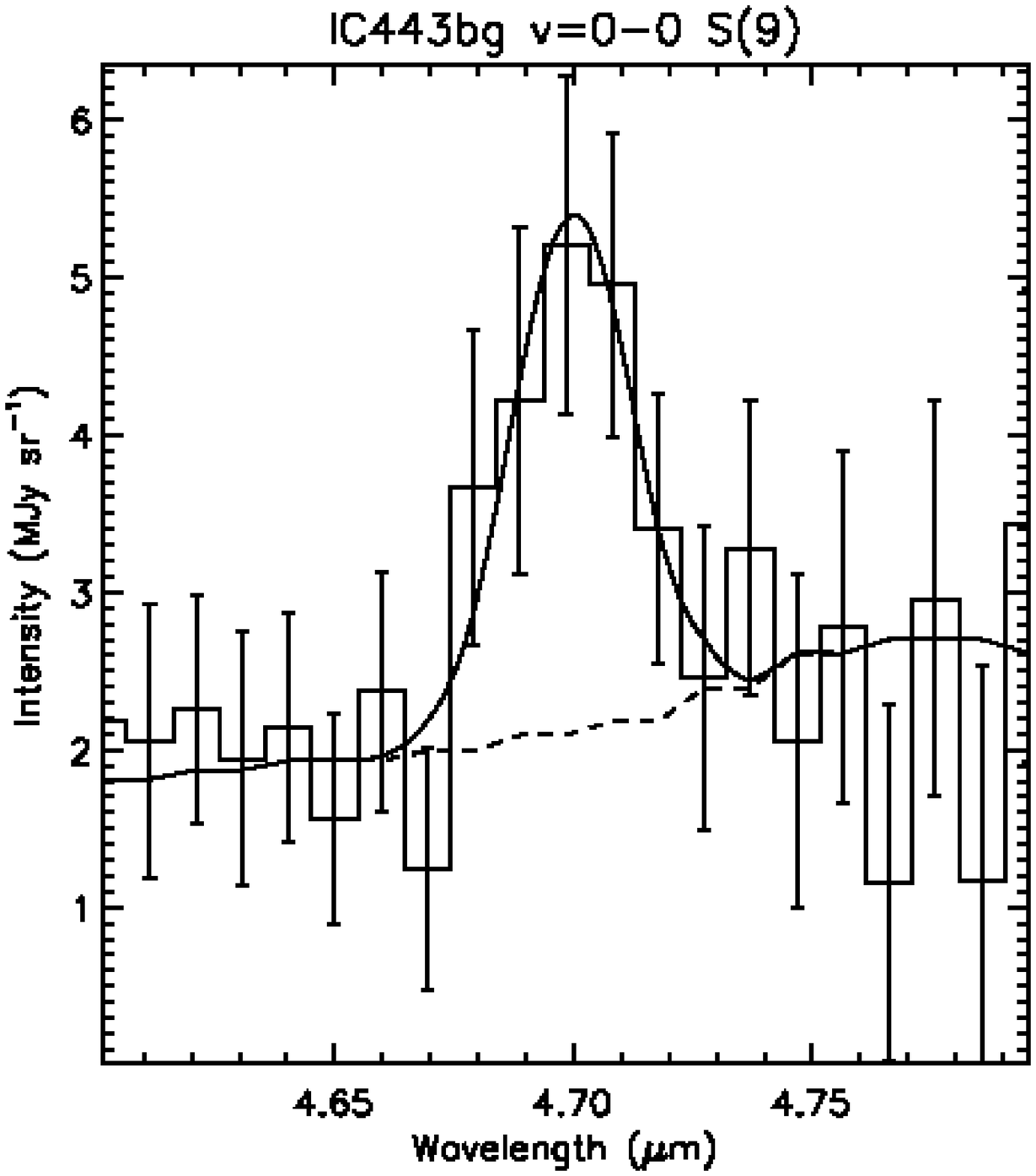}
}
\caption{Fitting for the \Htwo{} emission lines, observed toward the background (BG). The rest is the same as Figure \ref{fig-fit-B}.} \label{fig-fit-bg}
\end{figure}

\clearpage
\begin{figure}
\center{
\includegraphics[scale=0.46]{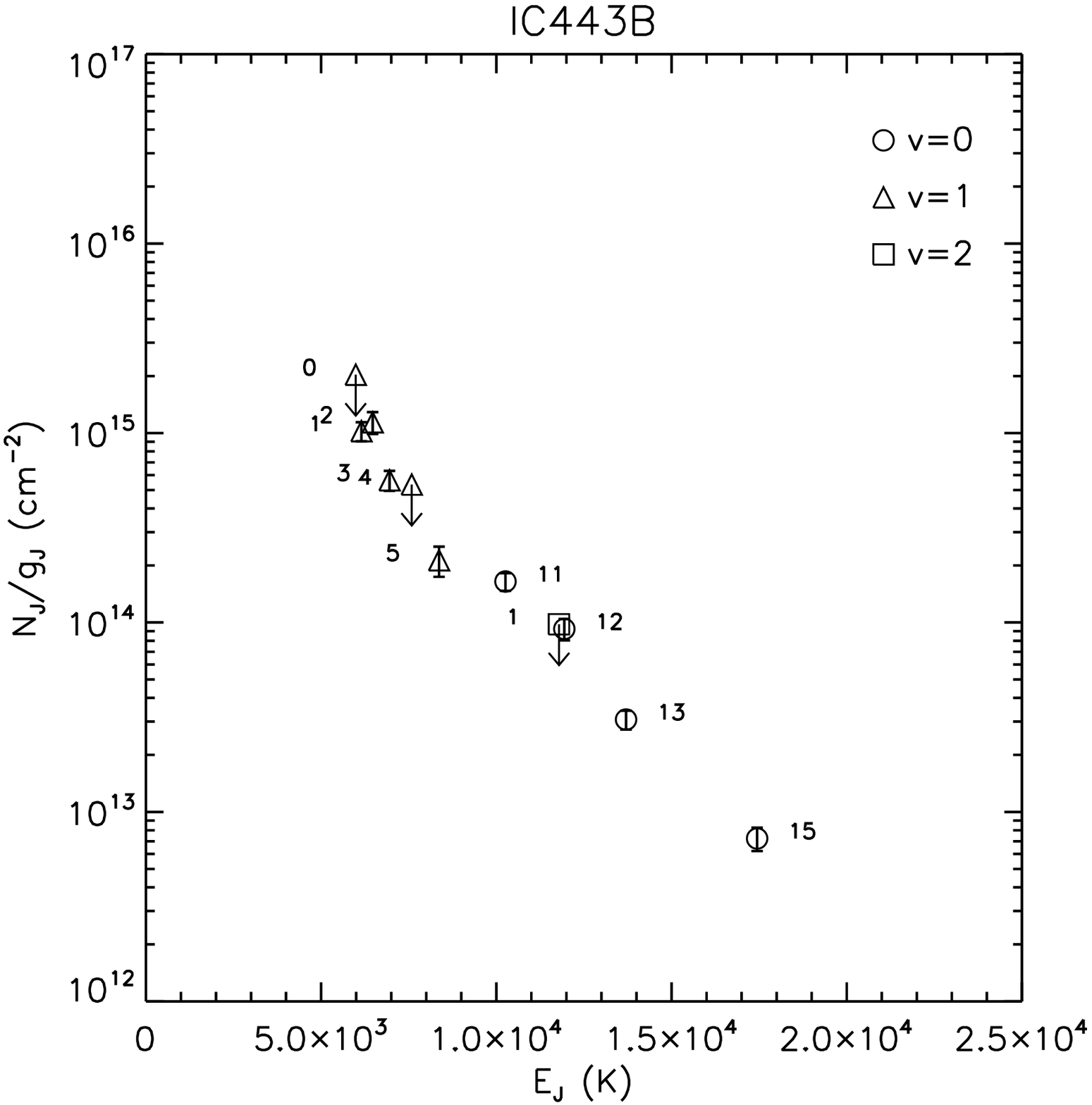}
\includegraphics[scale=0.46]{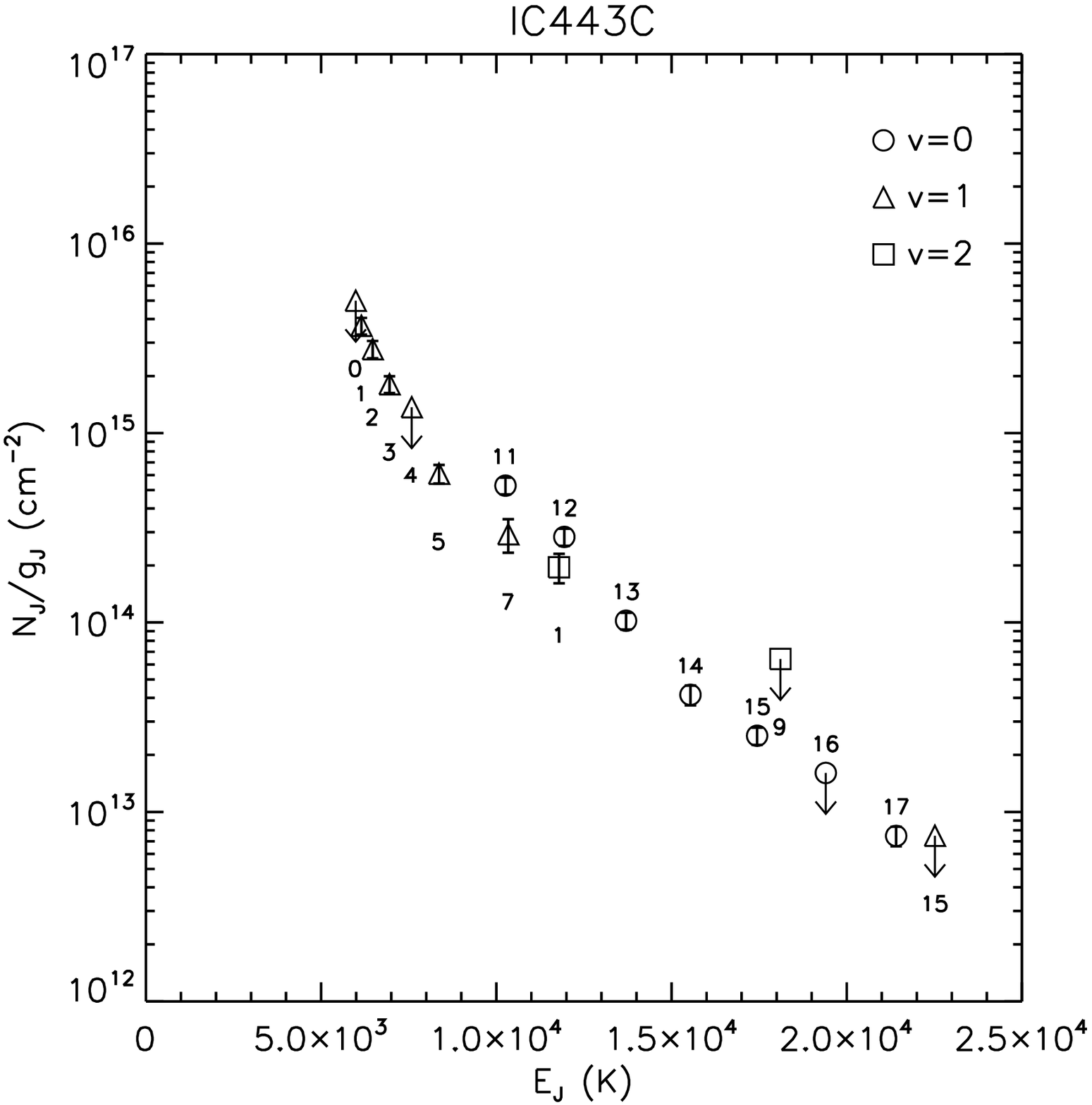}\\
\includegraphics[scale=0.46]{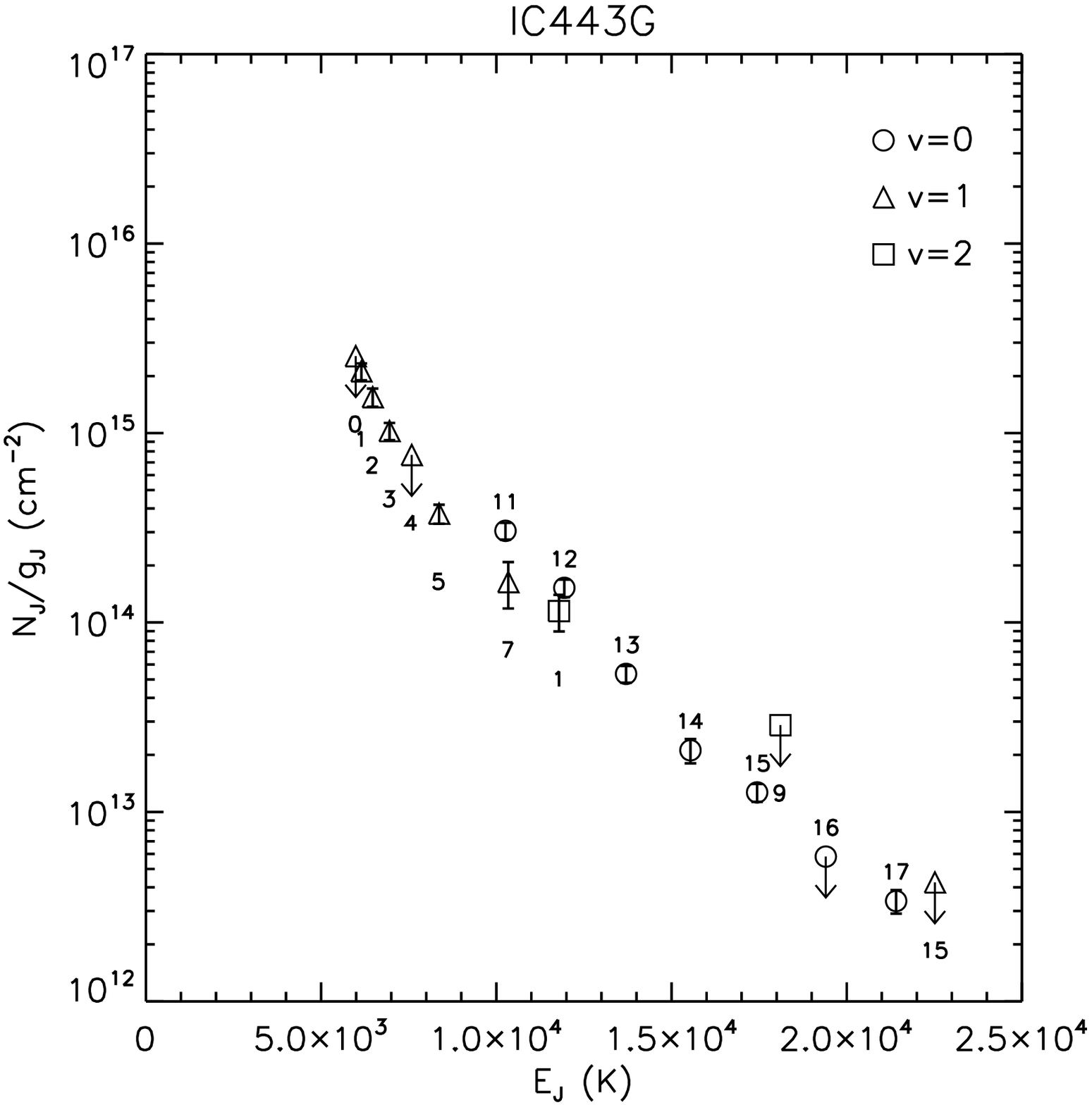}
\includegraphics[scale=0.46]{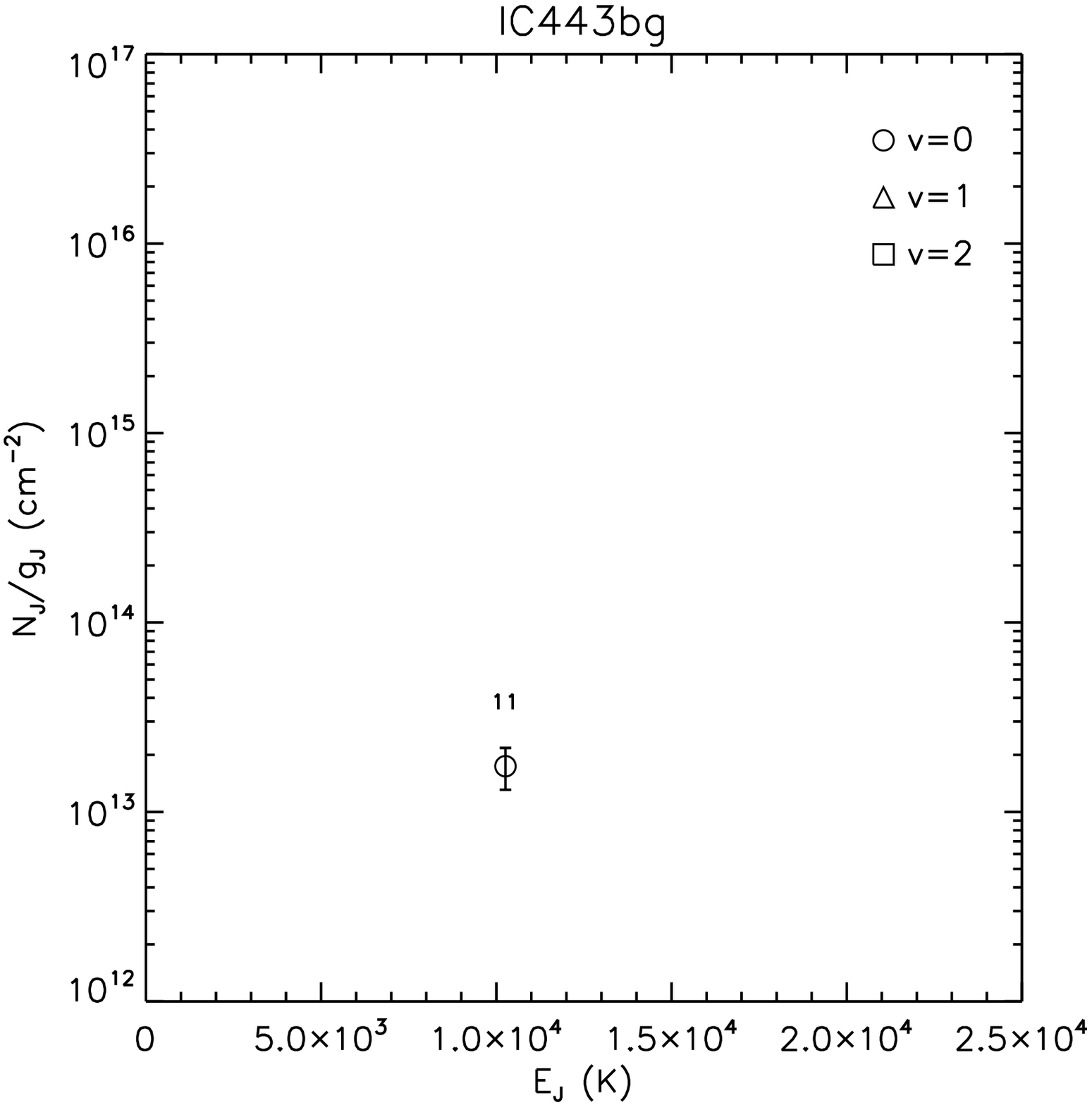}
}
\caption{The extinction-corrected population diagrams for the clumps B (\emph{top-left}), C (\emph{top-right}), G (\emph{bottom-left}), and the background (\emph{bottom-right}). The extinctions were corrected, using $A_V=13.5$ for the clumps B, C, and background \citep{Neufeld(2008)ApJ_678_974}, $A_V=10.8$ for the clump G \citep{Richter(1995)ApJ_454_277}, respectively. The rotational quantum number ($J$) is printed out near the corresponding point.} \label{fig-pop}
\end{figure}

\clearpage
\begin{figure}
\center{
\includegraphics[scale=0.45,viewport=0 -60 500 450]{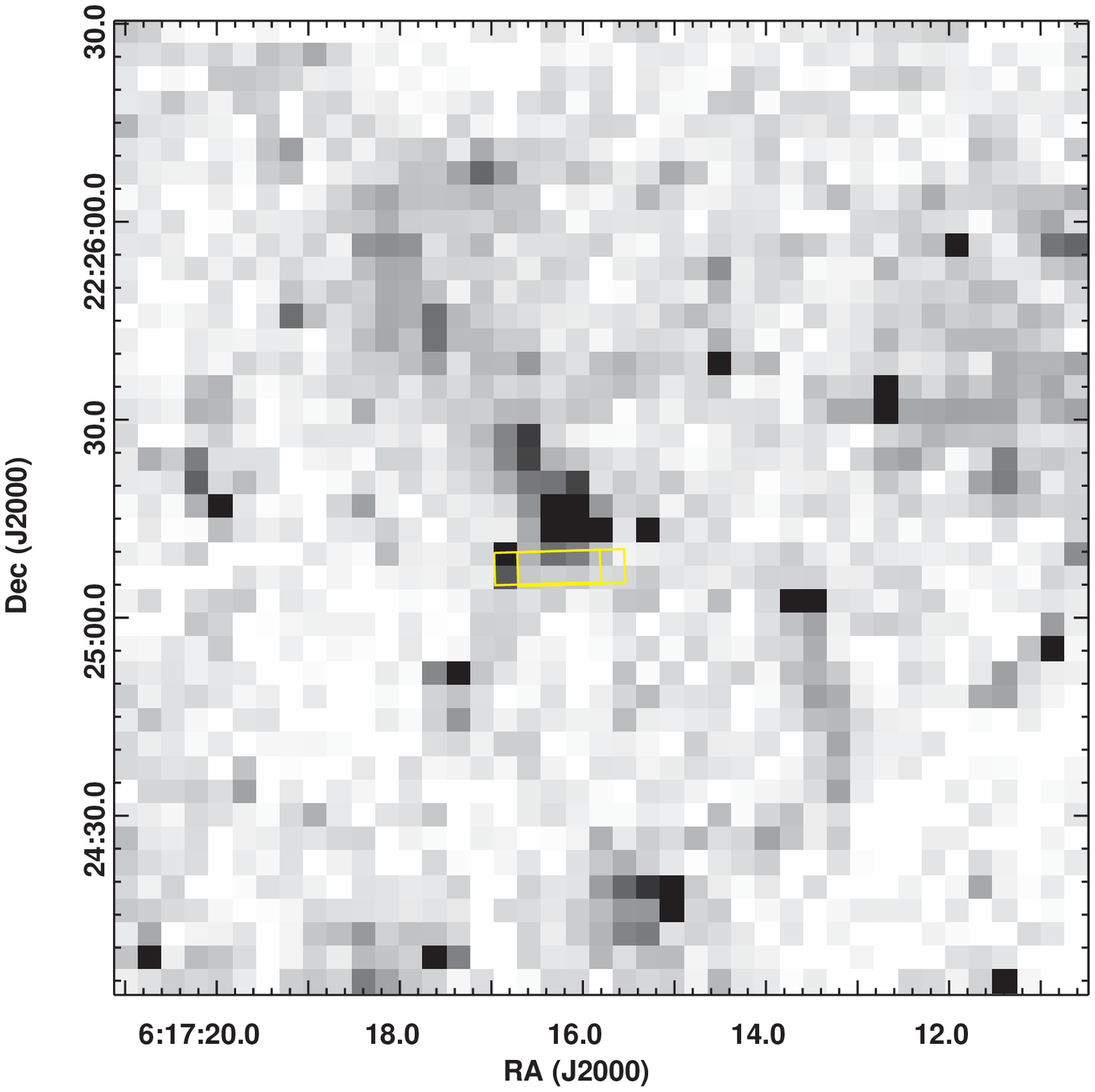} \\
\includegraphics[scale=0.45,viewport=0 0 500 450]{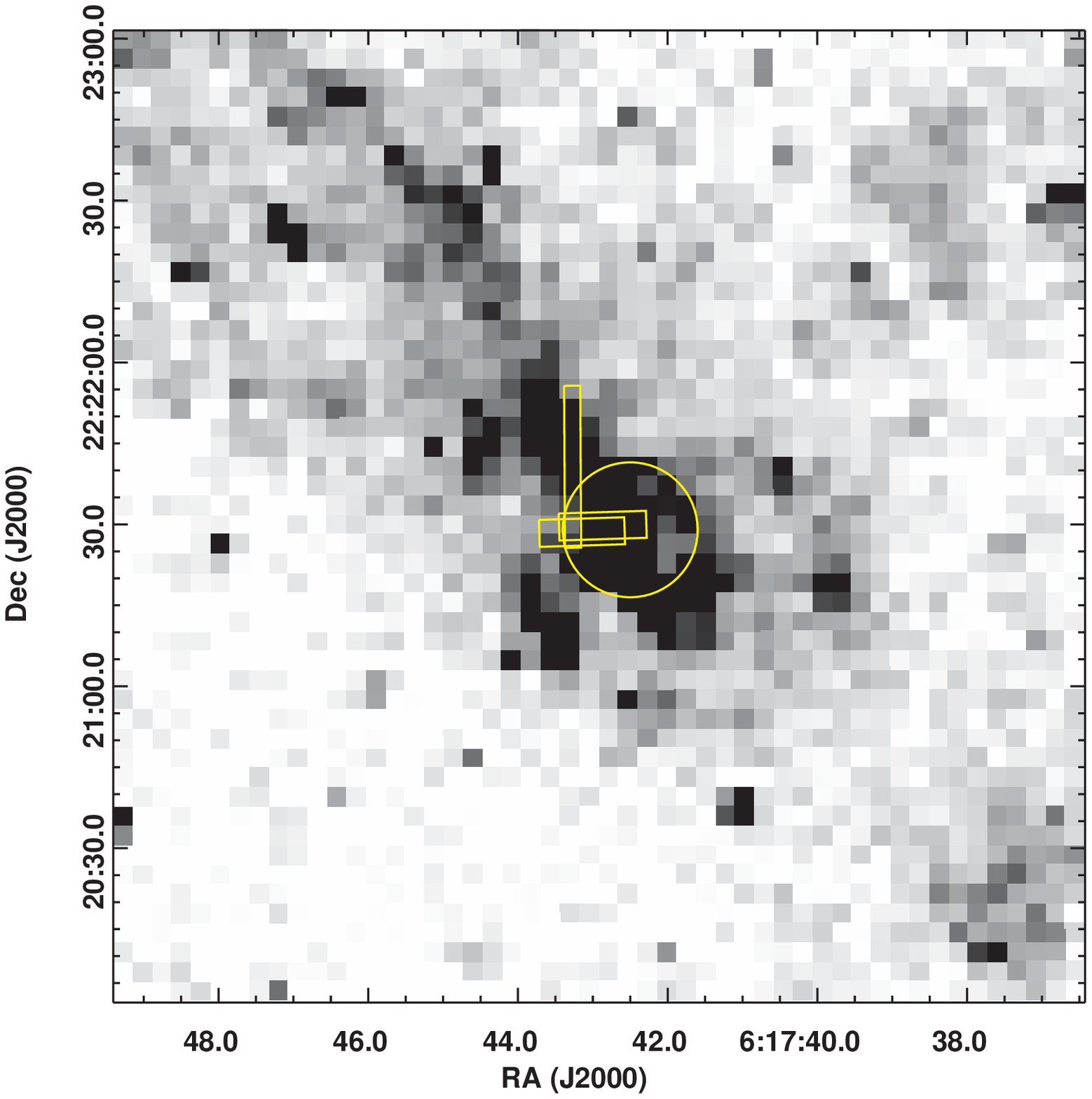}
\includegraphics[scale=0.45,viewport=0 0 500 450]{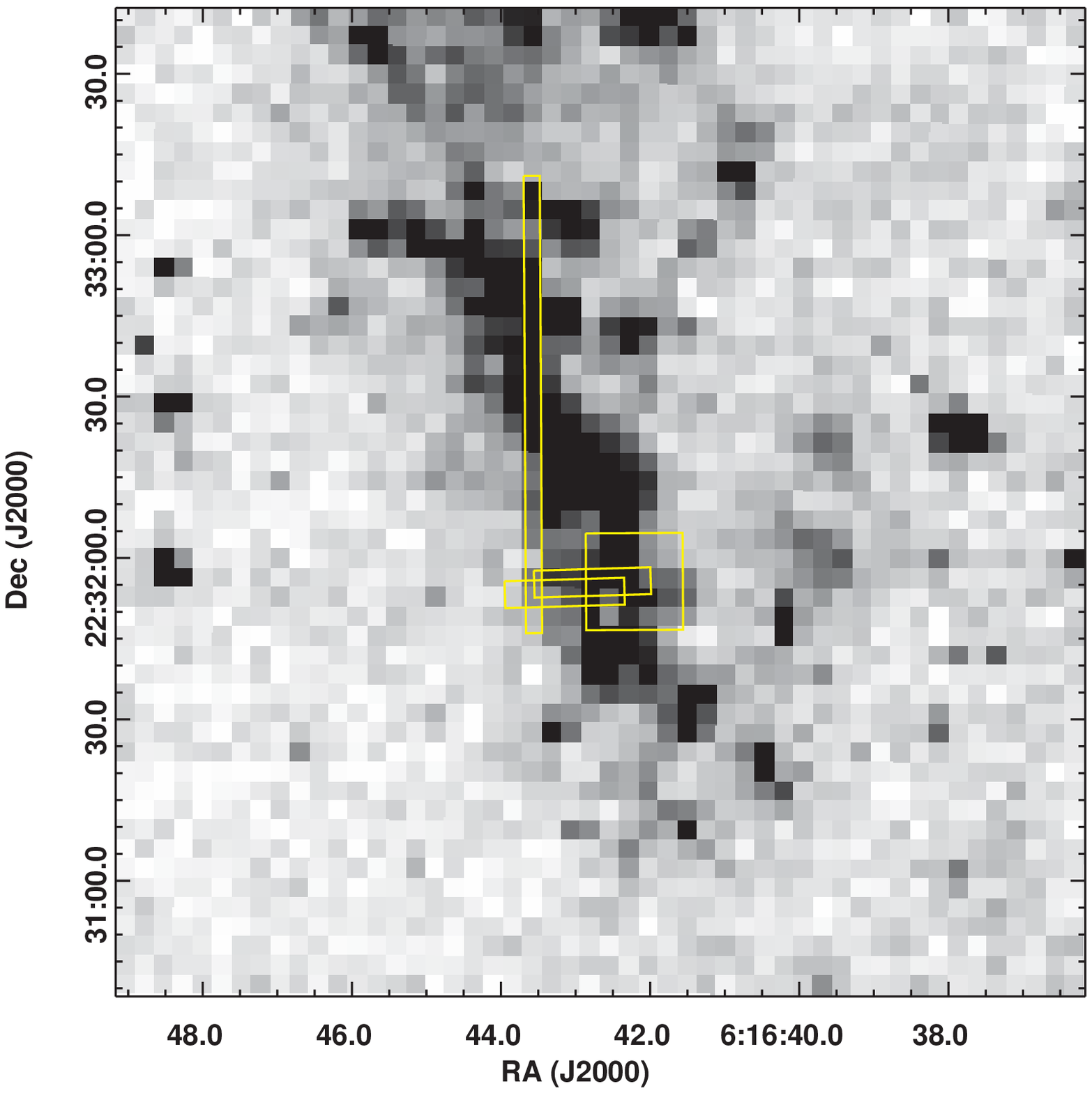}
}
\caption{The areas of spectra extraction around the clumps B (\emph{top}), C (\emph{bottom-left}), and G (\emph{bottom-right}). The background image is the same with Figure \ref{fig-slit}. The \emph{two horizontal rectangles} are the extraction areas for \akari's data. Each of the two indicates each exposure (cf. Table \ref{tbl-obs}). The \emph{vertical rectangles} are for the UKIRT CGS4 data \citep{Richter(1995)ApJ_454_277}; we adopted the ``position 1'' and ``position 3'' data for the clumps C and G, respectively. For the clump G, we indicate the whole slit area, since \cite{Richter(1995)ApJ_454_277} did not mention where the extraction area is. The \emph{circle} is the FWHM ($25''$) of the Gaussian taper, used for the spectra extraction of \spitzer{} data \citep{Neufeld(2007)ApJ_664_890}. The \emph{square} is the extraction area for \iso{} data \citep{Cesarsky(1999)A&A_348_945}; we adopted the $3\times3$ pixels area around the peak B.} \label{fig-specext} \label{fig-specext}
\end{figure}

\clearpage
\begin{figure}
\center{
\includegraphics[scale=0.46]{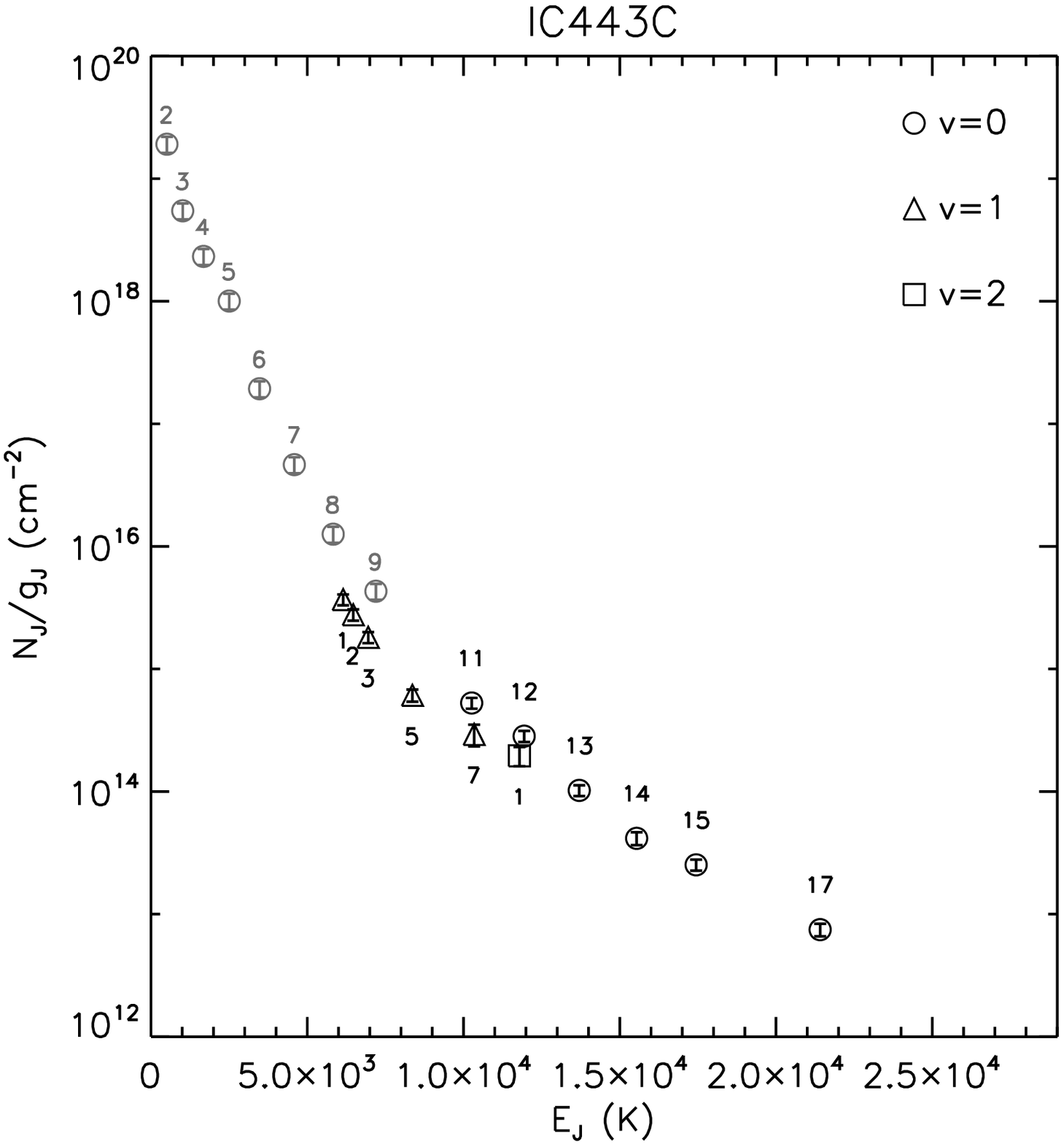}
\includegraphics[scale=0.46]{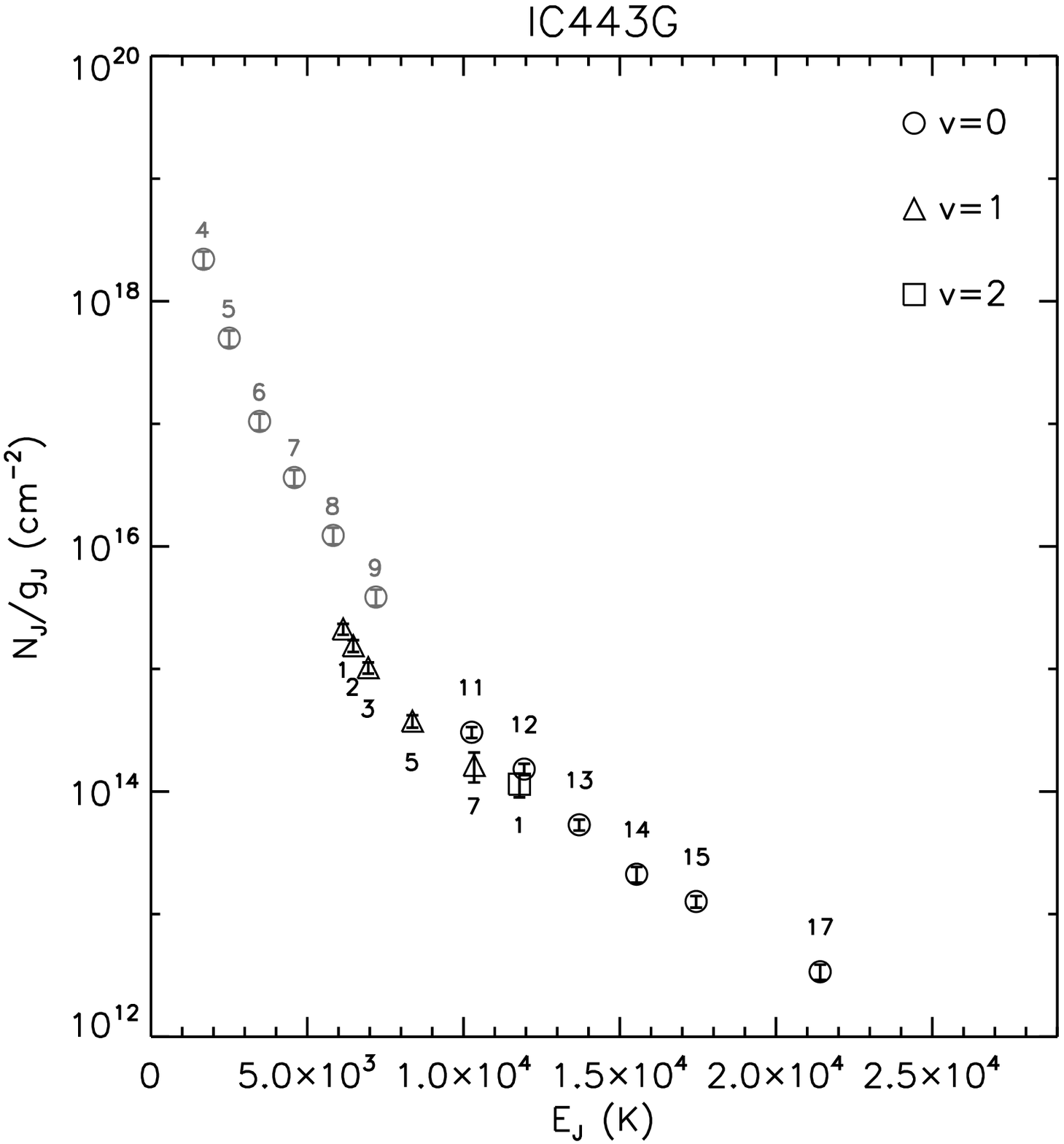} \\
\includegraphics[scale=0.46]{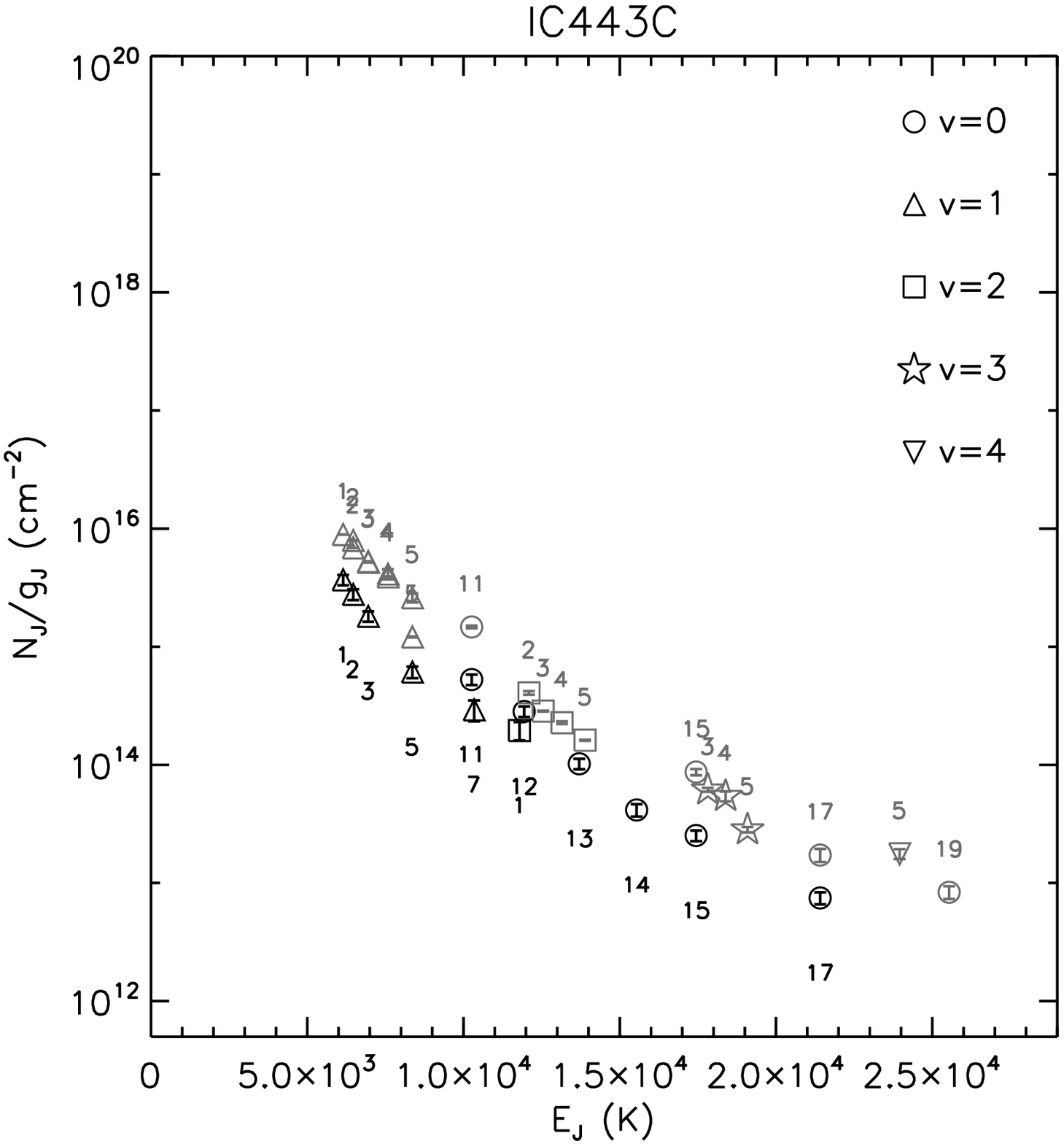}
\includegraphics[scale=0.46]{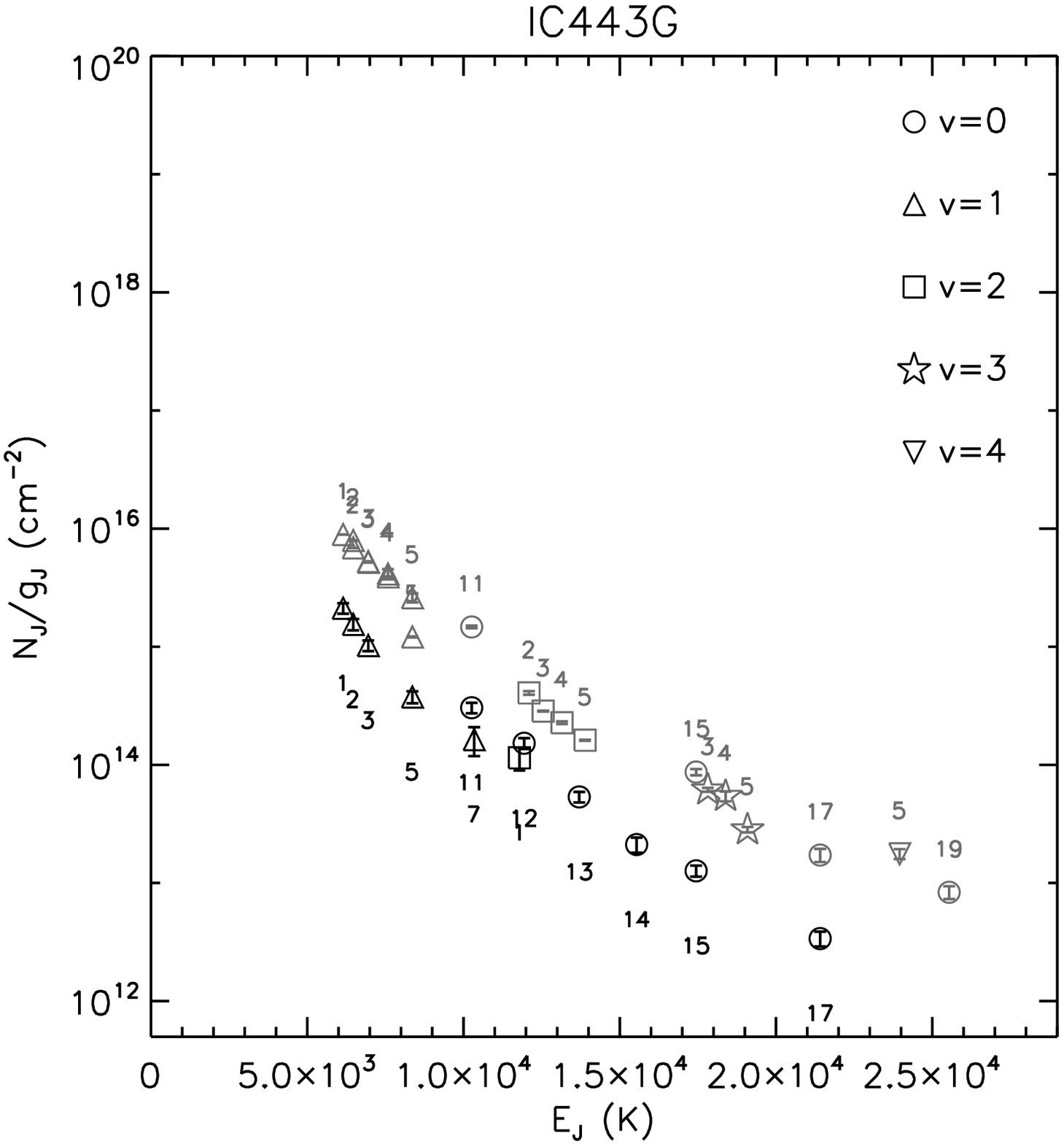} \\
}
\caption{The extinction-corrected population diagrams of \akari{} IRC data (\emph{black symbols}), in comparison with previous observations (\emph{grey symbols}), for the clump C (\emph{left panels}) and clump G (\emph{right panels}): (\emph{left-top}) \akari{} IRC and \spitzer{} IRS, (\emph{left-middle}) \akari{} IRC and UKIRT CGS4, (\emph{left-bottom}) \spitzer{} IRS and UKIRT CGS4, (\emph{right-top}) \akari{} IRC and \iso{} ISOCAM, (\emph{right-middle}) \akari{} IRC and UKIRT CGS4, (\emph{right-bottom}) \iso{} ISOCAM and UKIRT CGS4. (The \emph{bottom} figures are displayed in the following page.) The UKIRT CGS4 (near-IR), \iso{} ISOCAM (mid-IR), and \spitzer{} IRS data (mid-IR) are from \cite{Richter(1995)ApJ_454_277}, \cite{Cesarsky(1999)A&A_348_945}, and \cite{Neufeld(2007)ApJ_664_890}, respectively. The rotational quantum number ($J$) is printed out near the corresponding point.} \label{fig-popall}
\end{figure}

\clearpage
\begin{figure}
\figurenum{9}
\center{
\includegraphics[scale=0.46]{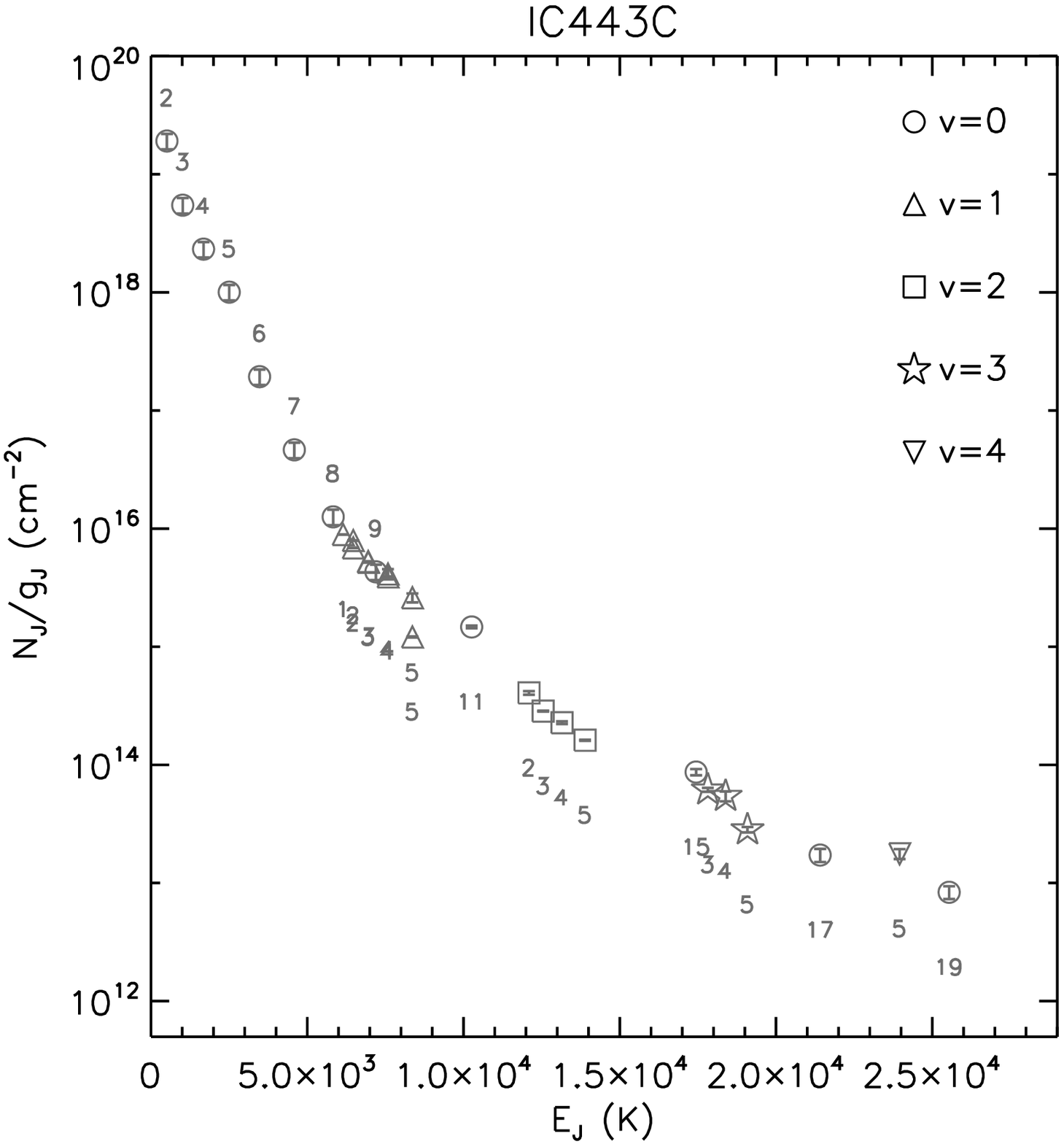}
\includegraphics[scale=0.46]{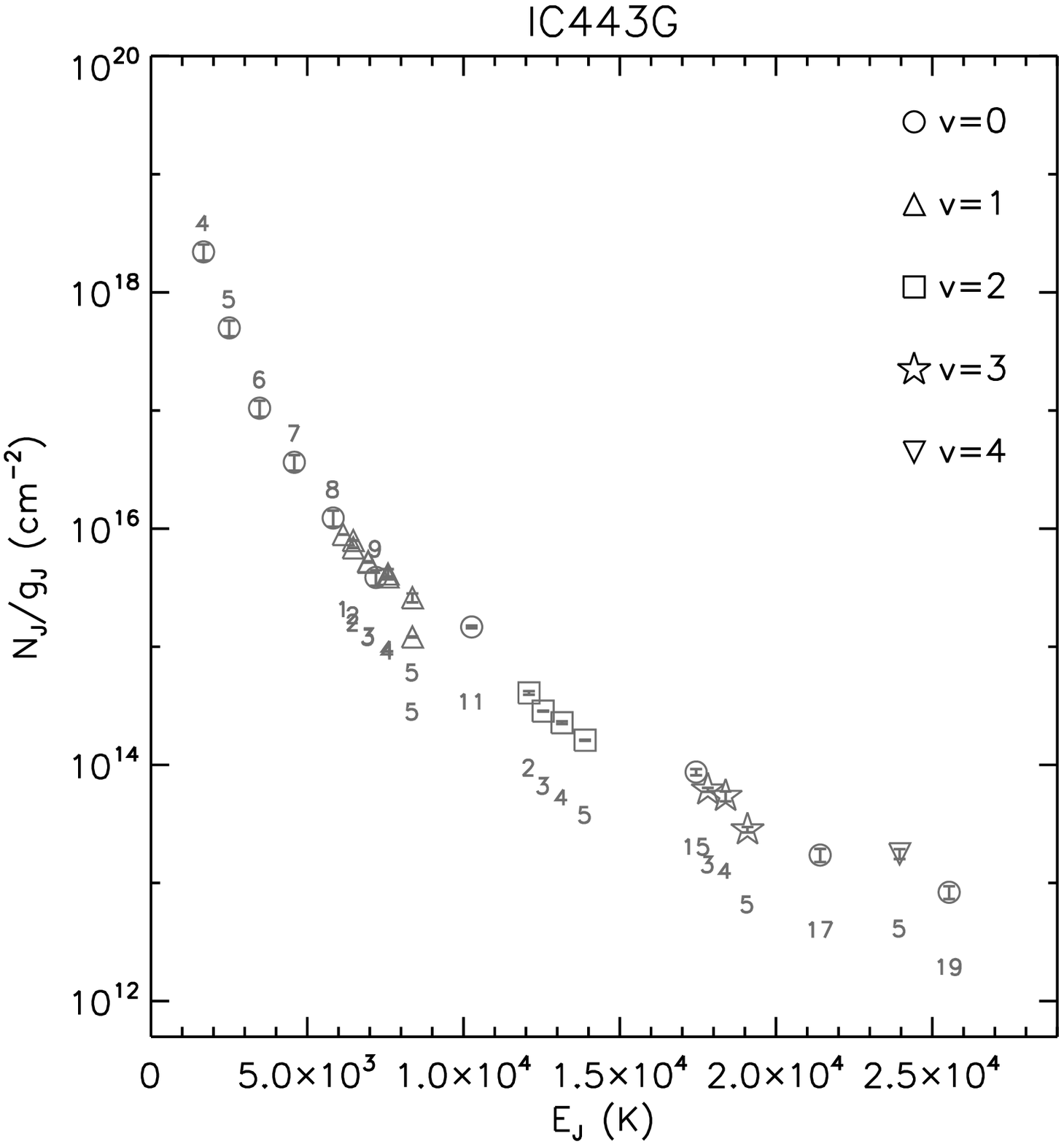}
}
\caption{Continued.}
\end{figure}

\clearpage
\begin{figure}
\center{
\includegraphics[scale=0.46]{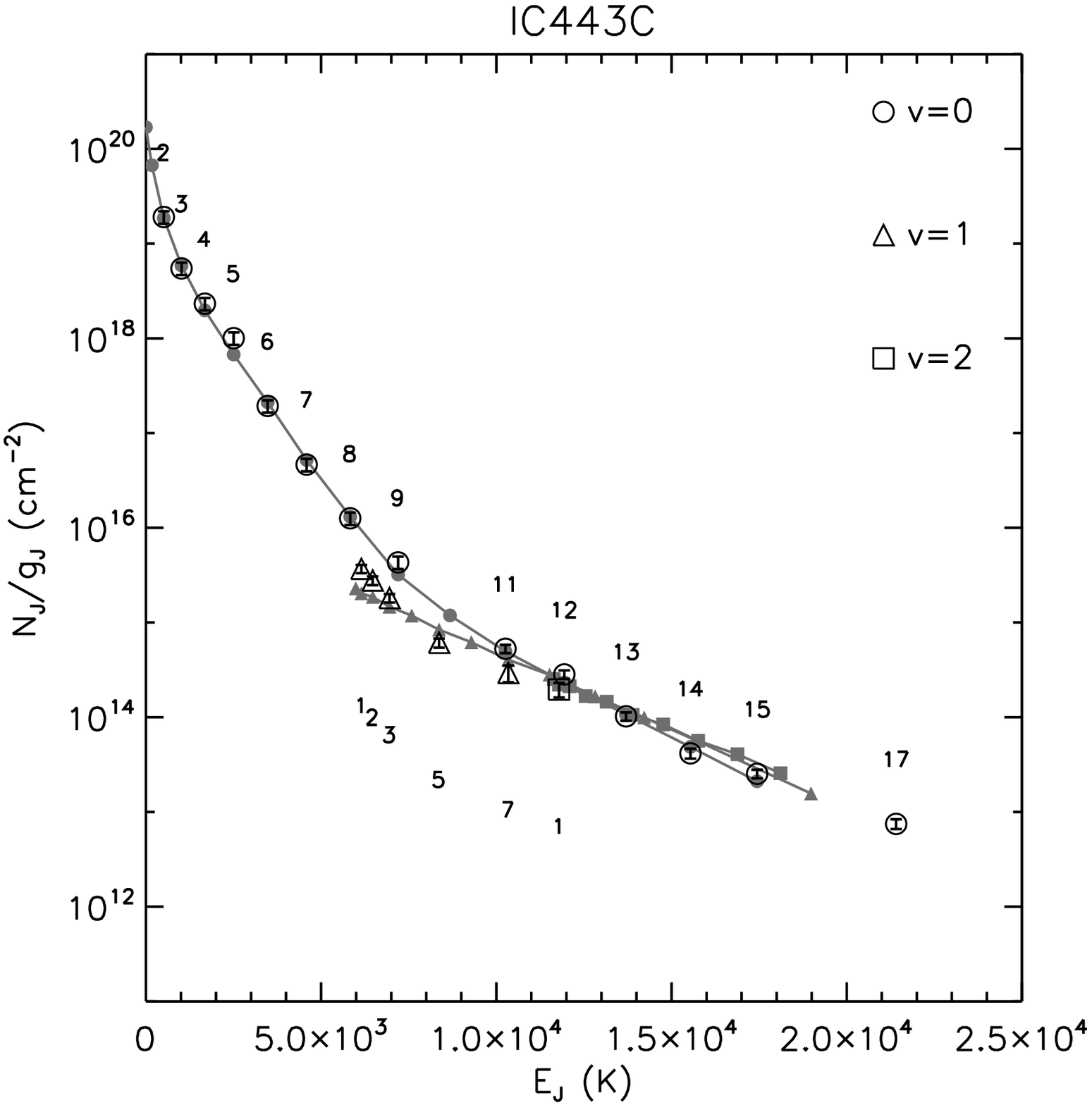}
\includegraphics[scale=0.46]{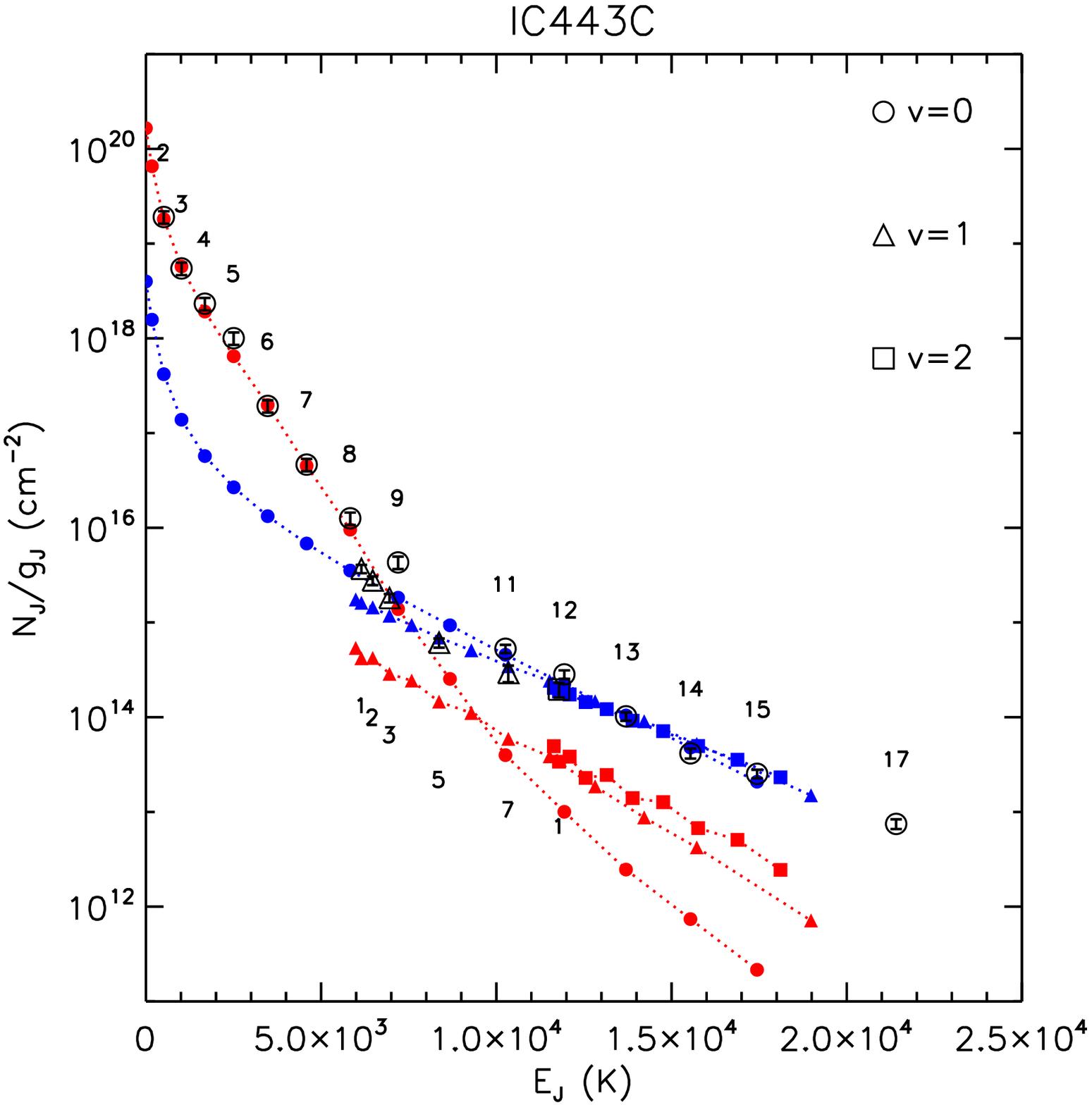}\\
\includegraphics[scale=0.46]{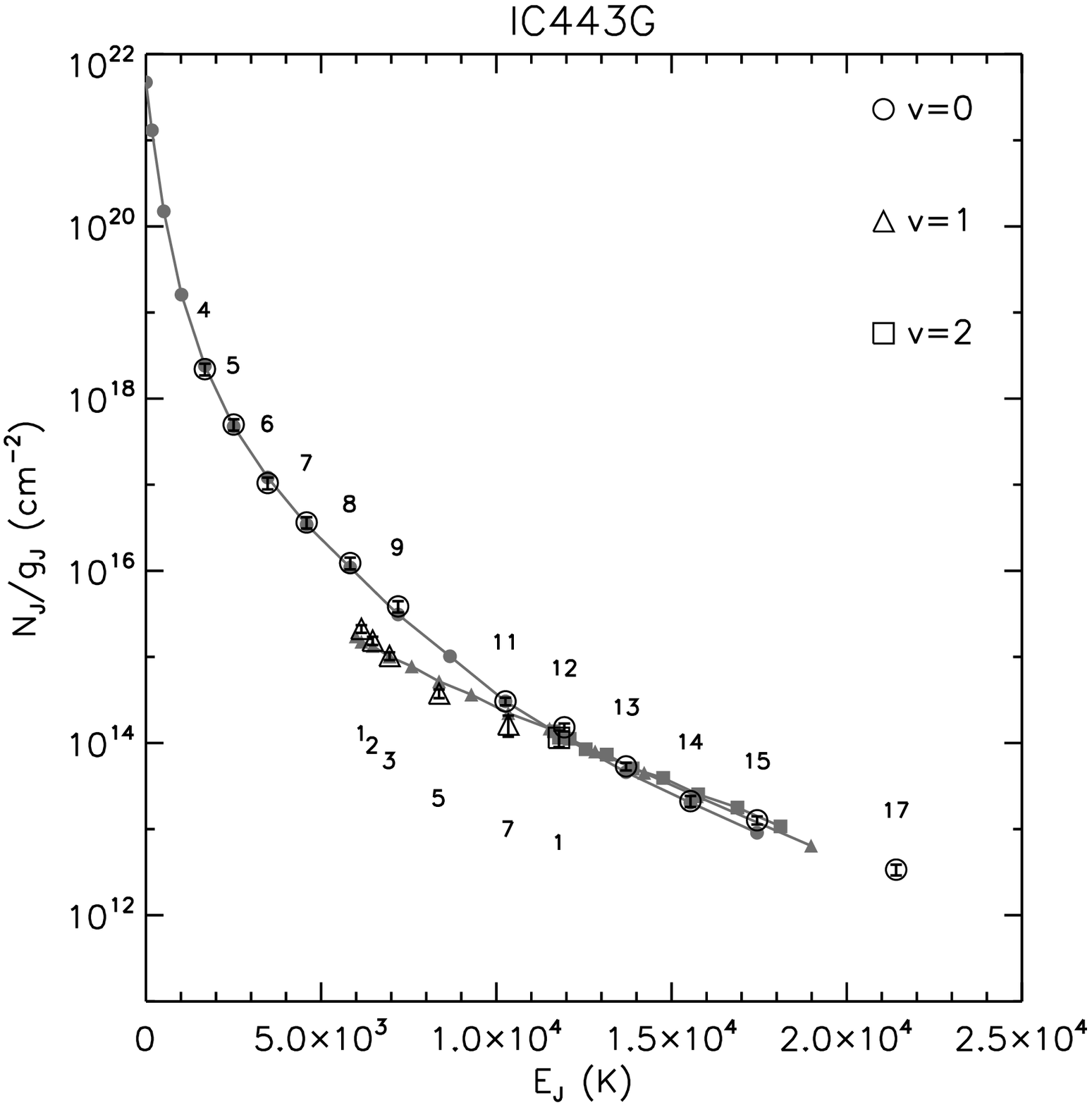}
\includegraphics[scale=0.46]{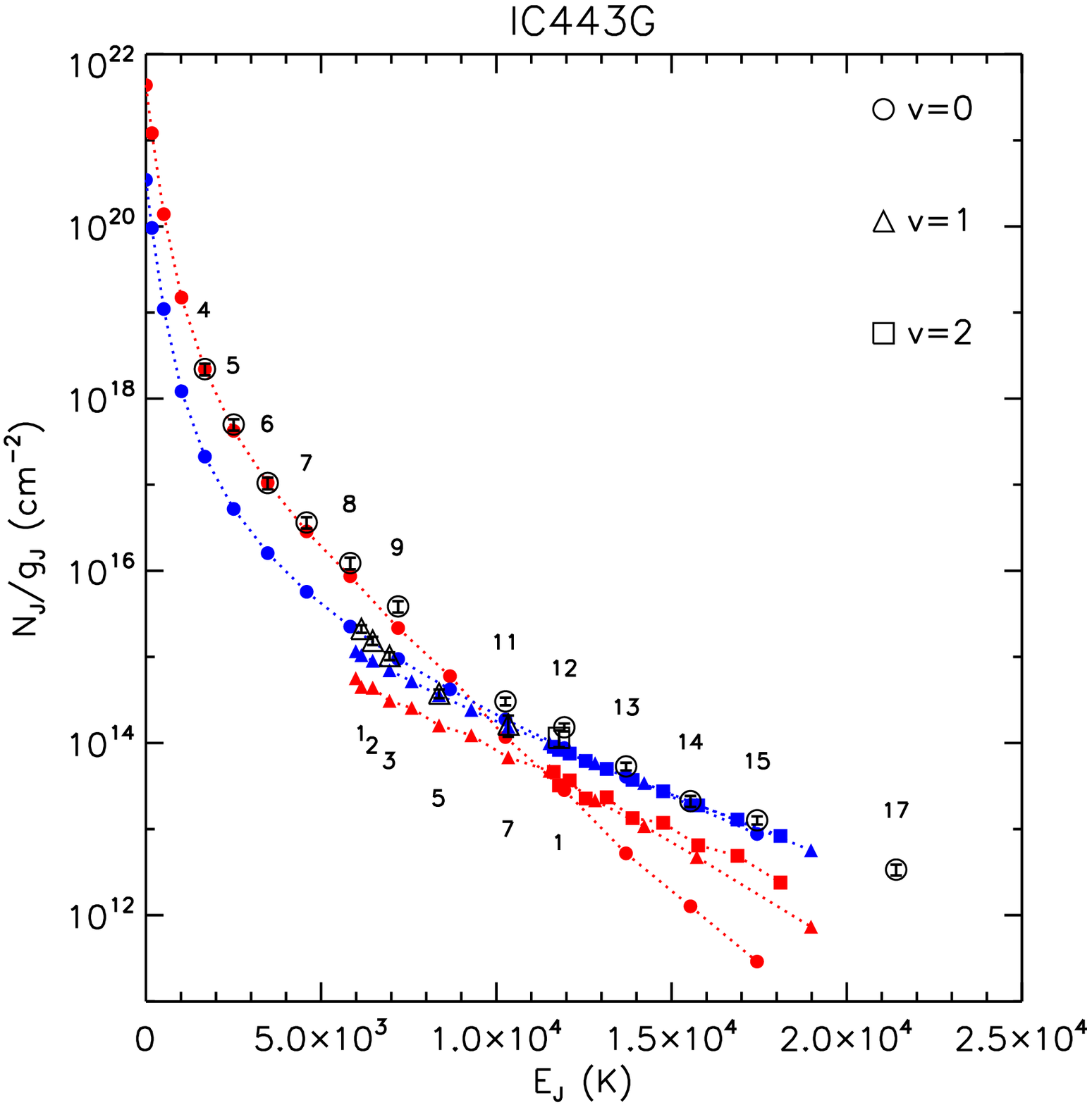}
}
\caption{The two-component model fitting results for the \Htwo{} population observed in the clump C (\emph{top panels}) and G (\emph{bottom panels}). The line-connected gray points in the \emph{left} panels are the total \Htwo{} population obtained from the model fitting, and their individual components are displayed in the \emph{right} panels as \emph{red (lower density)} and \emph{blue (higher density)} ones. (see text for the model description.) The rotational quantum number ($J$) is printed out near the corresponding point.} \label{fig-mfit}
\end{figure}

\clearpage
\begin{figure}
\center{
\includegraphics[scale=0.46]{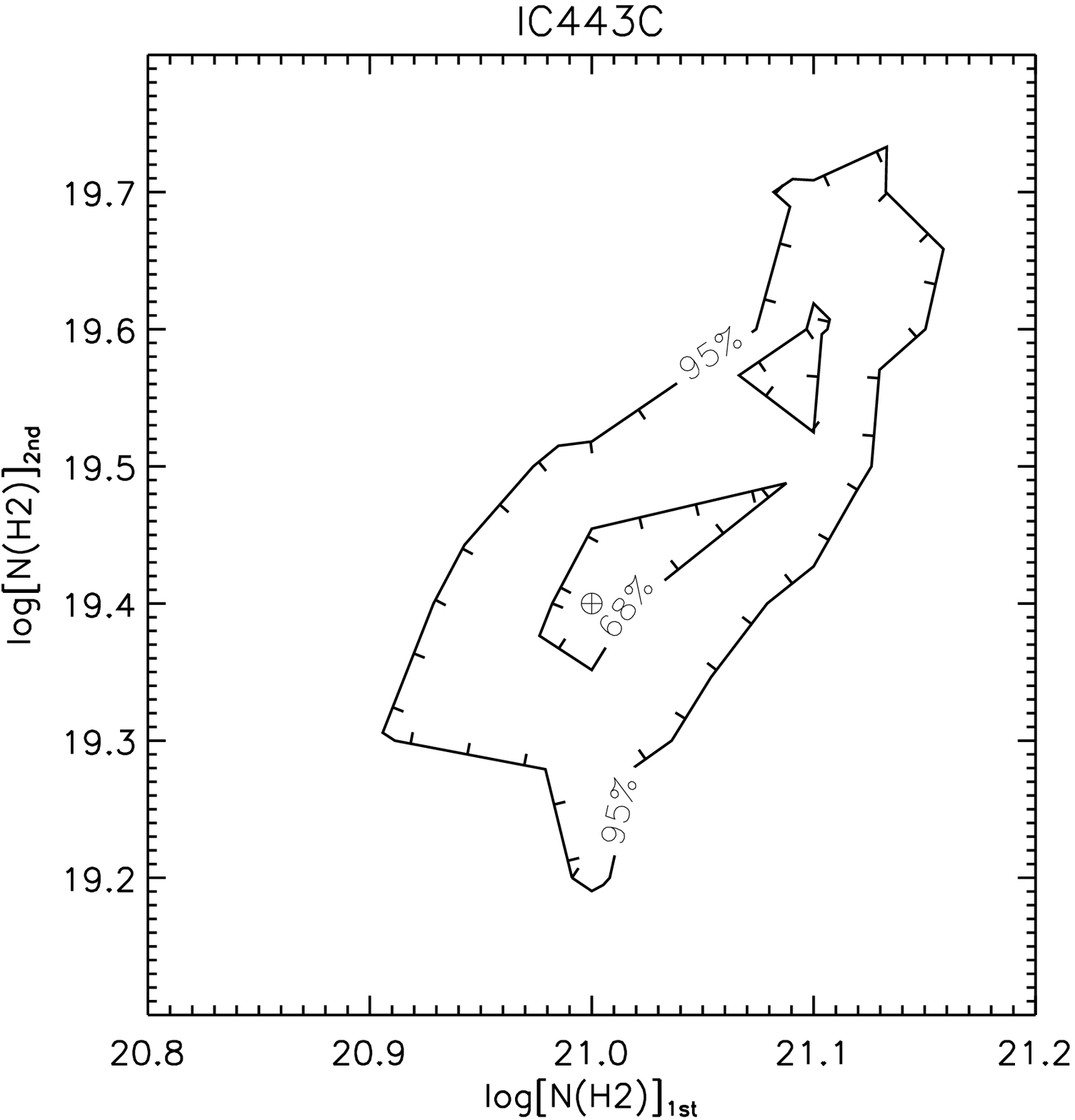}
\includegraphics[scale=0.46]{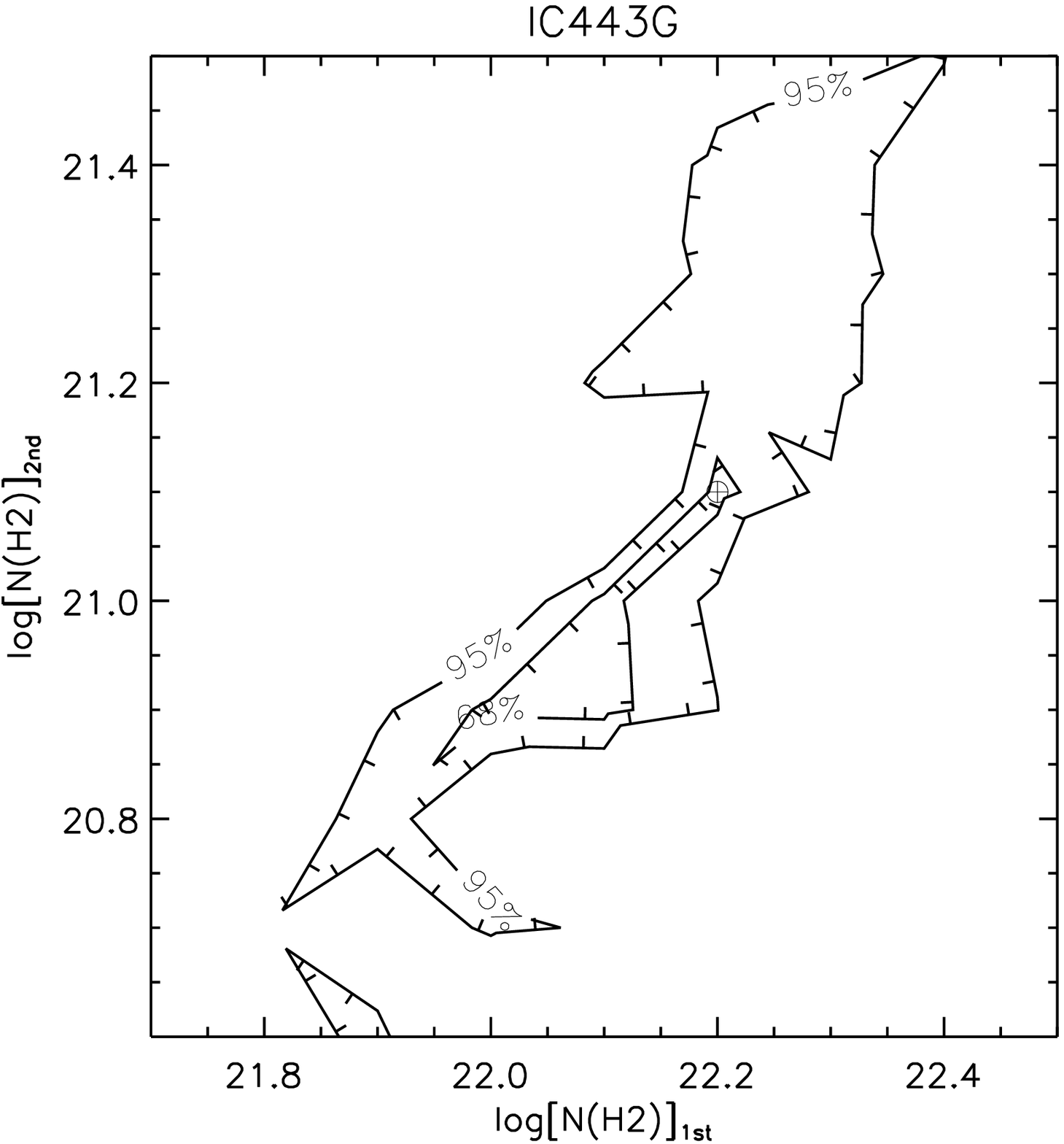} \\
\includegraphics[scale=0.46]{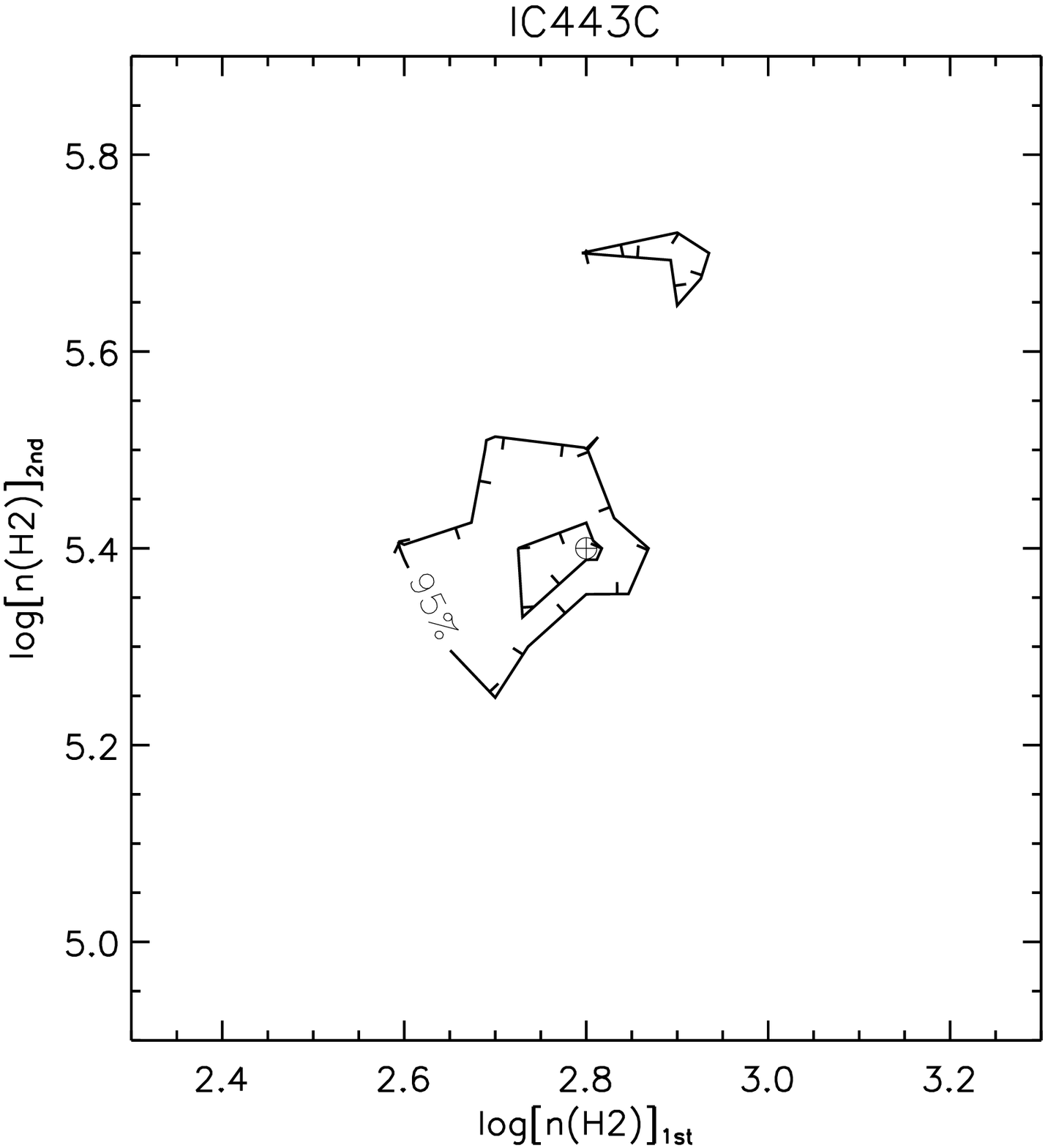}
\includegraphics[scale=0.46]{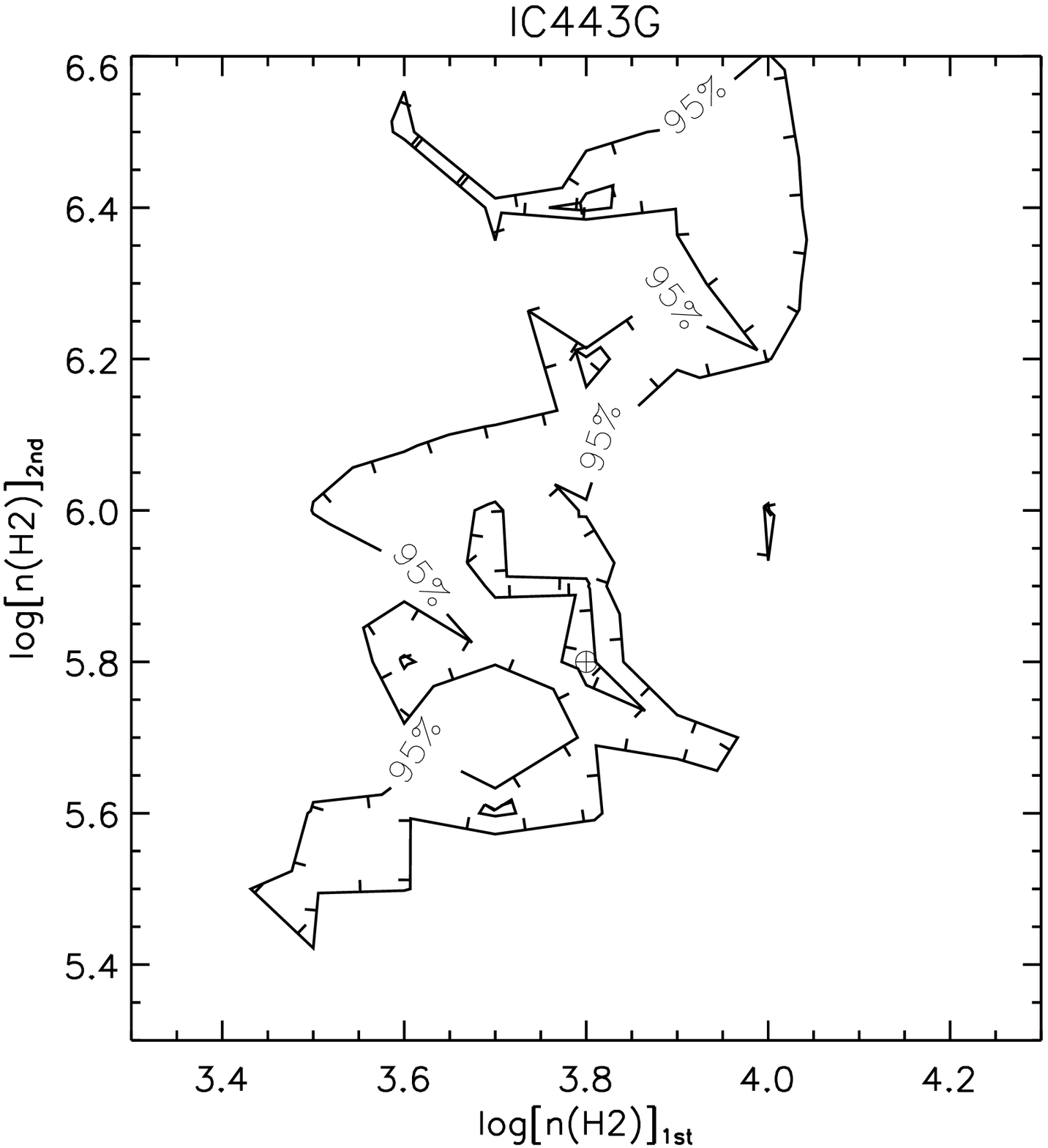} \\
}
\caption{The contour of $\chi^2$ in the plane of model parameters for the clump C (\emph{left panels}) and G (\emph{right panels}). The 68\% and 95\% confidence levels are outlined. The tick marks along the contours indicate the directions that $\chi^2$ values are decreasing. The `$\oplus$' indicates those model parameter values, whose $\chi^2$ is minimum, i.e. the best fit. In the \emph{right-bottom} panel, the $\chi^2$ values are plotted about the parameter $b$ only, since the $X_H$ is fixed (cf.~Table \ref{tbl-mfit}; the \emph{bottom} figures are displayed in the following page.). The \emph{dotted-line} indicates the best-fit parameter, and the \emph{dashed-lines} indicate the 68\% and 95\% confidence levels.} \label{fig-mconf}
\end{figure}

\clearpage
\begin{figure}
\figurenum{11}
\center{
\includegraphics[scale=0.46]{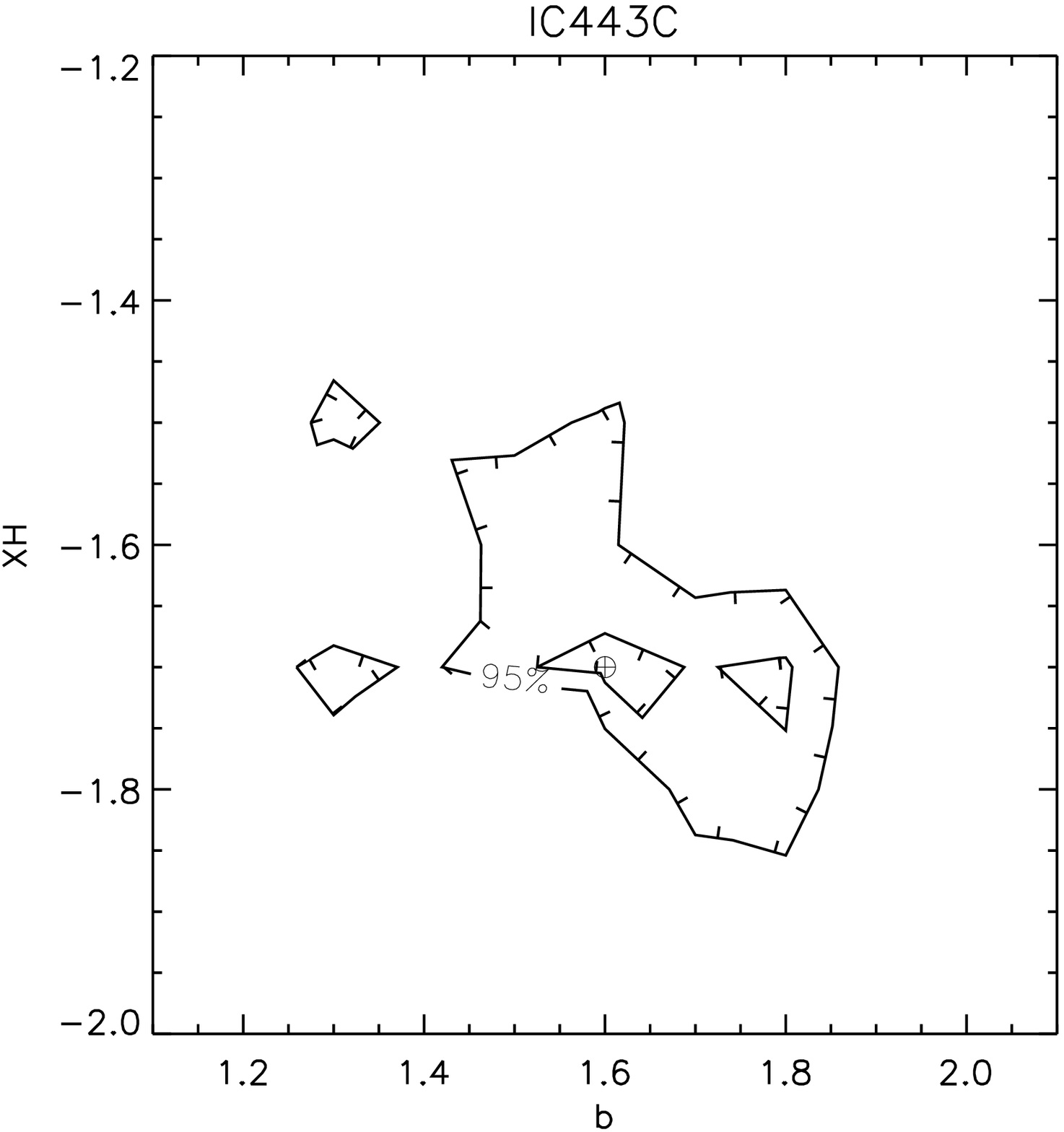}
\includegraphics[scale=0.46]{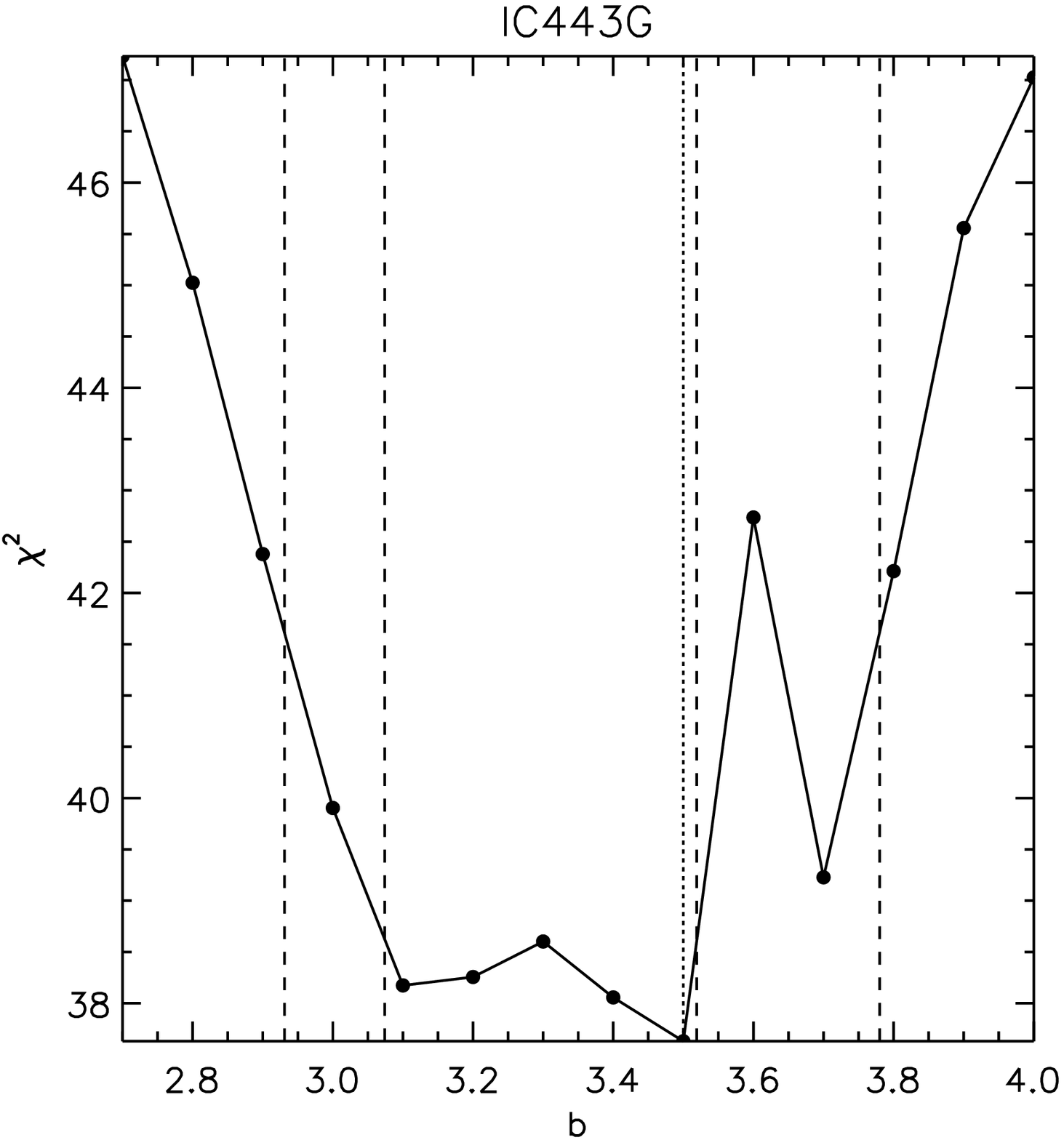}
}
\caption{Continued.}
\end{figure}

\clearpage
\begin{deluxetable}{cccccc}
\tablewidth{0pt}
\tablecaption{Summary of the \akari{} IRC Observations \label{tbl-obs}}
\tablehead{
\colhead{Region} &\colhead{Pointing Position} &\colhead{Observation ID} &\colhead{AOT\tablenotemark{a}} \\
&\colhead{(RA, Dec; J2000)} & }

\startdata
B	&(06:17:16.3, +22:25:41.0)	&1420803-00[1,2] &IRCZ4 \\
C	&(06:17:44.2, +22:21:49.1)	&1420804-00[1,2] &IRCZ4 \\
G	&(06:16:41.8, +22:31:41.0)	&1420805-00[1,2] &IRCZ4 \\
BG	&(06:17:30.0, +22:17:00.0)	&1420806-00[1,2] &IRCZ4 \\
\enddata
\tablenotetext{a}{Astronomical Observation Template. It is a pre-defined observation sequence. See the IRC Data User Manual for Post-Helium Mission. \citep{Onaka(2009)mana}.}
\end{deluxetable}

\begin{deluxetable}{lcrrrr}
\tabletypesize{\scriptsize}
\tablewidth{0pt}
\tablecaption{Observed \Htwo{} Emission Lines toward Each Region \label{tbl-result}}
\tablehead{
\colhead{Transition} & \colhead{Wavelength} & \multicolumn{4}{c}{Observed Intensity}\\
&($\mu$m)   &\multicolumn{4}{c}{($10^{-6}$ erg s$^{-1}$ cm$^{-2}$ sr$^{-1}$)}\\
\cline{3-6}
&   &\colhead{B}    &\colhead{C}    &\colhead{G}    &\colhead{BG}
}
\startdata

$\upsilon=$ 1-0 O(2)\tablenotemark{a}&   2.63&   $<$37.9$\pm$6.1&  $<$93.4$\pm$10.4&   $<$58.2$\pm$7.1&           \nodata\\
$\upsilon=$ 2-1 Q(9)\tablenotemark{a}&   2.72&           \nodata&   $<$25.1$\pm$5.3&   $<$13.5$\pm$4.2&           \nodata\\
$\upsilon=$ 1-0 O(3)&   2.80&      89.9$\pm$10.&     323.1$\pm$32.&     221.9$\pm$22.&           \nodata\\
$\upsilon=$ 2-1 O(3)&   2.97&           $<$13.5&      27.0$\pm$4.8&      18.5$\pm$4.1&           \nodata\\
$\upsilon=$ 1-0 O(4)&   3.00&      39.8$\pm$5.2&      96.7$\pm$10.&      63.0$\pm$6.9&           \nodata\\
$\upsilon=$ 1-0 O(5)&   3.23&      61.1$\pm$7.4&     196.2$\pm$20.&     127.0$\pm$13.&           \nodata\\
$\upsilon=$ 1-0 O(6)\tablenotemark{a}&   3.50&   $<$18.2$\pm$4.1&   $<$46.5$\pm$5.8&   $<$29.2$\pm$4.3&           \nodata\\
$\upsilon=$ 0-0 S(15)&   3.63&           \nodata&      47.8$\pm$5.6&      24.1$\pm$3.4&           \nodata\\
$\upsilon=$ 0-0 S(14)\tablenotemark{a}&   3.72&           \nodata&   $<$26.8$\pm$3.9&   $<$10.7$\pm$2.6&           \nodata\\
$\upsilon=$ 1-0 O(7)&   3.81&      18.9$\pm$3.4&      54.1$\pm$6.1&      36.7$\pm$4.2&           \nodata\\
$\upsilon=$ 0-0 S(13)&   3.85&      27.6$\pm$3.9&      96.0$\pm$10.&      53.1$\pm$5.7&           \nodata\\
$\upsilon=$ 0-0 S(12)&   4.00&           \nodata&      38.5$\pm$4.6&      21.5$\pm$3.1&           \nodata\\
$\upsilon=$ 1-1 S(13)\tablenotemark{a}&   4.07&           \nodata&   $<$24.1$\pm$3.5&   $<$14.9$\pm$2.7&           \nodata\\
$\upsilon=$ 0-0 S(11)&   4.18&      60.0$\pm$6.8&     199.3$\pm$20.&     113.4$\pm$11.&           \nodata\\
$\upsilon=$ 0-0 S(10)&   4.41&      40.1$\pm$5.2&     122.8$\pm$12.&      71.2$\pm$7.6&           \nodata\\
$\upsilon=$ 1-0 O(9)&   4.58&           \nodata&      16.0$\pm$3.2&       9.6$\pm$2.6&           \nodata\\
$\upsilon=$ 0-0 S(9)&   4.69&     134.4$\pm$14.&     430.8$\pm$43.&     265.7$\pm$27.&      14.2$\pm$3.5\\

\enddata
\tablenotetext{a}{These lines may be blended with nearby lines, hence we indicated with the `$<$' sign.}
\tablecomments{For those lines whose significance is lower than 3.0, the intensities are expressed with 90\% upper confidence limits. The error includes both statistical and systematic components. See text for detail.}
\end{deluxetable}

\begin{deluxetable}{lrrrrr}
\tabletypesize{\scriptsize}
\tablewidth{0pt}
\tablecaption{Extinction Corrected \Htwo{} Column Density toward Each Region \label{tbl-h2col}}
\tablehead{
\colhead{State} & \colhead{Energy Level} & \multicolumn{4}{c}{log N(\Htwo; $\upsilon,J$)}\\
($\upsilon,J$)  & \colhead{(K)} &\multicolumn{4}{c}{(cm$^{-2}$)}\\
\cline{3-6}
&   &\colhead{B}    &\colhead{C}    &\colhead{G}    &\colhead{BG}
}
\startdata

(0,11)&    10261.&          16.06$\pm$0.05&          16.56$\pm$0.04&          16.32$\pm$0.04&          15.08$\pm$0.11\\
(0,12)&    11940.&          15.36$\pm$0.06&          15.85$\pm$0.04&          15.58$\pm$0.05&                 \nodata\\
(0,13)&    13703.&          15.40$\pm$0.05&          15.92$\pm$0.04&          15.64$\pm$0.04&                 \nodata\\
(0,14)&    15540.&                 \nodata&          15.08$\pm$0.05&          14.79$\pm$0.06&                 \nodata\\
(0,15)&    17443.&          14.83$\pm$0.06&          15.37$\pm$0.05&          15.07$\pm$0.05&                 \nodata\\
(0,16)\tablenotemark{a}&    19403.&                 \nodata&       $<$14.72$\pm$0.06&       $<$14.28$\pm$0.10&                 \nodata\\
(0,17)&    21411.&                 \nodata&          14.89$\pm$0.05&          14.55$\pm$0.06&                 \nodata\\
(1, 0)\tablenotemark{a}&     5987.&       $<$15.31$\pm$0.07&       $<$15.70$\pm$0.05&       $<$15.41$\pm$0.05&                 \nodata\\
(1, 1)&     6149.&          15.96$\pm$0.05&          16.52$\pm$0.04&          16.28$\pm$0.04&                 \nodata\\
(1, 2)&     6471.&          15.76$\pm$0.06&          16.14$\pm$0.05&          15.89$\pm$0.05&                 \nodata\\
(1, 3)&     6951.&          16.07$\pm$0.05&          16.58$\pm$0.04&          16.33$\pm$0.05&                 \nodata\\
(1, 4)\tablenotemark{a}&     7584.&       $<$15.68$\pm$0.10&       $<$16.09$\pm$0.05&       $<$15.84$\pm$0.06&                 \nodata\\
(1, 5)&     8365.&          15.85$\pm$0.08&          16.30$\pm$0.05&          16.09$\pm$0.05&                 \nodata\\
(1, 7)&    10341.&                 \nodata&          16.12$\pm$0.09&          15.87$\pm$0.12&                 \nodata\\
(1,15)\tablenotemark{a}&    22516.&                 \nodata&       $<$14.84$\pm$0.06&       $<$14.60$\pm$0.08&                 \nodata\\
(2, 1)&    11789.&                $<$14.94&          15.25$\pm$0.08&          15.01$\pm$0.10&                 \nodata\\
(2, 9)\tablenotemark{a}&    18107.&                 \nodata&       $<$15.56$\pm$0.09&       $<$15.21$\pm$0.14&                 \nodata\\

\enddata
\tablenotetext{a}{The population cannot be not determined because of the probable line blending with nearby lines. cf.~Table \ref{tbl-result}.}
\tablecomments{The extinctions are corrected, employing the ``Milky Way'' extinction curve ($R_V=3.1$, \citealt{Weingartner(2001)ApJ_548_296}), with these parameters: $A_V=13.5$ for clump B, clump C and BG \citep{Burton(1988)MNRAS_231_617,Neufeld(2008)ApJ_678_974} and $A_V=10.8$ for clump G \citep{Richter(1995)ApJ_454_277}, respectively.}
\end{deluxetable}

\begin{deluxetable}{ccccccccc}
\tabletypesize{\scriptsize}
\tablewidth{0pt}
\tablecaption{Fitting Results for the Model Parameters \label{tbl-mfit}}
\tablehead{
&\multicolumn{2}{c}{component 1} & &\multicolumn{2}{c}{component 2} \\
\cline{2-3} \cline{5-6}
\colhead{Region} & \colhead{log[$N$(H$_2$)]} & \colhead{log[$n$(H$_2$)]} & & \colhead{log[$N$(H$_2$)]} & \colhead{log[$n$(H$_2$)]} & \colhead{$b$} & \colhead{X$_H$} & \colhead{${\chi}^2_{\nu}$} \\
 & \colhead{(cm$^{-2}$)} & \colhead{(cm$^{-3}$)} & & \colhead{(cm$^{-2}$)} & \colhead{(cm$^{-3}$)} &  & \colhead{$\left(\equiv\textrm{log}\left[\frac{n(\textrm{\tiny H I})}{n(\textrm{\tiny H}_2)}\right]\right)$} & \colhead{$(\equiv\chi^2$/d.o.f)}
}
\startdata

IC443C&$21.0_{-0.1}^{+0.2}$&$2.8_{-0.2}^{+0.1}$&&$19.4_{-0.2}^{+0.3}$&$5.4_{-0.2}^{+0.3}$&$1.6_{-0.3}^{+0.3}$&$-1.7_{-0.2}^{+0.2}$&5.1 (=66.1/13.0)
\\
IC443G&$22.2_{-0.4}^{+0.2}$&$3.8_{-0.4}^{+0.2}$&&$21.1_{-0.5}^{+0.4}$&$5.8_{-0.4}^{+0.8}$&$3.5_{-0.6}^{+0.3}$&$-1.7$\tablenotemark{a}&3.4 (=40.6/12.0)
\\

\enddata
\tablenotetext{a}{In this case, the parameter $X_H$ is fixed as $-1.7$, to increase the degree of freedom. See section \ref{ana-res-plmod} for more detail.}
\tablecomments{The confidence limits are given with a 95\% significance (cf.~Fig.~\ref{fig-mconf}). \NHtwo{} means $N$(\Htwo; $T>100$ K). See section \ref{ana-res-plmod} for the detailed description about the parameters. \NHtwo{} of the component 2 is not firmly determined value, since it sensitively depends on the $T_{min}$ of the model (see section \ref{dis-plmod-cmp}).}
\end{deluxetable}

\begin{deluxetable}{cccccc}
\tablewidth{0pt}
\tablecaption{Obtained \nHtwo{} and $b$ from Previous Studies \label{tbl-nh2b}}
\tablehead{
\colhead{Target} & \colhead{log[$n$(H$_2$)]} & \colhead{$b$} & \colhead{Instrument} & \colhead{Estimated Levels} & \colhead{Ref.} \\
 & \colhead{(cm$^{-3}$)} &  & & \colhead{}
}
\startdata
SNR IC 443			&$5.0-7.0$	&$3.0-6.0$	&\spitzer{} IRAC	&$\upsilon=0,J=6-15$	&1	\\
SNR HB 21 - Cloud N	&$2.7-3.3$	&$2.9$		&\akari{} IRC		&$\upsilon=0,J=3-8$		&2	\\
SNR HB 21 - Cloud S	&$4.6$		&$4.2$		&\akari{} IRC		&$\upsilon=0,J=4-13$	&3	\\
OMC-1				&$3.5$		&$1.9$		&\akari{} IRC		&$\upsilon=0,J=3-8$		&3	\\
OMC-1				&$6.0$		&$3.7$		&\akari{} IRC		&$\upsilon=0,J=4-13$	&3	\\
outflows of YSOs	&$3.5-3.8$	&$2.3-3.3$	&\spitzer{} IRS		&$\upsilon=0,J=2-9$		&4	\\
outflows of YSOs	&$\gtrsim7.0$	&$3.0-5.5$	&\spitzer{} IRAC	&$\upsilon=0,J=6-15$&5\tablenotemark{a}	\\
outflows of YSOs	&$\gtrsim5.0$	&$3.0-6.0$	&\spitzer{} IRAC	&$\upsilon=0,J=6-15$&6	\\
SNRs and outflows of YSOs	&$3.3-3.6$	&$2.3-3.1$	&\spitzer{} IRS	&$\upsilon=0,J=2-9$&7	\\
\enddata
\tablenotetext{a}{The central sources (position C and E) are excluded since \Htwo{} may not be the only contribution to the IRAC bands.}
\tablecomments{1: \cite{Neufeld(2008)ApJ_678_974}, 2: \cite{Shinn(2009)ApJ_693_1883}, 3: \cite{Shinn(2010)AdSpR_45_445}, 4: \cite{Neufeld(2009)ApJ_706_170}, 5: \cite{Lee(2010)ApJ_709_L74},  6: \cite{Takami(2010)ApJ_720_155}, 7: \cite{Yuan(2011)ApJ_726_76}. The modeled gas consists of \Htwo{} and He only, with \nHtwo=0.2 \nHe.}
\end{deluxetable}

\end{document}